\documentclass[12pt]{article}

\pdfoutput=1

\usepackage{jheppub}
\usepackage{graphicx,epsfig,amsmath,amssymb}
\usepackage{subcaption}
\usepackage{multirow}
\usepackage{slashed}
\usepackage{cleveref}
\usepackage{enumitem}
\usepackage{calc}
\usepackage{placeins}
\usepackage[compat=1.1.0]{tikz-feynman}
\usetikzlibrary{positioning}
\usepackage{xcolor}

\newcommand{\amp}{\ensuremath{\mathcal{M}}}

\newcommand{\msbar}{\ensuremath{\overline{\mathrm{MS}}}}

\def\eps{\epsilon}

\def\be{\begin{equation}}
\def\ee{\end{equation}}
\def\bea{\begin{eqnarray}}
\def\eea{\end{eqnarray}}

\newcommand{\mhh}{m_{hh}}

\pgfmathsetmacro\sizedot{1.1}
\pgfmathsetmacro\sizesqdot{2.0}
\pgfmathsetmacro\sizerectangle{0.8}
\pgfmathsetmacro\sizecirc{0.55}
\pgfmathsetmacro\sizecrodot{1.0}
\newcommand{\lentriborn}{28}
\newcommand{\lenboxborn}{40}

\newcommand{\lentrichromo}{28}
\newcommand{\lenboxchromo}{40}

\newcommand{\lentrismall}{20}

\newcommand{\lentriftop}{28}
\newcommand{\lenboxftop}{40}

\newcommand{\lentrieq}{20}
\newcommand{\lenboxeq}{28}

\title{Combining chromomagnetic and four-fermion operators with leading
SMEFT operators for $gg\to hh$ at NLO QCD}
\author[a]{Gudrun Heinrich,}
\author[a]{Jannis Lang}

\affiliation[a]{Institute for Theoretical Physics, Karlsruhe Institute of Technology (KIT), \\Wolfgang-Gaede-Str.\,1, 76131 Karlsruhe, Germany}

\emailAdd{gudrun.heinrich@kit.edu}
\emailAdd{jannis.lang@kit.edu}

\preprint{{\small KA-TP-29-2023\\
    \hphantom{.}\hfill  P3H-23-095}}

\abstract{
  We present the calculation of the contribtuions from the
  chromomagnetic and four-top-quark-operators within Standard Model Effective Field Theory (SMEFT)
  to Higgs boson pair production in gluon fusion, 
  combined with QCD corrections that are at NLO with full $m_t$-dependence for the leading operators. 
  We study the effects of  these operators on the total cross section and the invariant mass distribution of the Higgs-boson pair, at $\sqrt{s}=13.6$\,TeV. 
 These subleading operators are implemented in the generator
 \texttt{ggHH\_SMEFT}, in the same \texttt{Powheg-Box-V2} framework as
 the leading operators, such that their effects can be easily studied
 in a unified setup.
}


\begin{document}

\maketitle

\section{Introduction}

Where is New Physics? If it resides at energy scales well separated from the electroweak scale, our ignorance about its exact nature can be parametrised within an Effective Field Theory (EFT) framework~\cite{Weinberg:1978kz,Buchmuller:1985jz,Brivio:2017vri,Isidori:2023pyp}.

Predictions for key LHC processes within Standard Model Effective Field Theory (SMEFT)~\cite{Grzadkowski:2010es} up to the level of dimension-6 operators, in combination with NLO QCD corrections, have become available in the last few years, see e.g. Refs.~\cite{Mimasu:2015nqa,Alioli:2018ljm,Degrande:2020evl,Baglio:2020oqu,Battaglia:2021nys,Dawson:2021ofa,Brivio:2021xhp,Heinrich:2022idm,Buchalla:2022igv}.
In addition, the importance of renormalisation group running effects of the Wilson coefficients, calculated up to one loop in Refs.~\cite{Jenkins:2013zja,Jenkins:2013wua,Alonso:2013hga}, has gained increasing attention~\cite{Battaglia:2021nys,Aoude:2022aro,Chala:2021wpj,Dawson:2022ewj,Fuentes-Martin:2023ljp,DiNoi:2023onw,DasBakshi:2024krs} and is implemented in dedicated tools~\cite{Aebischer:2018bkb,Fuentes-Martin:2020zaz,DiNoi:2022ejg,Fuentes-Martin:2022jrf,Machado:2022ozb,Martin:2023fad,Dedes:2023zws,terHoeve:2023pvs}.
The effect of double insertions of dimension-6 operators at the level of squared amplitudes also has been studied in the literature~\cite{Dawson:2021xei,Ellis:2022zdw,Alioli:2022fng,GomezAmbrosio:2022mpm,Allwicher:2022gkm,Asteriadis:2022ras}.

Here we will focus on Higgs boson pair production in gluon fusion, combining the NLO QCD corrections with full top quark mass dependence with anomalous couplings within SMEFT.
The full NLO QCD corrections have been calculated in Refs.~\cite{Borowka:2016ehy,Borowka:2016ypz,Baglio:2018lrj,Baglio:2020ini}, based on numerical evaluations of the two-loop integrals entering the virtual corrections.
The results of \cite{Borowka:2016ehy} have been implemented into the {\tt Powheg-Box-V2} event generator~\cite{Nason:2004rx,Frixione:2007vw,Alioli:2010xd}, first for the SM only~\cite{Heinrich:2017kxx}, then also for $\kappa_\lambda$ variations~\cite{Heinrich:2019bkc} as well as for the leading operators contributing to this process in non-linear EFT (HEFT)~\cite{Buchalla:2018yce,Heinrich:2020ckp} and SMEFT~\cite{Heinrich:2022idm}.
Recently, the NLO QCD corrections obtained from the combination of a $p_T$-expansion and an expansion in the high-energy regime have been calculated analytically and implemented in the {\tt Powheg-Box-V2}~\cite{Bagnaschi:2023rbx}, allowing to study top mass scheme uncertainties in an event generator framework.

In Ref.~\cite{deFlorian:2017qfk} the combination of NNLO corrections in an $m_t$-improved heavy top limit (HTL)
has been performed including anomalous couplings, extending earlier work at NLO in the $m_t$-improved HTL~\cite{Grober:2015cwa,Grober:2017gut}.
The work of~\cite{deFlorian:2017qfk}  has been combined with the full NLO corrections within non-linear EFT of
Ref.~\cite{Buchalla:2018yce}  to provide approximate NNLO predictions in Ref.~\cite{deFlorian:2021azd}, dubbed
NNLO$^\prime$, which include the full top-quark mass dependence up to NLO and
higher order corrections up to NNLO in the $m_t$-improved HTL, combined with operators related
to the five most relevant anomalous couplings for the process $gg\to hh$.
Recently, full NLO electroweak corrections have been computed in Ref.~\cite{Bi:2023bnq}, following the emergence of previous partial results,
i.e.\ the full NLO electroweak corrections in the large-$m_t$ limit~\cite{Davies:2023npk}, the NLO Yukawa corrections in the high-energy limit~\cite{Davies:2022ram} and Yukawa corrections in the (partial) large-$m_t$ limit~\cite{Muhlleitner:2022ijf}.

In this paper, we investigate the effect of two classes of operators 
contributing at dimension-6 level to the process $gg\to hh$, which however are suppressed by loop factors compared to the leading operators considered in Ref.~\cite{Heinrich:2022idm} 
when the potential UV completion is assumed to be a weakly coupling and renormalisable quantum field theory. 
These are the chromomagnetic operator and 4-top-operators.
As has been shown in Ref.~\cite{DiNoi:2023ygk} for the case of single Higgs production, the latter are intricately related since they are individually $\gamma_5$-scheme dependent, the scheme dependence only dropping out when they are consistently combined in a renormalised amplitude.
Apart from the $\gamma_5$ continuation scheme, other  sources of scheme differences in bottom-up SMEFT calculations also have been studied recently~\cite{Corbett:2021cil,Martin:2023fad,Aebischer:2023djt}.

The subsequent sections are organised as follows:
in Section~\ref{sec:operators}, we describe these contributions and their scheme dependence in detail.
Their implementation into the {\tt POWHEG ggHH\_SMEFT} generator is described in Section~\ref{sec:usage}, together with instructions for the user how to turn them on or off.
Section~\ref{sec:results} contains our phenomenological results, focusing on the effects of these newly included operators on the total cross section and on the Higgs boson pair invariant mass distribution, before we summarise and conclude.


\section{Contributions of the chromomagnetic and  four-top operators}
\label{sec:operators}

In this section we describe our selection of contributing operators. Subsequently we recapitulate the power counting scheme for SMEFT and discuss the
new contributions in detail, which will be identified as subleading 
based on the minimal assumption of a weakly coupling and renormalisable UV theory.

Any bottom-up EFT  is defined by its degrees of freedom, 
the imposed symmetries and a power counting scheme.
Since SMEFT builds upon the SM, the above specifications are given by the field content and gauge symmetries of the SM and the main power counting, 
which relies on the counting of the canonical (mass) dimension. 
Due to strong experimental constraints 
it is common to exclude baryon and lepton number violating operators,
hence only operators of even dimension are considered. 
Therefore, the dominant contributions are expected to be described by dimension-6 operators, on which we focus our attention in this paper. 
To further cut down the number of operators,\footnote{
    A complete basis for the dimension-6 operators in full generality of the flavour sector includes 2499 real parameters~\cite{Grzadkowski:2010es}, 
    with a large subset potentially contributing to the considered process. 
    Thus, for a first study as presented here, making a further selection based on phenomenologically motivated flavour assumptions  appears to be necessary.
}
we impose an exact flavour symmetry $U(2)_q\times U(2)_u\times U(3)_d$ in the quark sector for a first investigation, which forbids chirality flipping bilinears involving light quarks ($b$-quarks included) and right-handed charged currents~\cite{Greljo:2022cah,Ethier:2021bye,Degrande:2020evl}. 
This effectively makes the CKM matrix diagonal and sets all fermion masses and Yukawa couplings to zero, with the top quark as the only exception, 
thus being well compatible with a 5-flavour scheme in QCD which we employ.
In addition, this flavour choice reflects the expected prominent role of the top quark in many BSM scenarios and 
could be a starting point for a 
spurion expansion as in
minimal flavour violation~\cite{DAmbrosio:2002vsn,Greljo:2022cah}.

We also neglect operators whose contributions involve only diagrams with electroweak particles propagating in the loop.
In principle, electroweak corrections and such electroweak-like operator contributions
can be of the same order in the power counting as the subleading contributions studied in this paper.
    In addition, the close connection between operators of class $\psi^2\phi^2D$ of Ref.~\cite{Grzadkowski:2010es} 
    and $C_{tG}$, observed by the structure of the $\gamma_5$-scheme dependence in Ref.~\cite{DiNoi:2023ygk},
    demonstrates that our subset does not fully comprise a consistent subleading order in a systematic power counting.
    Nevertheless, we expect it to be useful to investigate the sensitivity of the process $gg\to hh$ to the chromomagnetic operator and 4-top operators in
    the presented form, especially since even in the simpler case of the SM, full electroweak effects to $gg\to hh$ have only become available very recently~\cite{Bi:2023bnq}.
 
With these restrictions, all dimension-6 CP even operators that contribute to $gg\to hh$ are given by
\begin{equation}
\begin{split}
\mathcal{L}_{\text{SMEFT}}&\supset
\frac{C_{H\Box}}{\Lambda^2} (\phi^{\dagger} \phi)\Box (\phi^{\dagger } \phi)
+ \frac{C_{H D}}{\Lambda^2}(\phi^{\dagger} D_{\mu}\phi)^*(\phi^{\dagger}D^{\mu}\phi)
+ \frac{C_H}{\Lambda^2} (\phi^{\dagger}\phi)^3 
\\ &
+\frac{C_{tH}}{\Lambda^2} \left( \phi^{\dagger}{\phi}\bar{Q}_L\tilde{\phi} t_R + {\rm h.c.}\right)
+\frac{C_{H G}}{\Lambda^2} \phi^{\dagger} \phi G_{\mu\nu}^a G^{\mu\nu,a}
\\ &
+\frac{C_{tG}}{\Lambda^2} 
\left(\bar Q_L\sigma^{\mu\nu}T^aG_{\mu\nu}^a\tilde\phi t_R +{\rm h.c.}\right)
\\ &
+\frac{C_{Qt}^{(1)}}{\Lambda^2}\bar{Q}_L\gamma^\mu Q_L\bar{t}_R\gamma_\mu t_R
+\frac{C_{Qt}^{(8)}}{\Lambda^2}\bar{Q}_L\gamma^\mu T^aQ_L\bar{t}_R\gamma_\mu T^a t_R
\\ &
+\frac{C_{QQ}^{(1)}}{\Lambda^2}\bar{Q}_L\gamma^\mu Q_L\bar{Q}_L\gamma_\mu Q_L
+\frac{C_{QQ}^{(8)}}{\Lambda^2}\bar{Q}_L\gamma^\mu T^aQ_L\bar{Q}_L\gamma_\mu T^a Q_L
\\ &
+\frac{C_{tt}}{\Lambda^2}\bar{t}_R\gamma^\mu t_R\bar{t}_R\gamma_\mu t_R
\;,  \label{eq:LagSMEFT}
\end{split}
\end{equation}
where $\sigma^{\mu \nu}=\frac{i}{2}\left[\gamma^{\mu},\gamma^{\nu}\right]$ and $\tilde{\phi}=i\sigma_2\phi$ is the charge conjugate of the Higgs doublet. 
For the covariant derivative, we use the sign convention\footnote{
    Note that the sign of $C_{tG}$ is sensitive to the convention of the covariant 
    derivative. This is more apparent when a factor of $g_s$ is extracted, i.e.\ 
    $C_{tG}=g_s\tilde{C}_{tG}$, which is for example the case in the basis definition of Ref.~\cite{Ethier:2021bye}.
}
\begin{equation}
D_\mu=\partial_\mu-i g_s T^a G_\mu^a\;,
\end{equation}
in order to be compatible with {\tt FeynRules}~\cite{Christensen:2008py,Alloul:2013bka} 
conventions and tools relying on {\tt UFO}~\cite{Degrande:2011ua,Darme:2023jdn} models.
The first two lines in Eq.~\eqref{eq:LagSMEFT} comprise the leading EFT contribution which has been studied in Ref.~\cite{Heinrich:2022idm}. 
For convenience of the reader and later reference, we show the Born-level diagrams related to those operators in Fig.~\ref{fig:feyndiag_born}.
\begin{figure}[ht]
\begin{center}
\begin{minipage}[c]{0.275\textwidth}
    \centering
\begin{tikzpicture} 
    \begin{feynman}[small]
        \vertex  (g1)  {$g$};
        \vertex  (gtt1) [dot,scale=\sizedot,right=\lentriborn pt of g1] {};
        \vertex  (htt1) [dot,scale=\sizesqdot,right=\lenboxborn pt of gtt1,color=gray]  {};
        \vertex  (gtt2) [dot,scale=\sizedot,below=\lenboxborn pt of gtt1] {};
        \vertex  (htt2) [dot,scale=\sizesqdot,below=\lenboxborn pt of htt1,color=gray] {};
        \vertex  (g2) [left=\lentriborn pt of gtt2]  {$g$};
        \vertex (h1) [right =\lentriborn pt of htt1] {$h$};
        \vertex (h2) [right =\lentriborn pt of htt2] {$h$};
        \diagram* {
            (g1)  -- [gluon] (gtt1),
            (g2) -- [gluon] (gtt2),
            (h2)  -- [scalar] (htt2),
            (h1)  -- [scalar] (htt1),
            (gtt1) -- [fermion, line width = 1.5 pt] (htt1)  -- [fermion, line width = 1.5 pt] (htt2)
            -- [fermion, line width = 1.5 pt] (gtt2)
             -- [fermion, line width = 1.5 pt] (gtt1), 
        };
    \end{feynman}
\end{tikzpicture}
    \caption*{(a)}
\end{minipage}
\begin{minipage}[c]{0.275\textwidth}
    \centering
\begin{tikzpicture} 
    \begin{feynman}[small]
        \vertex  (g1)  {$g$};
        \vertex  (gtt1) [dot, scale=\sizedot,right=\lentriborn pt of g1] {};
        \vertex (htt) [dot,scale=\sizesqdot,below right = \lentriborn pt of gtt1,color=gray]  {};
        \vertex  (gtt2) [dot,scale=\sizedot,below left =\lentriborn pt of htt] {};
        \vertex  (g2) [left=\lentriborn pt of gtt2]  {$g$};
        \vertex (hhh) [dot,scale=\sizesqdot,right =\lentriborn pt of htt,color=gray]  {};
        \vertex (h1) [above right =\lenboxborn pt of hhh] {$h$};
        \vertex (h2) [below right =\lenboxborn pt of hhh] {$h$};
        \diagram* {
            (g1)  -- [gluon] (gtt1),
            (g2) -- [gluon] (gtt2),
            (htt) -- [scalar] (hhh) --[scalar] (h1),
            (h2)  -- [scalar] (hhh),
            (gtt1) -- [fermion, line width = 1.5 pt] (htt) 
            -- [fermion, line width = 1.5 pt] (gtt2)
             -- [fermion, line width = 1.5 pt] (gtt1), 
        };
    \end{feynman}
\end{tikzpicture}
\caption*{(b)}
\end{minipage}
\begin{minipage}[c]{0.275\textwidth}
    \centering
\begin{tikzpicture} 
    \begin{feynman}[small]
        \vertex  (g1)  {$g$};
        \vertex  (gtt1) [dot, scale=\sizedot,right=\lentriborn pt of g1] {};
        \vertex (hhtt) [dot,scale=\sizesqdot,below right = \lentriborn pt of gtt1,color=gray]  {};
        \vertex  (gtt2) [dot,scale=\sizedot,below left =\lentriborn pt of htt] {};
        \vertex  (g2) [left=\lentriborn pt of gtt2]  {$g$};
        \vertex (h1) [above right =\lenboxborn pt of hhtt] {$h$};
        \vertex (h2) [below right =\lenboxborn pt of hhtt] {$h$};
        \diagram* {
            (g1)  -- [gluon] (gtt1),
            (g2) -- [gluon] (gtt2),
            (h2)  -- [scalar] (hhtt) --[scalar] (h1),
            (gtt1) -- [fermion, line width = 1.5 pt] (hhtt) 
            -- [fermion, line width = 1.5 pt] (gtt2)
             -- [fermion, line width = 1.5 pt] (gtt1), 
        };
    \end{feynman}
\end{tikzpicture}
    \caption*{(c)}
\end{minipage}
\\
\begin{minipage}[c]{0.275\textwidth}
    \centering
\begin{tikzpicture} 
    \begin{feynman}[small]
        \vertex  (g1)  {$g$};
        \vertex  (ggh) [square dot, scale=\sizesqdot,below right=\lenboxborn pt of g1,color=gray] {};
        \vertex  (g2) [below left=\lenboxborn pt of ggh]  {$g$};
        \vertex (hhh) [dot,scale=\sizedot,right =\lentriborn pt of ggh]  {};
        \vertex (h1) [above right =\lenboxborn pt of hhh] {$h$};
        \vertex (h2) [below right =\lenboxborn pt of hhh] {$h$};
        \diagram* {
            (g1)  -- [gluon] (ggh) -- [gluon] (g2),
            (h2)  -- [scalar] (hhh) --[scalar] (h1),
            (ggh) -- [scalar] (hhh),
        };
    \end{feynman}
\end{tikzpicture}
    \caption*{(d)}
\end{minipage}
\begin{minipage}[c]{0.275\textwidth}
    \centering
\begin{tikzpicture} 
    \begin{feynman}[small]
        \vertex  (g1)  {$g$};
        \vertex  (gghh) [square dot, scale=\sizesqdot,below right=\lenboxborn pt of g1,color=gray] {};
        \vertex  (g2) [below left=\lenboxborn pt of gghh]  {$g$};
        \vertex (h1) [above right =\lenboxborn pt of gghh] {$h$};
        \vertex (h2) [below right =\lenboxborn pt of gghh] {$h$};
        \diagram* {
            (g1)  -- [gluon] (gghh) -- [gluon] (g2),
            (h2)  -- [scalar] (gghh) --[scalar] (h1),
        };
    \end{feynman}
\end{tikzpicture}
    \caption*{(d)}
\end{minipage}
    \caption{\label{fig:feyndiag_born} Feynman diagrams of the leading SMEFT contributions to $gg\to hh$ (Born level). 
    Black dots denote insertions of SM couplings, gray dots
    (potentially) tree-induced EFT operators, gray squares denote insertions of loop-induced couplings (here $C_{HG}$).}
\end{center}
\end{figure}
The third line in Eq.~\eqref{eq:LagSMEFT} contains the chromomagnetic operator and lines 4-6 show the relevant 4-top operators.
The operator ${\cal O}_{qq,\,\text{Warsaw}}^{(3)\,3333}$ of the Warsaw basis~\cite{Grzadkowski:2010es} has been replaced by ${\cal O}_{QQ}^{(8)}$ where the relation
in terms of the Wilson coefficients has the form~\cite{Aguilar-Saavedra:2018ksv}
\begin{equation}
\begin{aligned}
    C_{QQ}^{(1)}&=2C_{qq,\,\text{Warsaw}}^{(1)\,3333}-\frac{2}{3}C_{qq,\,\text{Warsaw}}^{(3)\,3333}\\
    C_{QQ}^{(8)}&=8C_{qq,\,\text{Warsaw}}^{(3)\,3333}\;,
\end{aligned}
\end{equation}
the other 4-top operators are already present in the 3rd generation 4-fermion operators of the Warsaw basis.

The chromomagnetic operator and the 4-top operators of Eq.~\eqref{eq:LagSMEFT} together form the 
subleading contribution that will be the focus of this work. 
Below the scale of electroweak symmetry breaking, and after performing a field redefinition for the physical Higgs field in unitary gauge~\cite{Heinrich:2022idm}, 
the relevant interaction terms of the Lagrangian have the form
\begin{equation}
\begin{aligned}
    {\cal L}_{\text{SMEFT}}&\supset
-\left(\frac{m_t}{v}\left(1+v^2\frac{C_{H,\text{kin}}}{\Lambda^2}\right)-\frac{v^2}{\sqrt{2}}\frac{C_{tH}}{\Lambda^2}\right)\,h\,\bar{t}\,t
-\left(m_t \frac{C_{H,\text{kin}}}{\Lambda^2}-\frac{3v}{2\sqrt{2}}\frac{C_{tH}}{\Lambda^2}\right)\,h^2\,\bar{t}\,t
\\
&-\left(\frac{m_h^2}{2v}\left(1+3v^2\frac{C_{H,\text{kin}}}{\Lambda^2}\right)-v^3\frac{C_H}{\Lambda^2}\right)\,  h^3
+\frac{C_{HG}}{\Lambda^2}\left(v\,h+\frac{1}{2}\,h^2\right)\, G^a_{\mu \nu} G^{a,\mu \nu}
\\
&+g_s\bar{t}\,\gamma^{\mu}T^a\,t\,G_{\mu}^a
+\frac{C_{tG}}{\Lambda^2}\sqrt{2}\left(h+v\right)\left(\bar{t}\,\sigma^{\mu\nu}T^a\,t\,G_{\mu\nu}^a\right)
\\ &
+\frac{C_{Qt}^{(1)}}{\Lambda^2}\bar{t}_L\gamma^\mu t_L\bar{t}_R\gamma_\mu t_R
+\frac{C_{Qt}^{(8)}}{\Lambda^2}\bar{t}_L\gamma^\mu T^at_L\bar{t}_R\gamma_\mu T^a t_R
\\ &
+\frac{C_{QQ}^{(1)}}{\Lambda^2}\bar{t}_L\gamma^\mu t_L\bar{t}_L\gamma_\mu t_L
+\frac{C_{QQ}^{(8)}}{\Lambda^2}\bar{t}_L\gamma^\mu T^at_L\bar{t}_L\gamma_\mu T^a t_L
\\ &
+\frac{C_{tt}}{\Lambda^2}\bar{t}_R\gamma^\mu t_R\bar{t}_R\gamma_\mu t_R\;,\label{eq:LagSMEFT_expanded}
\end{aligned}
\end{equation}
which is valid up to ${\cal O}(\Lambda^{-4})$ differences.
Here $v$ denotes the full vacuum expectation value including 
a higher dimensional contribution of $\frac{C_H}{\Lambda^2}$ and\footnote{
    For more details on the definition of physical quantities in SMEFT 
    we refer to Chapter 5 of Ref.~\cite{Alonso:2013hga}.} 
\begin{equation}
m_t=\frac{v}{\sqrt{2}}\left(y_t -\frac{v^2}{2}\frac{C_{tH}}{\Lambda^2}\right)\;,
\end{equation} where
$y_t$ is the top-Yukawa parameter of the dimension-4 Lagrangian.

In the following, we will briefly comment on the notions of `leading' and `subleading' we have used above.
In SMEFT, the operators are ordered by their canonical dimension, i.e.\ the
expansion is based on powers in $E/\Lambda$.
However, in a perturbative expansion, in particular in the combination
of EFT expansions with expansions in a SM coupling, loop suppression
factors also play a role. Therefore, a classification of operators
into {\em potentially} tree-level induced  and necessarily 
loop-generated operators~\cite{Arzt:1994gp}, the latter thus carrying an implicit loop factor $L=\left(16\pi^2\right)^{-1}$,
leads to a more refined counting scheme, corroborated by observations from
renormalisation and the cancellation of scheme-dependent
terms~\cite{DiNoi:2023ygk}.  The same loop factors can be derived by
supplementing the SMEFT expansion by a chiral counting of
operators~\cite{Buchalla:2022vjp}, see also
\cite{Guedes:2023azv}. Such a classification can only
be made when making some minimal UV assumptions, which are however
quite generic, assuming renormalisability and weak coupling of the underlying UV complete theory\footnote{
    Non-renormalisable contributions, for example due to an intermediate new physics sector that is not the UV complete theory,  would introduce a stronger suppression due to factors of an even higher NP scale  $\Lambda^\prime$, that is likely to overcompensate the loop factor. 
  The RGE flow of the Wilson coefficients can mix potentially tree-level induced and loop suppressed coefficients. However, coefficients of the RGE flow also carry a loop factor and therefore such mixings are suppressed.
  Furthermore, in our selection $C_{tG}$ is the only loop suppressed coefficient that could be affected by a mixing of $C_{QtQb}^{(1/8)}$, see (A.21) of Ref.~\cite{Jenkins:2013wua}, thus the mixing is suppressed by $y_b/y_t$, i.e.\ not allowed by our flavour assumption.}. 
  Therefore, under these assumptions and if the Wilson coefficients $C_i$ in the SMEFT expansion are considered to be of similar magnitude, it makes sense to expand in $C_i\times 1/\Lambda^a \times 1/(16\pi^2)^b$. 
  Fixing $a=2$ (dimension-6 operators) we call the operator contributions with $b=0$ `leading' and those with $b>0$ `subleading'.
 The above factors are to be combined with {\em explicit} loop factors $ 1/(16\pi^2)^c$ from the SM perturbative expansion. 
Nonetheless, one has to keep in mind that this approach does not cover UV effects in full generality and 
that this classification is not invariant under field redefinitions and thus necessarily basis dependent~\cite{Brivio:2017vri}.

  Applying those rules to the Born contributions of Fig.~\ref{fig:feyndiag_born} and collecting loop factors of QCD origin together with associated powers of $g_s$ leads to 
${\cal M}_\text{Born}\sim {\cal O}\left((g_s^2L)\Lambda^{-2}\right)$. Here we identify both types of contributions:
explicit diagrammatic loop factors combined with tree-generated operator insertions (first line, grey dots, $b=0$, $c=1$ in the above classification), 
and tree diagrams combined with implicitly loop-generated operators (second line, grey squares, $b=1$, $c=0$ in the above classification).
The power counting of the subleading contributions is addressed in Sections \ref{sec:MEchromo} and \ref{sec:ME4t}.

At cross section level, we therefore have
\begin{equation}
\sigma_\text{EFT}\sim 
\sigma_\text{EFT}^\text{Born}+\sigma_\text{EFT}^\text{NLO}\;,\label{eq:XS_expansion}
\end{equation}
where\footnote{
    We associate a factor of $g_s$ with each Wilson coefficient where a  field-strength tensor is contained in the corresponding operator.
}
\begin{equation}
  \begin{aligned}
\sigma_\text{EFT}^\text{Born}&\sim 
\sigma_\text{SM}\left[(g_s^2L)^{2}\right]
+\sigma_\text{dim6}^\text{lead}\left[(g_s^2L)^{2}\Lambda^{-2}\right] 
+\sigma_\text{dim6}^{C_{tG},\text{4-top}}\left[(g_s^2L)^{2}{\mathbf L}\,\Lambda^{-2}\right] 
\\[+10pt]
&
\quad\left\{
  +\sigma_{\text{dim6}^2}^{\text{lead}\times\text{lead}}\left[(g_s^2L)^{2}\Lambda^{-4}\right] 
  +\sigma_{\text{dim6}^2}^{(C_{tG},\text{4-top})\times\text{lead}}\left[(g_s^2L)^{2}{\mathbf L}\,\Lambda^{-4}\right]\right\}\;,
  \end{aligned}\label{eq:XSBorn_expansion}
\end{equation}
and
\begin{equation}
  \begin{aligned}
\sigma_\text{EFT}^\text{NLO}&\sim 
\sigma_\text{SM}^\text{NLO}\left[(g_s^2L)^{3}\right]
+\sigma_\text{dim6}^{\text{NLO, }\text{lead}}\left[(g_s^2L)^{3}\Lambda^{-2}\right] 
\quad\left\{
  +\sigma_{\text{dim6}^2}^{\text{NLO, }\text{lead}\times\text{lead}}\left[(g_s^2L)^{3}\Lambda^{-4}\right] \right\}\;,
  \end{aligned}\label{eq:XSNLO_expansion}
\end{equation}
with ${\mathbf L}=\frac{1}{16\pi^2}$ originating from subleading operator contributions.
Here $\sigma_\text{dim6}^{(\dots)}$ denotes the interference of the dimension-6 amplitude with the SM amplitude and
the terms inside $\left\{\dots\right\}$ are the $|\amp_\text{dim6}|^2$ 
parts of the cross section, which can be switched on or off in the {\tt ggHH\_SMEFT} code. 
The EFT contribution only based on leading operators is denoted by $\sigma_{\text{dim6}}^\text{lead}$ 
and $\sigma_{\text{dim6}^2}^{\text{lead}\times\text{lead}}$, while
$\sigma_{\text{dim6}}^{C_{tG},\text{4-top}}$ and $\sigma_{\text{dim6}^2}^{(C_{tG},\text{4-top})\times\text{lead}}$ 
contain the contributions with a single insertion of $C_{tG}$ and/or 4-top operators.
Values inside the square brackets in Eqs.~\eqref{eq:XSBorn_expansion} and \eqref{eq:XSNLO_expansion} denote the
order in power counting of the respective contribution at cross section level.

In the subsequent parts of this section, we discuss the structure of the contributions to the amplitude which involve single insertions of the chromomagnetic operator 
and the 4-top operators of eq.~(\ref{eq:LagSMEFT}).
All relevant diagrams were generated with {\tt QGraf}~\cite{Nogueira:1991ex} and the calculation was performed 
analytically using {\tt FeynCalc}~\cite{Shtabovenko:2020gxv,Shtabovenko:2016sxi,Mertig:1990an}.
UV divergences are absorbed in a mixed on-shell-$\msbar$ renormalisation scheme, 
where the mass of the top-quark is renormalised on-shell and 
the dimension-6 Wilson coefficients are renormalised in the $\msbar$ scheme.
The contribution of the chromomagnetic operator has been checked against a private version of {\tt GoSam}~\cite{GoSam:2014iqq,Cullen:2011ac}. 
Moreover, we compared with the results of Ref.~\cite{Maltoni:2016yxb} for the total cross section 
evaluated at the central scale and,  after adjustment to the described conventions, found complete agreement.
The amplitude involving 4-top operators has been checked in $D$ dimensions against {\tt alibrary}~\cite{alibrary} in combination with 
{\tt Kira}~\cite{Klappert:2020nbg,Maierhofer:2017gsa}.  
The renormalised 4-top amplitudes were tested numerically in four dimensions by comparing the analytic implementation in the 
{\tt Powheg-Box-V2}~\cite{Nason:2004rx,Frixione:2007vw,Alioli:2010xd} 
against the result obtained with {\tt alibrary} and evaluated with {\tt pySecDec}~\cite{Heinrich:2021dbf,Borowka:2018goh,Borowka:2017idc}
for several phase-space points.
The chiral structure of the 4-top couplings is treated in the Naive Dimensional Regularisation (NDR) scheme~\cite{Chanowitz:1979zu} 
assuming the cyclicity of traces of strings of gamma matrices. 
This is possible since 
(after reduction of loop integrals onto the integral basis of 't Hooft-Passarino-Veltman scalar integrals~\cite{Passarino:1978jh,tHooft:1978jhc})
all appearing traces with an odd number of $\gamma_5$ matrices can be explicitly brought into the form 
$\sum \gamma^{\mu_1}\dots\gamma^{\mu_n}\gamma_5$ with $n<4$ 
through anti-commutation and therefore vanish.
In addition, the analytic calculation of the 4-top contributions in {\tt FeynCalc} 
is repeated in the Breitenlohner-Maison-t'Hooft-Veltman (BMHV) scheme~\cite{tHooft:1972tcz,Breitenlohner:1977hr}, with the symmetric definition for chiral vertices
\begin{equation}
    \gamma^\mu\mathbb{P}_{L/R}\to\mathbb{P}_{R/L}\gamma^\mu\mathbb{P}_{L/R}\;,\label{eq:chiral_vert}
\end{equation} 
and the translation between the Lagrangian parameters
obtained in Ref.~\cite{DiNoi:2023ygk} is verified. 
For convenience, the explicit form of the translation is also presented in Eq.~\eqref{eq:scheme_translation}.

\subsection{\label{sec:MEchromo}Amplitude structure of chromomagnetic operator insertions}
The contribution of the chromomagnetic operator to the amplitude leads to the diagram types shown in Fig.~\ref{fig:feyndiag_chromo}. 
\begin{figure}[h]
\begin{center}
\begin{minipage}[c]{0.24\textwidth}
    \centering
\begin{tikzpicture} 
    \begin{feynman}[small]
        \vertex  (g1)  {$g$};
        \vertex  (gtt1) [square dot, scale=\sizesqdot,right=\lentrichromo pt of g1,color=gray] {};
        \vertex (htt) [dot,scale=\sizedot,below right =\lentrichromo pt of gtt1]  {};
        \vertex  (gtt2) [dot,scale=\sizedot,below left =\lentrichromo pt of htt] {};
        \vertex  (g2) [left=\lentrichromo pt of gtt2]  {$g$};
        \vertex (hhh) [dot,scale=\sizedot,right =\lentrichromo pt of htt]  {};
        \vertex (h1) [above right =\lentrichromo pt of hhh] {$h$};
        \vertex (h2) [below right =\lentrichromo pt of hhh] {$h$};
        \diagram* {
            (g1)  -- [gluon] (gtt1),
            (g2) -- [gluon] (gtt2),
            (h1)  -- [scalar] (hhh) -- [scalar] (h2),
            (htt) -- [scalar] (hhh),
            (gtt1) -- [fermion, line width = 1.5 pt] (htt) 
            -- [fermion, line width = 1.5 pt] (gtt2)
             -- [fermion, line width = 1.5 pt] (gtt1), 
            
        };
    \end{feynman}
\end{tikzpicture}
    \caption*{(a)}
\end{minipage} 
\begin{minipage}[c]{0.24\textwidth}
    \centering
\begin{tikzpicture} 
    \begin{feynman}[small]
        \vertex  (g1)  {$g$};
        \vertex  (gtt1) [square dot, scale=\sizesqdot,right= of g1,color=gray] {};
         \vertex (htt) [dot,scale=0.01,below right =\lentrichromo pt of gtt1,color=white]  {};
        \vertex  (gtt2) 
        [dot,scale=\sizedot,below left =\lentrichromo pt of htt] {};
        \vertex  (g2) [left=of gtt2]  {$g$};
        \vertex (hhh) [dot,scale=\sizedot,right =7 pt of htt] {};
        \vertex (h1) [above right =\lentrichromo pt of hhh] {$h$};
        \vertex (h2) [below right =\lentrichromo pt of hhh] {$h$};
        \diagram* {
            (g1)  -- [gluon] (gtt1),
            (g2) -- [gluon] (gtt2),
            (hhh)  -- [scalar] (gtt1),
            (h1) -- [scalar] (hhh) -- [scalar] (h2),
            (gtt1) -- [fermion, line width = 1.5 pt,quarter right] (gtt2)
             -- [fermion, line width = 1.5 pt, quarter right] (gtt1), 
            
        };
    \end{feynman}
\end{tikzpicture}
    \caption*{(b)}
\end{minipage}
\begin{minipage}[c]{0.24\textwidth}
    \centering
\begin{tikzpicture} 
    \begin{feynman}[small]
        \vertex  (g1)  {$g$};
        \vertex  (gtt1) [square dot,scale=\sizesqdot,right=\lentrichromo pt of g1,color=gray] {};
        \vertex  (htt1) [dot,scale=\sizedot,right=\lenboxchromo pt of gtt1]  {};
        \vertex  (gtt2) [dot,scale=\sizedot,below=\lenboxchromo pt of gtt1] {};
        \vertex  (htt2) [dot,scale=\sizedot,below=\lenboxchromo pt of htt1] {};
        \vertex  (g2) [left=\lentrichromo pt of gtt2]  {$g$};
        \vertex (h1) [right =\lentrichromo pt of htt1] {$h$};
        \vertex (h2) [right =\lentrichromo pt of htt2] {$h$};
        \diagram* {
            (g1)  -- [gluon] (gtt1),
            (g2) -- [gluon] (gtt2),
            (h2)  -- [scalar] (htt2),
            (h1)  -- [scalar] (htt1),
            (gtt1) -- [fermion, line width = 1.5 pt] (htt1)  -- [fermion, line width = 1.5 pt] (htt2)
            -- [fermion, line width = 1.5 pt] (gtt2)
             -- [fermion, line width = 1.5 pt] (gtt1), 
        };
    \end{feynman}
\end{tikzpicture}
    \caption*{(c)}
\end{minipage}
\begin{minipage}[c]{0.24\textwidth}
    \centering
\begin{tikzpicture} 
    \begin{feynman}[small]
        \vertex  (g1)  {$g$};
        \vertex  (ghtt1) [square dot, scale=\sizesqdot,right=\lentrichromo pt of g1,color=gray] {};
        \vertex (htt) [dot,scale=\sizedot,below right =\lentrichromo pt of gtt1]  {};
        \vertex  (gtt2) [dot,scale=\sizedot,below left =\lentrichromo pt of htt] {};
        \vertex  (g2) [left=\lentrichromo pt of gtt2]  {$g$};
        \vertex (h1) [right =\lenboxchromo pt of ghtt1] {$h$};
        \vertex (h2) [below right =\lentrichromo pt of htt] {$h$};
        \diagram* {
            (g1)  -- [gluon] (ghtt1),
            (g2) -- [gluon] (gtt2),
            (h1)  -- [scalar] (ghtt1),
            (h2) -- [scalar] (htt),
            (ghtt1) -- [fermion, line width = 1.5 pt] (htt) 
            -- [fermion, line width = 1.5 pt] (gtt2)
             -- [fermion, line width = 1.5 pt] (ghtt1), 
        };
    \end{feynman}
\end{tikzpicture}
    \caption*{(d)}
\end{minipage}
    \caption{\label{fig:feyndiag_chromo} Feynman diagrams involving insertions of the chromomagnetic operator. 
    The gray squares denote insertions of the (loop-suppressed) chromomagnetic operator.}
\end{center}
\end{figure}
At first sight, the diagrams are at one-loop order, such that,
together with the explicit dimensional factor, the prefactor of the Wilson coefficient appears at ${\cal O}\left((g_s^2L)\Lambda^{-2}\right)$.
However, the chromomagnetic operator 
belongs to the class of operators that, in 
renormalisable
UV completions, can only be generated at loop level~\cite{Buchalla:2022vjp,Arzt:1994gp}.
Hence, the implicit loop factor of its Wilson coefficient promotes the order in power counting to ${\cal M}_{tG}\sim {\cal O}\left((g_s^2L){\mathbf L}\,\Lambda^{-2}\right)$, 
which is in that sense subleading with regards to the leading Born diagrams of Fig.~\ref{fig:feyndiag_born}.

The diagrams of type (a), (b) and (d) are UV divergent even though they constitute the leading order contribution of $C_{tG}$ to the gluon fusion process. 
However, this behaviour is well known~\cite{Degrande:2012gr,Maltoni:2016yxb,Deutschmann:2017qum} and leads to a renormalisation of 
$C_{HG}^0=\mu^{2\eps}\left(C_{HG}+\delta_{C_{HG}}^{C_{i}}\right)$   ($\mu$ being the renormalisation scale)
which in the $\msbar$ scheme takes the form~\cite{Alonso:2013hga,Maltoni:2016yxb,Deutschmann:2017qum}
\begin{equation}
    \delta _{C_{HG}}^{C_{tG}}=\frac{\left(4\pi e^{-\gamma_E}\right)^\eps}{16\pi^2\eps}\frac{4\sqrt{2}g_s m_t}{v}T_F\,C_{tG}\;.
\end{equation}
With this renormalisation term the finiteness of the amplitude is restored, and it can be numerically evaluated using standard integral libraries.

\subsection{\label{sec:ME4t}Amplitude structure involving four-top operators}

Four-top operators appear first at two-loop order in gluon-fusion Higgs- or di-Higgs production.
Thus, their contribution is of the same order in the power counting as the one of the chromomagnetic operator,
i.e.\ ${\cal M}_\text{4-top}\sim {\cal O}\left((g_s^2L){\mathbf L}\,\Lambda^{-2}\right)$.
Following the reasoning of Ref.~\cite{Alasfar:2022zyr} in single Higgs production, 
we separate the contribution into different diagram classes, which are shown in Fig.~\ref{fig:feyndiag_4top}.
\begin{figure}[h]
\begin{center}
\begin{tabular}{ c  c  c }
\begin{minipage}[c]{0.3\textwidth}
    \centering
\begin{tikzpicture} 
    \begin{feynman}[small]
        \vertex  (g1)  {$g$};
        \vertex  (gtt1) [dot,scale=\sizedot,right =\lentrismall+\lentriftop/2 pt  of g1] {};
        \vertex (htt) [dot,scale=\sizedot,below right =\lentriftop pt of gtt1] {};
        \vertex (4F) [dot, scale = \sizesqdot,color=gray,below=\lenboxftop/2 pt of gtt1] {};
        \vertex (inv1) [scale=0.01,left =\lentriftop/2 pt of 4F] {};
        \vertex  (gtt2) [dot,scale=\sizedot,below left =\lentriftop pt of htt] {};
        \vertex  (g2) [left=\lentrismall+\lentriftop/2 pt of gtt2]  {$g$};
        \vertex (hhh) [dot,scale=\sizedot,right =\lentriftop pt  of htt] {};
        \vertex (h1) [above right =\lentriftop pt of hhh] {$h$};
        \vertex (h2) [below right =\lentriftop pt of hhh] {$h$};
        \diagram* {
            (g1)  -- [gluon] (gtt1),
            (g2) -- [gluon] (gtt2),
            (hhh)  -- [scalar] (htt),
            (h1)  -- [scalar] (hhh) -- [scalar] (h2),
            (gtt1) -- [fermion, line width = 1.5 pt] (htt) 
            -- [fermion, line width = 1.5 pt] (gtt2) -- [fermion, line width = 1.5 pt] (4F)
             -- [fermion, line width = 1.5 pt] (gtt1), 
             (4F) -- [fermion, line width = 1.5 pt, half left] (inv1) -- [fermion, line width = 1.5 pt, half left] (4F)
};
    \end{feynman}
\end{tikzpicture}
    \caption*{(a)}
\end{minipage}
&
\begin{minipage}[c]{0.3\textwidth}
    \centering
\begin{tikzpicture} 
    \begin{feynman}[small]
        \vertex  (g1)  {$g$};
        \vertex  (gtt1) [dot,scale=\sizedot,right =\lentrismall+\lentriftop/2 pt  of g1] {};
        \vertex (htt1) [dot,scale=\sizedot,right =\lenboxftop pt of gtt1] {};
        \vertex (htt2) [dot,scale=\sizedot,below =\lenboxftop pt of htt1] {};
        \vertex (4F) [dot, scale = \sizesqdot,color=gray,below=\lenboxftop/2 pt of gtt1] {};
        \vertex (inv1) [scale=0.01,left =\lentriftop/2 pt of 4F] {};
        \vertex  (gtt2) [dot,scale=\sizedot,left =\lenboxftop pt of htt2] {};
        \vertex  (g2) [left=\lentrismall+\lentriftop/2 pt of gtt2]  {$g$};
        \vertex (h1) [right =\lentrismall+\lentriftop/2 pt of htt1] {$h$};
        \vertex (h2) [right =\lentrismall+\lentriftop/2 pt of htt2] {$h$};
        \diagram* {
            (g1)  -- [gluon] (gtt1),
            (g2) -- [gluon] (gtt2),
            (h1)  -- [scalar] (htt1),
            (h2)  -- [scalar] (htt2),
            (gtt1) -- [fermion, line width = 1.5 pt] (htt1) -- [fermion, line width = 1.5 pt] (htt2)
            -- [fermion, line width = 1.5 pt] (gtt2) -- [fermion, line width = 1.5 pt] (4F)
             -- [fermion, line width = 1.5 pt] (gtt1), 
             (4F) -- [fermion, line width = 1.5 pt, half left] (inv1) -- [fermion, line width = 1.5 pt, half left] (4F)
};
    \end{feynman}
\end{tikzpicture}
    \caption*{(b)}
\end{minipage}
&
\\
\begin{minipage}[c]{0.3\textwidth}
    \centering
\begin{tikzpicture} 
    \begin{feynman}[small]
        \vertex  (g1)  {$g$};
        \vertex  (gtt1) [dot,scale=\sizedot,right =\lentrismall+\lentriftop/2 pt of g1] {};
        \vertex (4F) [dot, scale=\sizesqdot,below right=\lentriftop pt of gtt1,color=gray] {};
        \vertex  (gtt2) [dot,scale=\sizedot,below left =\lentriftop pt  of 4F] {};
        \vertex  (g2) [left=\lentrismall+\lentriftop/2 pt of gtt2]  {$g$};
        \vertex (htt) [dot,scale=\sizedot,right = \lentriftop/2 pt of 4F] {};
        \vertex (hhh) [dot,scale=\sizedot,right =\lentriftop/2 pt of htt] {};
        \vertex (h1) [above right =\lentriftop pt  of hhh] {$h$};
        \vertex (h2) [below right =\lentriftop pt  of hhh] {$h$};

        \diagram* {
            (g1)  -- [gluon] (gtt1),
            (g2) -- [gluon] (gtt2),
            (hhh)  -- [scalar] (htt),
            (h1) -- [scalar] (hhh) -- [scalar] (h2),
            (gtt1) -- [fermion, line width = 1.5 pt] (4F) 
            -- [fermion, line width = 1.5 pt] (gtt2)
             -- [fermion, line width = 1.5 pt] (gtt1), 
             (4F) -- [fermion, line width = 1.5 pt, half left] (htt) -- [fermion, line width = 1.5 pt, half left] (4F)
        };
    \end{feynman}
\end{tikzpicture}
    \caption*{(c)}
\end{minipage}
&
\begin{minipage}[c]{0.3\textwidth}
    \centering
\begin{tikzpicture} 
    \begin{feynman}[small]
        \vertex  (g1)  {$g$};
        \vertex  (gtt1) [dot,scale=\sizedot,right =\lentrismall+\lentriftop/2 pt of g1] {};
        \vertex (4F) [dot, scale=\sizesqdot,right=\lenboxftop pt of gtt1,color=gray] {};
        \vertex  (gtt2) [dot,scale=\sizedot,below =\lenboxftop pt  of gtt1] {};
        \vertex  (g2) [left=\lentrismall+\lentriftop/2 pt of gtt2]  {$g$};
        \vertex (htt1) [dot,scale=\sizedot,right = \lentriftop/2 pt of 4F] {};
        \vertex (htt2) [dot,scale=\sizedot,below = \lenboxftop pt of 4F] {};
        \vertex (h1) [right =\lentrismall pt  of htt1] {$h$};
        \vertex (h2) [right =\lentrismall+\lentriftop/2 pt  of htt2] {$h$};

        \diagram* {
            (g1)  -- [gluon] (gtt1),
            (g2) -- [gluon] (gtt2),
            (h1)  -- [scalar] (htt1),
            (h2) -- [scalar] (htt2),
            (gtt1) -- [fermion, line width = 1.5 pt] (4F) -- [fermion, line width = 1.5 pt] (htt2)
            -- [fermion, line width = 1.5 pt] (gtt2)
             -- [fermion, line width = 1.5 pt] (gtt1), 
             (4F) -- [fermion, line width = 1.5 pt, half left] (htt1) -- [fermion, line width = 1.5 pt, half left] (4F)
        };
    \end{feynman}
\end{tikzpicture}
    \caption*{(d)}
\end{minipage}
&
\begin{minipage}[c]{0.3\textwidth}
    \centering
\begin{tikzpicture} 
    \begin{feynman}[small]
        \vertex  (g1)  {$g$};
        \vertex  (gtt1) [dot,scale=\sizedot,right =\lentrismall+\lentriftop/2 pt of g1] {};
        \vertex (4F) [dot, scale=\sizesqdot,below right=\lentriftop pt of gtt1,color=gray] {};
        \vertex  (gtt2) [dot,scale=\sizedot,below left=\lentriftop pt of 4F] {};
        \vertex  (g2) [left=\lentrismall+\lentriftop/2 pt of gtt2]  {$g$};
        \vertex (htt1) [dot,scale=\sizedot,above right = \lentriftop pt of 4F] {};
        \vertex (htt2) [dot,scale=\sizedot,below right= \lentriftop pt of 4F] {};
        \vertex (h1) [right =\lentrismall+\lentriftop/2 pt  of htt1] {$h$};
        \vertex (h2) [right =\lentrismall+\lentriftop/2 pt  of htt2] {$h$};

        \diagram* {
            (g1)  -- [gluon] (gtt1),
            (g2) -- [gluon] (gtt2),
            (h1)  -- [scalar] (htt1),
            (h2) -- [scalar] (htt2),
            (gtt1) -- [fermion, line width = 1.5 pt] (4F) -- [fermion, line width = 1.5 pt] (gtt2) -- [fermion, line width = 1.5 pt] (gtt1), 
            (htt1) -- [fermion, line width = 1.5 pt] (4F) -- [fermion, line width = 1.5 pt] (htt2) -- [fermion, line width = 1.5 pt] (htt1), 
        };
    \end{feynman}
\end{tikzpicture}
    \caption*{(e)}
\end{minipage}
\\
\begin{minipage}[c]{0.3\textwidth}
    \centering
\begin{tikzpicture} 
    \begin{feynman}[small]
        \vertex  (g1)  {$g$};
        \vertex  (gtt1) [dot,scale=\sizedot,right =\lentrismall pt  of g1] {};
        \vertex (4F) [dot, scale = \sizesqdot,color=gray,right=\lentriftop/2 pt of gtt1] {};
        \vertex (htt) [dot,scale=\sizedot,below right =\lentriftop pt of 4F] {};
        \vertex  (gtt2) [dot,scale=\sizedot,below left =\lentriftop pt of htt] {};
        \vertex  (g2) [left=\lentrismall+\lentriftop/2 pt of gtt2]  {$g$};
        \vertex (hhh) [dot,scale=\sizedot,right =\lentriftop pt  of htt] {};
        \vertex (h1) [above right =\lentriftop pt of hhh] {$h$};
        \vertex (h2) [below right =\lentriftop pt of hhh] {$h$};
        \diagram* {
            (g1)  -- [gluon] (gtt1),
            (g2) -- [gluon] (gtt2),
            (hhh)  -- [scalar] (htt),
            (h1)  -- [scalar] (hhh) -- [scalar] (h2),
            (4F) -- [fermion, line width = 1.5 pt] (htt) 
            -- [fermion, line width = 1.5 pt] (gtt2) -- [fermion, line width = 1.5 pt] (4F), 
             (4F) -- [fermion, line width = 1.5 pt, half left] (gtt1) -- [fermion, line width = 1.5 pt, half left] (4F)
};
    \end{feynman}
\end{tikzpicture}
    \caption*{(f)}
\end{minipage}
&
\begin{minipage}[c]{0.3\textwidth}
    \centering
\begin{tikzpicture} 
    \begin{feynman}[small]
        \vertex  (g1)  {$g$};
        \vertex  (gtt1) [dot,scale=\sizedot,right =\lentrismall pt  of g1] {};
        \vertex (4F) [dot, scale = \sizesqdot,color=gray,right=\lentriftop/2 pt of gtt1] {};
        \vertex (htt1) [dot,scale=\sizedot,right =\lenboxftop pt of 4F] {};
        \vertex (htt2) [dot,scale=\sizedot,below =\lenboxftop pt of htt1] {};
        \vertex  (gtt2) [dot,scale=\sizedot,left =\lenboxftop pt of htt2] {};
        \vertex  (g2) [left=\lentrismall+\lentriftop/2 pt of gtt2]  {$g$};
        \vertex (h1) [right =\lentrismall+\lentriftop/2  pt of htt1] {$h$};
        \vertex (h2) [right =\lentrismall+\lentriftop/2  pt of htt2] {$h$};
        \diagram* {
            (g1)  -- [gluon] (gtt1),
            (g2) -- [gluon] (gtt2),
            (h1)  -- [scalar] (htt1),
            (htt2) -- [scalar] (h2),
            (4F) -- [fermion, line width = 1.5 pt] (htt1) -- [fermion, line width = 1.5 pt] (htt2) 
            -- [fermion, line width = 1.5 pt] (gtt2) -- [fermion, line width = 1.5 pt] (4F), 
             (4F) -- [fermion, line width = 1.5 pt, half left] (gtt1) -- [fermion, line width = 1.5 pt, half left] (4F)
};
    \end{feynman}
\end{tikzpicture}
    \caption*{(g)}
\end{minipage}
&
\begin{minipage}[c]{0.3\textwidth}
    \centering
\begin{tikzpicture} 
    \begin{feynman}[small]
        \vertex  (g1)  {$g$};
        \vertex  (gtt1) [dot,scale=\sizedot,right =\lentrismall+\lentriftop/2 pt of g1] {};
        \vertex (4F) [dot, scale=\sizesqdot,below right=\lentriftop pt of gtt1,color=gray] {};
        \vertex  (gtt2) [dot,scale=\sizedot,below left=\lentriftop pt of 4F] {};
        \vertex  (g2) [left=\lentrismall+\lentriftop/2 pt of gtt2]  {$g$};
        \vertex (htt1) [dot,scale=\sizedot,above right = \lentriftop pt of 4F] {};
        \vertex (htt2) [dot,scale=\sizedot,below right= \lentriftop pt of 4F] {};
        \vertex (h1) [right =\lentrismall+\lentriftop/2 pt  of htt1] {$h$};
        \vertex (h2) [right =\lentrismall+\lentriftop/2 pt  of htt2] {$h$};

        \diagram* {
            (g1)  -- [gluon] (gtt1),
            (g2) -- [gluon] (gtt2),
            (h1)  -- [scalar] (htt1),
            (h2) -- [scalar] (htt2),
            (gtt1) -- [fermion, line width = 1.5 pt] (4F) -- [fermion, line width = 1.5 pt] (htt1) -- [fermion, line width = 1.5 pt] (gtt1), 
            (gtt2) -- [fermion, line width = 1.5 pt] (4F) -- [fermion, line width = 1.5 pt] (htt2) -- [fermion, line width = 1.5 pt] (gtt2), 
        };
    \end{feynman}
\end{tikzpicture}
    \caption*{(h)}
\end{minipage}
\end{tabular}
    \caption{\label{fig:feyndiag_4top} Feynman diagrams involving insertions of 4-top operators.
    The gray dots denote insertions of 4-top operators.}
\end{center}
\end{figure}
The ordering in columns is chosen in order to group in underlying Born topologies (i.e.\ triangles and boxes), the rows combine the type of 
one-loop correction (if applicable).
The first column is thus analogous to single Higgs production as in Ref.~\cite{Alasfar:2022zyr}, with one Higgs splitting into two, however we do not include bottom quark loops 
(and loops of other light quarks), 
since we apply a more restrictive flavour assumption in which the bottom quark remains massless and diagrams with bottom loops 
vanish in an explicit calculation, either due to the  bottom-Yukawa coupling being zero or
due to vanishing scaleless integrals. 

The categories of diagrams in Fig.~\ref{fig:feyndiag_4top} can be structured in the following way: 
(a) and (b): loop corrections to top propagators,
(c) and (d): loop corrections to the Yukawa interaction, 
(e): loop correction to the $tthh$ vertex, (f) and (g): loop corrections to the gauge interaction 
(more precisely, a contraction of a one-loop subdiagram of (f) leads
 to the topologies of Fig.~\ref{fig:feyndiag_chromo} (a) or (b)),
 and (h) without clear correspondence to a vertex correction of a Born structure 
(but related to type (d) diagrams of Fig.~\ref{fig:feyndiag_chromo} after contraction of a one-loop subdiagram). 

In the following we sketch the calculation of the contribution of those classes 
and then refer to the $\gamma_5$-scheme dependence of the calculation, which first has been investigated in Ref.~\cite{DiNoi:2023ygk}.
We represent the results in terms of master integrals
that are given by Passarino-Veltman scalar functions $N_0,\, N\in\{A,B,C, \ldots\}$ in the convention of 
{\tt FeynCalc}~\cite{Shtabovenko:2020gxv,Shtabovenko:2016sxi,Mertig:1990an} 
(which is equivalent to the {\tt LoopTools}~\cite{Hahn:1998yk} convention),
%
such that loop factors are kept manifest in the formulas.

We begin with propagator corrections 
which have no momentum dependence and therefore contribute only
proportional to a mass insertion 
\begin{equation}\begin{aligned}
\begin{tikzpicture}[baseline=(t1)]
\begin{feynman}[small]
    \vertex  (t1) {$t$}; 
    \vertex (4F) [dot, scale=\sizesqdot, right = 25 pt of t1,color=gray] {};
    \vertex (t2) [right= 25 pt of 4F] {$t$};
    \vertex  (inv) [scale = 0.01,above = 20 pt of 4F] {$t$};
    \diagram* {
    (t1)  -- [fermion, line width = 1.5 pt] (4F) -- [fermion, line width = 1.5 pt] (t2),
    (inv)  -- [fermion, line width = 1.5 pt, half right] (4F) -- [fermion, line width = 1.5 pt,half right] (inv),
}; 
\end{feynman}
\end{tikzpicture} &= 
\frac{C_{Qt}^{(1)}+c_F C_{Qt}^{(8)}}{8\pi^2\Lambda^2}
\left(2A_0\left(m_t^2\right)-m_t^2\right) \times 
\begin{tikzpicture}[baseline=(t1)]
\begin{feynman}[small]
    \vertex  (t1) [] {$t$}; 
    \vertex (4F) [dot, scale=0.01, right = 25 pt of t1,color=black] {};
    \vertex (t2) [right= 25 pt of 4F] {$t$};
    \node[shape=star,star points=4,star point ratio = 5,fill=black, draw,scale = 0.15, rotate=45] at (4F) {};
    \diagram* {
    (t1)  -- [fermion, line width = 1.5 pt] (4F) -- [fermion, line width = 1.5 pt] (t2),
     };
\end{feynman}
\end{tikzpicture}\;.
\label{eq:diag_mt}
\end{aligned}
 \end{equation}
Hence, after applying an on-shell renormalisation of the top quark 
mass $m_t^0=m_t+\delta m_t$ 
 with
\begin{equation}
    \begin{split}
     \delta{m_t^\text{4-top}}=-m_t\frac{C_{Qt}^{(1)}+c_F C_{Qt}^{(8)}}{8\pi^2\Lambda^2}
      \left(2A_0\left(m_t^2\right)-m_t^2\right)\;,
    \end{split}\label{eq:mt-4F-ct}
\end{equation}
the diagrams of class (a) and (b) are completely removed.

Next, we consider loop corrections to Yukawa-type interactions. 
The explicit expression for $h\to \bar{t}t$ for an off-shell Higgs is proportional to the SM Yukawa coupling
\begin{equation} 
\begin{aligned}
    \vcenter{\hbox{\begin{tikzpicture}[baseline=(4F)]
            \begin{feynman}[small]
                \vertex  (g1)  {$h$};
                 \vertex (gtt1) [dot, scale=\sizedot, right= 25 pt of g1] {};
                \vertex  (4F) [dot,scale=\sizesqdot,right = 20 pt of gtt1, color = gray] {};
                \vertex  (t1) [above right= 25 pt of 4F] {$t$};
                \vertex (t2) [below right= 25 pt of 4F] {$t$};

                \diagram* {
                    (g1)  -- [scalar] (gtt1),  
                    (gtt1) -- [anti fermion, line width = 1.5 pt, half right] (4F) 
                    -- [anti fermion, line width = 1.5 pt,half right] (gtt1),
                    (t2) -- [fermion, line width = 1.5 pt]  (4F) -- [fermion, line width = 1.5 pt] (t1)
                };
            \end{feynman}
        \end{tikzpicture}}} &= 
        \left(\frac{C_{Qt}^{(1)}+c_F C_{Qt}^{(8)}}{\Lambda^2}\frac{4m_t^2-q^2}{16\pi^2}\left(2B_0\left(q^2,m_t^2,m_t^2\right)-1\right) -\frac{\delta{m_t^\text{4-top}}}{m_t} \right)  
       \\
        &\qquad\times  \vcenter{\hbox{\begin{tikzpicture}[baseline=(gtt1)]
            \begin{feynman}[small]
                \vertex  (g1)  {$h$};
                 \vertex (gtt1) [dot, scale=\sizedot, right= 25 pt of g1, color=black] {};
                \vertex  (t1) [above right= 25 pt of gtt1] {$t$};
                \vertex (t2) [below right= 25 pt of gtt1] {$t$};

                \diagram* {
                    (g1)  -- [scalar] (gtt1),  
                    (t2) -- [fermion, line width = 1.5 pt] (gtt1) -- [fermion, line width = 1.5 pt] (t1),};
            \end{feynman}
        \end{tikzpicture}}},
        \label{eq:diag_ghtt}
        \end{aligned}
\end{equation}
where $q$ denotes the momentum of the Higgs. The part involving the one-loop tadpole integral in Eq.~\eqref{eq:diag_ghtt}
is expressed in terms of the on-shell mass counter term $\delta{m_t^\text{4-top}}$ such
that the effect of on-shell $m_t$ renormalisation on the correction of the 
Yukawa interaction is made obvious.
In order to derive the necessary counter term for $C_{tH}$, 
it is sufficient to consider the case of the Higgs being on-shell.
Renormalising
$C_{tH}^{(0)}=\mu^{3\epsilon}\left(C_{tH}+\delta_{C_{tH}}^{C_i}\right)$ in the $\msbar$ scheme then leads to
\begin{equation}
    \begin{split}
       \delta_{C_{tH}}^\text{4-top}=\frac{\left(4\pi e^{-\gamma_E}\right)^\eps}{16\pi^2\eps}\frac{2\sqrt{2}m_t\left(4m_t^2-m_h^2\right)}{v^3}\left(C_{Qt}^{(1)}+c_F C_{Qt}^{(8)}\right)\;,
    \end{split}
\end{equation}
which coincides with $\delta_{C_{i}}^{C_j}=\frac{\left(4\pi e^{-\gamma_E}\right)^\eps}{16\pi^2\eps}\frac{\gamma_{C_i,C_j}}{2}C_j$ using 
the respective part of the anomalous dimension matrix $\gamma_{C_i,C_j}$ of Refs.~\cite{Jenkins:2013wua,Jenkins:2013zja}.\footnote{
    Cf. Appendix B of Ref.~\cite{DiNoi:2023ygk} for the derivation of the factor $\frac{1}{2}$ in the relation between anomalous dimension and counter term.}
With the additional counter term diagrams of $\delta{m_t^\text{4-top}}$ and $\delta_{C_{tH}}^\text{4-top}$ 
the diagram classes (a), (b) and (c) of Fig.~\ref{fig:feyndiag_4top} are made finite, and
we write schematically
\begin{equation}
    \begin{aligned}
    &\vcenter{\hbox{\begin{tikzpicture} 
        \begin{feynman}[small]
            \vertex  (g1)  {$g$};
            \vertex  (gtt1) [dot,scale=\sizedot,right =\lentrismall+\lentrieq/2 pt of g1] {};
            \vertex (4F) [dot, scale=\sizesqdot,below right=\lentrieq pt of gtt1,color=gray] {};
            \vertex  (gtt2) [dot,scale=\sizedot,below left =\lentrieq pt  of 4F] {};
            \vertex  (g2) [left=\lentrismall+\lentrieq/2 pt of gtt2]  {$g$};
            \vertex (htt) [dot,scale=\sizedot,right = \lentrieq/2 pt of 4F] {};
            \vertex (hhh) [dot,scale=\sizedot,right =\lentrieq/2 pt of htt] {};
            \vertex (h1) [above right =\lenboxeq pt  of hhh] {$h$};
            \vertex (h2) [below right =\lenboxeq pt  of hhh] {$h$};
    
            \diagram* {
                (g1)  -- [gluon] (gtt1),
                (g2) -- [gluon] (gtt2),
                (hhh)  -- [scalar] (htt),
                (h1) -- [scalar] (hhh) -- [scalar] (h2),
                (gtt1) -- [fermion, line width = 1.5 pt] (4F) 
                -- [fermion, line width = 1.5 pt] (gtt2)
                 -- [fermion, line width = 1.5 pt] (gtt1), 
                 (4F) -- [fermion, line width = 1.5 pt, half left] (htt) -- [fermion, line width = 1.5 pt, half left] (4F)
            };
        \end{feynman}
    \end{tikzpicture}}}
    +
    \vcenter{\hbox{\begin{tikzpicture} 
        \begin{feynman}[small]
            \vertex  (g1)  {$g$};
            \vertex  (gtt1) [dot,scale=\sizedot,right =\lentrismall+\lentrieq/2 pt of g1] {};
            \vertex (4F) [dot, scale=\sizesqdot,below right=\lentrieq pt of gtt1,color=gray] {};
            \vertex  (gtt2) [dot,scale=\sizedot,below left=\lentrieq pt of 4F] {};
            \vertex  (g2) [left=\lentrismall+\lentrieq/2 pt of gtt2]  {$g$};
            \vertex (htt1) [dot,scale=\sizedot,above right = \lentrieq pt of 4F] {};
            \vertex (htt2) [dot,scale=\sizedot,below right= \lentrieq pt of 4F] {};
            \vertex (h1) [right =\lentrismall+\lentrieq/2 pt  of htt1] {$h$};
            \vertex (h2) [right =\lentrismall+\lentrieq/2 pt  of htt2] {$h$};
    
            \diagram* {
                (g1)  -- [gluon] (gtt1),
                (g2) -- [gluon] (gtt2),
                (h1)  -- [scalar] (htt1),
                (h2) -- [scalar] (htt2),
                (gtt1) -- [fermion, line width = 1.5 pt] (4F) -- [fermion, line width = 1.5 pt] (gtt2) -- [fermion, line width = 1.5 pt] (gtt1), 
                (htt1) -- [fermion, line width = 1.5 pt] (4F) -- [fermion, line width = 1.5 pt] (htt2) -- [fermion, line width = 1.5 pt] (htt1), 
            };
        \end{feynman}
    \end{tikzpicture}}}&
    \\&\qquad\qquad+
    \vcenter{\hbox{\begin{tikzpicture} 
        \begin{feynman}[small]
            \vertex  (g1)  {$g$};
            \vertex  (gtt1) [dot,scale=\sizedot,right =\lentrismall+\lentrieq/2 pt of g1] {};
            \vertex (htt) [dot,scale=\sizesqdot,below right=\lentrieq pt of gtt1,color=gray!50] {};
            \node[shape=star,star points=4,star point ratio = 5,fill=black, draw,scale = 0.15, rotate=45] at (htt) {};
            \node[shape=circle,scale = \sizecirc,color=black,draw] at (htt) {};
            \vertex  (gtt2) [dot,scale=\sizedot,below left =\lentrieq pt  of htt] {};
            \vertex  (g2) [left=\lentrismall+\lentrieq/2 pt of gtt2]  {$g$};
            \vertex (hhh) [dot,scale=\sizedot,right =\lentrieq pt of htt] {};
            \vertex (h1) [above right =\lenboxeq pt  of hhh] {$h$};
            \vertex (h2) [below right =\lenboxeq pt  of hhh] {$h$};
    
            \diagram* {
                (g1)  -- [gluon] (gtt1),
                (g2) -- [gluon] (gtt2),
                (hhh)  -- [scalar] (htt),
                (h1) -- [scalar] (hhh) -- [scalar] (h2),
                (gtt1) -- [fermion, line width = 1.5 pt] (htt) 
                -- [fermion, line width = 1.5 pt] (gtt2)
                 -- [fermion, line width = 1.5 pt] (gtt1), 
            };
        \end{feynman}
    \end{tikzpicture}}}
    +
    \vcenter{\hbox{\begin{tikzpicture} 
        \begin{feynman}[small]
            \vertex  (g1)  {$g$};
            \vertex  (gtt1) [dot,scale=\sizedot,right =\lentrismall+\lentrieq/2 pt of g1] {};
            \vertex (htt) [dot,scale=\sizesqdot,below right=\lentrieq pt of gtt1,color=gray!50] {};
            \node[shape=star,star points=4,star point ratio = 5,fill=black, draw,scale = 0.15, rotate=45] at (htt) {};
            \node[shape=circle,scale = \sizecirc,color=black,draw] at (htt) {};
            \vertex  (gtt2) [dot,scale=\sizedot,below left =\lentrieq pt  of htt] {};
            \vertex  (g2) [left=\lentrismall+\lentrieq/2 pt of gtt2]  {$g$};
            \vertex (h1) [above right =\lenboxeq pt  of htt] {$h$};
            \vertex (h2) [below right =\lenboxeq pt  of htt] {$h$};
    
            \diagram* {
                (g1)  -- [gluon] (gtt1),
                (g2) -- [gluon] (gtt2),
                (h1) -- [scalar] (htt) -- [scalar] (h2),
                (gtt1) -- [fermion, line width = 1.5 pt] (htt) 
                -- [fermion, line width = 1.5 pt] (gtt2)
                 -- [fermion, line width = 1.5 pt] (gtt1), 
            };
        \end{feynman}
    \end{tikzpicture}}}
    &&=\frac{C_{Qt}^{(1)}+c_F C_{Qt}^{(8)}}{\Lambda^2}{\cal F}_{\bar{t}t\to hh}^\text{4-top}{\cal M}_\text{SM}^{gg\to h}
\\
&\qquad\qquad
\vcenter{\hbox{\begin{tikzpicture} 
    \begin{feynman}[small]
        \vertex  (g1)  {$g$};
        \vertex  (gtt1) [dot,scale=\sizedot,right =\lentrismall+\lentrieq/2 pt of g1] {};
        \vertex (4F) [dot, scale=\sizesqdot,right=\lenboxeq pt of gtt1,color=gray] {};
        \vertex  (gtt2) [dot,scale=\sizedot,below =\lenboxeq pt  of gtt1] {};
        \vertex  (g2) [left=\lentrismall+\lentrieq/2 pt of gtt2]  {$g$};
        \vertex (htt1) [dot,scale=\sizedot,right = \lentrieq/2 pt of 4F] {};
        \vertex (htt2) [dot,scale=\sizedot,below = \lenboxeq pt of 4F] {};
        \vertex (h1) [right =\lentrismall pt  of htt1] {$h$};
        \vertex (h2) [right =\lentrismall+\lentrieq/2 pt  of htt2] {$h$};

        \diagram* {
            (g1)  -- [gluon] (gtt1),
            (g2) -- [gluon] (gtt2),
            (h1)  -- [scalar] (htt1),
            (h2) -- [scalar] (htt2),
            (gtt1) -- [fermion, line width = 1.5 pt] (4F) -- [fermion, line width = 1.5 pt] (htt2)
            -- [fermion, line width = 1.5 pt] (gtt2)
             -- [fermion, line width = 1.5 pt] (gtt1), 
             (4F) -- [fermion, line width = 1.5 pt, half left] (htt1) -- [fermion, line width = 1.5 pt, half left] (4F)
        };
    \end{feynman}
\end{tikzpicture}}}
+
\vcenter{\hbox{\begin{tikzpicture} 
    \begin{feynman}[small]
        \vertex  (g1)  {$g$};
        \vertex  (gtt1) [dot,scale=\sizedot,right =\lentrismall+\lentrieq/2 pt of g1] {};
        \vertex (htt1) [dot,scale=\sizesqdot,right=\lenboxeq pt of gtt1,color=gray!50] {};
        \node[shape=star,star points=4,star point ratio = 5,fill=black, draw,scale = 0.15, rotate=45] at (htt1) {};
        \node[shape=circle,scale = \sizecirc,color=black,draw] at (htt1) {};
        \vertex  (gtt2) [dot,scale=\sizedot,below =\lenboxeq pt  of gtt1] {};
        \vertex  (g2) [left=\lentrismall+\lentrieq/2 pt of gtt2]  {$g$};
        \vertex (htt2) [dot,scale=\sizedot,below = \lenboxeq pt of 4F] {};
        \vertex (h1) [right =\lentrismall+\lentrieq/2 pt  of htt1] {$h$};
        \vertex (h2) [right =\lentrismall+\lentrieq/2 pt  of htt2] {$h$};

        \diagram* {
            (g1)  -- [gluon] (gtt1),
            (g2) -- [gluon] (gtt2),
            (h1)  -- [scalar] (htt1),
            (h2) -- [scalar] (htt2),
            (gtt1) -- [fermion, line width = 1.5 pt] (htt1) -- [fermion, line width = 1.5 pt] (htt2)
            -- [fermion, line width = 1.5 pt] (gtt2)
             -- [fermion, line width = 1.5 pt] (gtt1), 
        };
    \end{feynman}
\end{tikzpicture}}}
&&=2\frac{C_{Qt}^{(1)}+c_F C_{Qt}^{(8)}}{\Lambda^2}{\cal F}_{\bar{t}t\to h}^\text{4-top}{\cal M}_{\Box,\text{SM}}^{gg\to hh}\;,
\end{aligned}\label{eq:amp_yukawatype}
\end{equation}
where 
\begin{equation}
    \begin{aligned}
        {\cal F}_{\bar{t}t\to h}^\text{4-top}&=\frac{4m_t^2-m_h^2}{16\pi^2}\left(2B_0^\text{fin}\left(m_h^2,m_t^2,m_t^2\right)-1\right)\;,
        \\
        {\cal F}_{\bar{t}t\to hh}^\text{4-top}&=\frac{1}{16\pi^2v}\times
        \\&\left(2\frac{4m_t^2s+8m_h^2m_t^2-3m_h^2s}{s-m_h^2}B_0^\text{fin}\left(s,m_t^2,m_t^2\right)
        +16m_t^2B_0^\text{fin}\left(m_h^2,m_t^2,m_t^2\right)
        \right.\\& \left.
        +4m_t^2\left(8m_t^2-2m_h^2-s\right)C_0\left(m_h^2,m_h^2,s,m_t^2,m_t^2,m_t^2\right)
        +3s\frac{m_h^2-4m_t^2}{s-m_h^2}\right)\;,
    \end{aligned}\label{eq:fac_yukawatype}
\end{equation} 
and ${\cal M}_\text{SM}^{gg\to h}$ and ${\cal M}_{\Box,\text{SM}}^{gg\to hh}$ denote the SM $gg\to h$ amplitude and
the SM box-type contribution to the $gg\to hh$ amplitude, respectively.

Subsequently, we investigate contributions to the gauge interaction, as they appear in diagram classes (d) and (e) of Fig.~\ref{fig:feyndiag_4top}. 
It is sufficient to consider the case of an on-shell external gluon. Thus, the vertex correction evaluates to
\begin{equation}
    \begin{tikzpicture}[baseline=(4F)]
        \begin{feynman}[small]
            \vertex  (g1)  {$g$};
             \vertex (gtt1) [dot, scale=\sizedot, right= 30 pt of g1] {};
            \vertex  (4F) [dot,scale=\sizesqdot,right = 20 pt of gtt1, color = gray] {};
            \vertex  (t1) [above right= 25 pt of 4F] {$t$};
            \vertex (t2) [below right= 25 pt of 4F] {$t$};

            \diagram* {
                (g1)  -- [gluon] (gtt1),  
                (gtt1) -- [anti fermion, line width = 1.5 pt, half right] (4F) 
                -- [anti fermion, line width = 1.5 pt,half right] (gtt1),
                (t2) -- [fermion, line width = 1.5 pt]  (4F) -- [fermion, line width = 1.5 pt] (t1)
            };
        \end{feynman}
    \end{tikzpicture} = \frac{C_{Qt}^{(1)}+\left(c_F-\frac{c_A}{2}\right) C_{Qt}^{(8)}}{C_{tG}} K_{tG} \times 
            \begin{tikzpicture}[baseline=(4F)]
        \begin{feynman}[small]
            \vertex  (g1)  {$g$};
             \vertex (gtt1) [square dot, scale=\sizesqdot, right= 30 pt of g1, color=gray] {};
            \vertex  (t1) [above right= 25 pt of gtt1] {$t$};
            \vertex (t2) [below right= 25 pt of gtt1] {$t$};

            \diagram* {
                (g1)  -- [gluon] (gtt1),  
                (t2) -- [fermion, line width = 1.5 pt] (gtt1) -- [fermion, line width = 1.5 pt] (t1),};
        \end{feynman}
    \end{tikzpicture}
    \label{eq:diag_tg},
\end{equation}
where we defined 
\begin{equation}
    K_{tG}=-\frac{\sqrt{2}m_tg_s}{16\pi^2v}\;.\label{eq:KtG}
\end{equation}
Since the Lorentz structure of the correction to the gauge vertex 
is similar to the insertion of a chromomagnetic operator, 
diagrams in class (d) of Fig.~\ref{fig:feyndiag_4top} acquire a UV divergence (class (e) remains finite)
which, analogous to the case of the chromomagnetic operator, can be absorbed by a (now two-loop) counter term of $C_{HG}$. 
In $\msbar$ the explicit form is
\begin{equation}
    \begin{split}
       \delta_{C_{HG}}^\text{4-top}=\frac{\left(4\pi e^{-\gamma_E}\right)^{2\eps}}{\left(16\pi^2\right)^2\eps}\frac{-4g_s^2m_t^2}{v^2}T_F
       \left(C_{Qt}^{(1)}+\left(c_F-\frac{c_A}{2}\right) C_{Qt}^{(8)}\right)\;.
    \end{split}\label{eq:HG-4ferm-ct}
\end{equation}
Schematically, we now have
\begin{equation}
\begin{split}
\vcenter{\hbox{\begin{tikzpicture} 
    \begin{feynman}[small]
        \vertex  (g1)  {$g$};
        \vertex  (gtt1) [dot,scale=\sizedot,right =\lentrismall pt  of g1] {};
        \vertex (4F) [dot, scale = \sizesqdot,color=gray,right=\lentrieq/2 pt of gtt1] {};
        \vertex (htt) [dot,scale=\sizedot,below right =\lentrieq pt of 4F] {};
        \vertex  (gtt2) [dot,scale=\sizedot,below left =\lentrieq pt of htt] {};
        \vertex  (g2) [left=\lentrismall+\lentrieq/2 pt of gtt2]  {$g$};
        \vertex (hhh) [dot,scale=\sizedot,right =\lentrieq pt  of htt] {};
        \vertex (h1) [above right =\lenboxeq pt of hhh] {$h$};
        \vertex (h2) [below right =\lenboxeq pt of hhh] {$h$};
        \diagram* {
            (g1)  -- [gluon] (gtt1),
            (g2) -- [gluon] (gtt2),
            (hhh)  -- [scalar] (htt),
            (h1)  -- [scalar] (hhh) -- [scalar] (h2),
            (4F) -- [fermion, line width = 1.5 pt] (htt) 
            -- [fermion, line width = 1.5 pt] (gtt2) -- [fermion, line width = 1.5 pt] (4F), 
             (4F) -- [fermion, line width = 1.5 pt, half left] (gtt1) -- [fermion, line width = 1.5 pt, half left] (4F)
};
    \end{feynman}
\end{tikzpicture}}}
+
\vcenter{\hbox{\begin{tikzpicture} 
    \begin{feynman}[small]
        \vertex  (g1)  {$g$};
        \vertex (hgg) [square dot, scale=\sizesqdot,below right =\lenboxeq pt of g1,color=gray!50] {};
        \node[shape=star,star points=4,star point ratio = 5,fill=black, draw,scale = 0.15, rotate=45] at (hgg) {};
        \node[shape=rectangle,scale = \sizerectangle,color=black,draw] at (hgg) {};
        \vertex  (g2) [below left=\lenboxeq pt of hgg]  {$g$};
        \vertex (hhh) [dot,scale=\sizedot,right =\lentrieq pt  of hgg] {};
        \vertex (h1) [above right =\lenboxeq pt of hhh] {$h$};
        \vertex (h2) [below right =\lenboxeq pt of hhh] {$h$};
        \diagram* {
            (g1)  -- [gluon] (hgg),
            (g2) -- [gluon] (hgg),
            (hhh)  -- [scalar] (hgg),
            (h1)  -- [scalar] (hhh) -- [scalar] (h2),
};
    \end{feynman}
\end{tikzpicture}}}
&=
\frac{C_{Qt}^{(1)}+\left(c_F-\frac{c_A}{2}\right) C_{Qt}^{(8)}}{C_{tG}}K_{tG}\left({\cal M}_{tG}^{(a)}+{\cal M}_{tG}^{(b)}\right)
\\
\vcenter{\hbox{\begin{tikzpicture} 
    \begin{feynman}[small]
        \vertex  (g1)  {$g$};
        \vertex  (gtt1) [dot,scale=\sizedot,right =\lentrismall pt  of g1] {};
        \vertex (4F) [dot, scale = \sizesqdot,color=gray,right=\lentrieq/2 pt of gtt1] {};
        \vertex (htt1) [dot,scale=\sizedot,right =\lenboxeq pt of 4F] {};
        \vertex (htt2) [dot,scale=\sizedot,below =\lenboxeq pt of htt1] {};
        \vertex  (gtt2) [dot,scale=\sizedot,left =\lenboxeq pt of htt2] {};
        \vertex  (g2) [left=\lentrismall+\lentrieq/2 pt of gtt2]  {$g$};
        \vertex (h1) [right =\lentrismall+\lentrieq/2  pt of htt1] {$h$};
        \vertex (h2) [right =\lentrismall+\lentrieq/2  pt of htt2] {$h$};
        \diagram* {
            (g1)  -- [gluon] (gtt1),
            (g2) -- [gluon] (gtt2),
            (h1)  -- [scalar] (htt1),
            (htt2) -- [scalar] (h2),
            (4F) -- [fermion, line width = 1.5 pt] (htt1) -- [fermion, line width = 1.5 pt] (htt2) 
            -- [fermion, line width = 1.5 pt] (gtt2) -- [fermion, line width = 1.5 pt] (4F), 
             (4F) -- [fermion, line width = 1.5 pt, half left] (gtt1) -- [fermion, line width = 1.5 pt, half left] (4F)
};
    \end{feynman}
\end{tikzpicture}}}
&=
\frac{C_{Qt}^{(1)}+\left(c_F-\frac{c_A}{2}\right) C_{Qt}^{(8)}}{C_{tG}}K_{tG}{\cal M}_{tG}^{(c)}\;,
\end{split}\label{eq:gghh-gauge-4F}
\end{equation}
where ${\cal M}_{tG}^{(a/b/c/d)}$ denote the amplitude of diagram types (a), (b), (c) and (d) of Fig.~\ref{fig:feyndiag_chromo}, respectively.
The remaining diagrams of class (h) of Fig.~\ref{fig:feyndiag_4top} are made UV finite by the $gghh$ counter term vertex using precisely the same 
value of $\delta_{C_{HG}}^\text{4-top}$ which is an indication that eq.~(\ref{eq:HG-4ferm-ct}) is indeed the correct two-loop counter term.
Finally, we obtain
\begin{equation}
    \begin{aligned}
        &
        \vcenter{\hbox{\begin{tikzpicture} 
            \begin{feynman}[small]
                \vertex  (g1)  {$g$};
                \vertex  (gtt1) [dot,scale=\sizedot,right =\lentrismall+\lentrieq/2 pt of g1] {};
                \vertex (4F) [dot, scale=\sizesqdot,below right=\lentrieq pt of gtt1,color=gray] {};
                \vertex  (gtt2) [dot,scale=\sizedot,below left=\lentrieq pt of 4F] {};
                \vertex  (g2) [left=\lentrismall+\lentrieq/2 pt of gtt2]  {$g$};
                \vertex (htt1) [dot,scale=\sizedot,above right = \lentrieq pt of 4F] {};
                \vertex (htt2) [dot,scale=\sizedot,below right= \lentrieq pt of 4F] {};
                \vertex (h1) [right =\lentrismall+\lentrieq/2 pt  of htt1] {$h$};
                \vertex (h2) [right =\lentrismall+\lentrieq/2 pt  of htt2] {$h$};
        
                \diagram* {
                    (g1)  -- [gluon] (gtt1),
                    (g2) -- [gluon] (gtt2),
                    (h1)  -- [scalar] (htt1),
                    (h2) -- [scalar] (htt2),
                    (gtt1) -- [fermion, line width = 1.5 pt] (4F) -- [fermion, line width = 1.5 pt] (htt1) -- [fermion, line width = 1.5 pt] (gtt1), 
                    (gtt2) -- [fermion, line width = 1.5 pt] (4F) -- [fermion, line width = 1.5 pt] (htt2) -- [fermion, line width = 1.5 pt] (gtt2), 
                };
            \end{feynman}
        \end{tikzpicture}}}
        +
        \vcenter{\hbox{\begin{tikzpicture} 
            \begin{feynman}[small]
                \vertex  (g1)  {$g$};
                \vertex (hhgg) [square dot, scale=\sizesqdot,below right =\lenboxeq pt of g1,color=gray!50] {};
                \node[shape=star,star points=4,star point ratio = 5,fill=black, draw,scale = 0.15, rotate=45] at (hhgg) {};
                \node[shape=rectangle,scale = \sizerectangle,color=black,draw] at (hhgg) {};
                \vertex  (g2) [below left=\lenboxeq pt of hgg]  {$g$};
                \vertex (h1) [above right =\lenboxeq pt of hhgg] {$h$};
                \vertex (h2) [below right =\lenboxeq pt of hhgg] {$h$};
                \diagram* {
                    (g1)  -- [gluon] (hhgg),
                    (g2) -- [gluon] (hhgg),
                    (h1)  -- [scalar] (hhgg) -- [scalar] (h2),
        };
            \end{feynman}
        \end{tikzpicture}}}
        =
        \frac{C_{Qt}^{(1)}+\left(c_F-\frac{c_A}{2}\right)C_{Qt}^{(8)}}{C_{tG}}K_{tG}{\cal M}_{tG}^{(d)}
        \\&\qquad+\left[\frac{C_{QQ}^{(1)}+C_{tt}+\left(c_F-\frac{c_A}{2}\right)C_{QQ}^{(8)}}{\Lambda^2}
        +T_F\frac{C_{QQ}^{(8)}+C_{Qt}^{(8)}}{\Lambda^2}\right]
        {\cal M}_{\Delta QQ,tt,(8)}^\text{4-top}\;,
    \end{aligned}\label{eq:pair-ghgh}
\end{equation}        
where ${\cal M}_{\Delta QQ,tt,(8)}^\text{4-top}$ is a remaining amplitude piece 
for which we could not identify an expression in terms of a one-loop subamplitude. 
Note that Eq.~\eqref{eq:pair-ghgh} is the only appearance of a non-vanishing contribution of the operators in the class
$(LL)(LL)$ and $(RR)(RR)$ of Ref.~\cite{Grzadkowski:2010es} with coefficients $C_{QQ}^{(1)}$, $C_{QQ}^{(8)}$ and $C_{tt}$. 
Evaluating the bubbles in Eqs.~\eqref{eq:diag_ghtt} and \eqref{eq:diag_tg} (for on-shell gluons) without attaching the 4-top vertex 
only leads to 
scalar respective rank-2 tensor structures in Dirac space and therefore induces a chirality flip,
which is incompatible with a 4-top interaction of the same chirality in both currents. 
Similarly, the tadpole in Eq.~\eqref{eq:diag_mt} and the triangle with two Higgs bosons attached in Eq.~\eqref{eq:amp_yukawatype} 
have a scalar structure. 
The triangles of Eq.~\eqref{eq:pair-ghgh}, each with one external gluon and one Higgs boson attached, are the only exception, 
since they also have contributions proportional to a single Dirac matrix. 
These parts lead to the combination $C_{QQ}^{(1)}+C_{tt}+\left(c_F-\frac{c_A}{2}\right)C_{QQ}^{(8)}$ for 
the single trace contraction and in addition 
allow a contribution with two trace contractions involving the octet operators, 
which leads to the combination $T_F(C_{QQ}^{(8)}+C_{Qt}^{(8)})$, 
both multiplying ${\cal M}_{\Delta QQ,tt,(8)}^\text{4-top}$ in Eq.~\eqref{eq:pair-ghgh}. 

A few comments about the difference between the NDR and BMHV schemes are in order. 
In our calculation, the treatment of $\gamma_5$ in the two schemes differs only by the $2\eps$-dimensional part of the Dirac algebra in $D$-dimensions.
In the limit $D\to~4$ the renormalised fixed order result between the two schemes therefore differs by terms
stemming from the $2\eps$-dimensional parts of the
Dirac algebra multiplying a pole of the loop integrals.
In the 4-top calculation of this work, the BMHV results are obtained by removing the finite pieces in 
Eqs.~\eqref{eq:diag_mt}, \eqref{eq:mt-4F-ct}, \eqref{eq:diag_ghtt} and \eqref{eq:fac_yukawatype} that do not multiply a Passarino-Veltman scalar function, i.e.\ the {\em rational parts},
and setting $K_{tG}=0$ in Eqs.~\eqref{eq:diag_tg}, \eqref{eq:gghh-gauge-4F} and \eqref{eq:pair-ghgh}. 
These differences only affect the terms dependent on $C_{Qt}^{(1)}$ and  $C_{Qt}^{(8)}$.
This scheme dependence has the same structure as the one in the process $gg\to h$ which was observed in Ref.~\cite{DiNoi:2023ygk},
thus 
the scheme dependent amplitude structure of $C_{Qt}^{(1)}$ and $C_{Qt}^{(8)}$ is compensated by scheme dependent values for the other parameters of the Lagrangian, 
resulting in an overall scheme independence of the EFT prediction. 
The $\gamma_5$ schemes hence represent equivalent parameterisations of the new physics effects 
and a translation between the two schemes can be achieved by means of finite shifts of the Lagrangian parameters.
The explicit form of the translation relation 
between the NDR and the BMHV scheme 
in terms of parameter shifts is as follows
\begin{equation}
\begin{aligned}
    \delta{m_t^\text{4-top;\,\text{BMHV}}}&=\delta{m_t^\text{4-top}}-\frac{m_t^3}{8\pi^2\Lambda^2}\left(C_{Qt}^{(1)}+c_F C_{Qt}^{(8)}\right)
    \\
    C_{tH}^\text{BMHV}&=C_{tH}+\frac{\sqrt{2}m_t\left(4m_t^2-m_h^2\right)}{16\pi^2v^3}\left(C_{Qt}^{(1)}+c_F C_{Qt}^{(8)}\right)
    \\
    C_{tG}^\text{BMHV}&=C_{tG}-\frac{\sqrt{2}m_tg_s}{16\pi^2v}\left(C_{Qt}^{(1)}+\left(c_F-\frac{c_A}{2}\right)
      C_{Qt}^{(8)}\right)\;,
\end{aligned}\label{eq:scheme_translation}
\end{equation}
which is equivalent to the relations presented in Eqs.~(45)--(47) of Ref.~\cite{DiNoi:2023ygk}.\footnote{
    Note the different sign for $K_{tG}$ in Eq.~\eqref{eq:KtG} as a consequence of different convention for the covariant derivative.
    } 
These relations are best understood in a top-down perspective:
in an explicit matching calculation, different choices of $\gamma_5$ schemes naturally lead to relations like Eq.~\eqref{eq:scheme_translation}. 
Moreover, Eq.~\eqref{eq:scheme_translation} describes a mutual relation. 
One could define
parameter combinations in which the scheme dependence is absorbed, 
however this would require to define a `canonical scheme' instead of using an intrinsically scheme independent form. 
In order to avoid such an arbitrary choice in physical predictions of the EFT, simultaneous contributions of 
several Wilson coefficients which allow to disentangle the scheme dependence at a given order, together with a 
documentation of the chosen scheme, would be necessary.


\section{Implementation and usage of the code within the {\tt Powheg-Box}}
\label{sec:usage}

The analytic formulas of the previous section are implemented as an extension
to {\tt ggHH\_SMEFT}~\cite{Heinrich:2022idm} that 
already includes the combination of NLO QCD corrections with the leading operators and is publicly available in the 
framework of the {\tt POWHEG-BOX-V2}~\cite{Nason:2004rx,Frixione:2007vw,Alioli:2010xd}. 
Therefore, the calculation of 
the cross section at fixed order is extended by the subleading contributions in the form
of Eqs.~\eqref{eq:XS_expansion}-\eqref{eq:XSNLO_expansion}.


The subleading contributions 
 enter the calculation as part of the Born contribution. 
Since the loop functions are expressed in terms of one-loop integrals, the evaluation time 
per phase-space point of the subleading contributions is of the order of the 
existing Born contribution,
thus does not significantly change the run-time of the code.


The usage of the program {\tt ggHH\_SMEFT} follows the existing version with the extension
by a few parameters in the input card. An example is given in the
folder {\tt testrun} in the input card {\tt powheg.input-save}. 
The new Wilson coefficients of the subleading operators in Eq.~\eqref{eq:LagSMEFT} can be set with:
\begin{description}[leftmargin=!,labelwidth=\widthof{\qquad{\tt Lambda=1.0}\,: }]
  \itemsep0em 
  \item[\qquad{\tt CtG}\,:] { Wilson coefficient of chromomagnetic operator $C_{tG}$,}
  \item[\qquad{\tt CQt}\,:] { Wilson coefficient  of 4-top operator $C_{Qt}^{(1)}$, }
  \item[\qquad{\tt CQt8}\,:] { Wilson coefficient of 4-top operator $C_{Qt}^{(8)}$,}
  \item[\qquad{\tt CQQtt}\,:] { sum of Wilson coefficients of 4-top operators $C_{QQ}^{(1)}+C_{tt}$,}
  \item[\qquad{\tt CQQ8}\,:] { Wilson coefficient of 4-top operator $C_{QQ}^{(8)}$.}
\end{description}

The available options for the selection of cross section contributions from EFT operators 
are visualized in Table~\ref{tab:sigma_gghh_smeft}.
\begin{scriptsize}
  \begin{table}[htb]
    \begin{tabular}{| c | p{150pt} l|}
      \hline
      & &
      \\[-12pt]
      truncation & (a) & (b) 
      \\[-12pt]
      & &
      \\
      \hline
      & &
      \\[-12pt]
      \multicolumn{3}{c}{$\sigma_\text{EFT}^\text{Born}$}
      \\[-12pt]
      & &
      \\
      \hline
      & &
      \\[-12pt]
      {\tt includesubleading} & & 
      \\
      0 
      & $\sigma_\text{dim6}^\text{lead}\left[(g_s^2L)^{2}\Lambda^{-2}\right]$
      & $\sigma_{\text{dim6}^2}^{\text{lead}\times\text{lead}}\left[(g_s^2L)^{2}\Lambda^{-4}\right]$
      \\[+10pt]
      1 
      &$\sigma_\text{dim6}^{C_{tG},\text{4-top}}\left[(g_s^2L)^{2}{\mathbf L}\,\Lambda^{-2}\right]$
      &$\sigma_{\text{dim6}^2}^{(C_{tG},\text{4-top})\times\text{lead}}\left[(g_s^2L)^{2}{\mathbf L}\,\Lambda^{-2}\right]$
      \\[+10pt]
      2 
      &
      &$\sigma_{\text{dim6}^2}^{C_{tG}\times C_{tG}}\left[(g_s^2L)^{2}{\mathbf L}^2\,\Lambda^{-4}\right]$
      \\[-12pt]
      & &
      \\
      \hline
      & &
      \\[-12pt]
      \multicolumn{3}{c}{$\sigma_\text{EFT}^\text{NLO}$}
      \\[-12pt]
      & &
      \\
      \hline
      & &
      \\[-12pt]
      & $\sigma_\text{dim6}^{\text{NLO, }\text{lead}}\left[(g_s^2L)^{3}\Lambda^{-2}\right]$
      & $\sigma_{\text{dim6}^2}^{\text{NLO, }\text{lead}\times\text{lead}}\left[(g_s^2L)^{3}\Lambda^{-4}\right]$
      \\[-12pt]
      & &
      \\
      \hline
    \end{tabular}
    \caption{\label{tab:sigma_gghh_smeft}Options to select EFT contributions for the calculation of the cross section. 
    Columns denote the truncation options for the $1/\Lambda$--expansion, rows show the 
    selection of subleading operator contributions for the Born cross section in the upper part 
    and the NLO cross section in the lower part which is untouched by the setting of {\tt includesubleading}. 
    The partial cross section contributions are understood to be added to the SM, 
    a higher setting for the selection always includes the previous contributions as well. 
    Note that {\tt includesubleading=2} requires the {\tt bornonly} mode.}
  \end{table}
  \end{scriptsize}  
The structure of the code still allows the user to choose all truncation options described in Ref.~\cite{Heinrich:2022idm}. 
However, including the subleading contributions, only options
(a) (SM+linear dimension-6) and (b) (SM+linear dimension-6+quadratic dimension-6) are available,
as the other options are not meaningful in combination with the subleading operators.
The subleading contributions are activated through the keyword {\tt includesubleading} which can 
be set to \texttt{0}, \texttt{1} or \texttt{2}. When {\tt includesubleading=0} the 
subleading contributions are not included and the program behaves as the previous {\tt ggHH\_SMEFT}
version, i.e.\ the values for {\tt CtG}, {\tt CQt}, {\tt CQt8}, {\tt CQQtt} and {\tt CQQ8} are ignored.
With {\tt includesubleading=1} the subleading contributions enter --
according to the power counting --
only in the interference with the leading LO matrix elements. The setting {\tt includesubleading=2} is only available 
in {\tt bornonly} mode.
This allows the user to remain completely agnostic about possible UV extensions such that $C_{tG}$
is treated as if it was part of the leading operator contribution,
i.e.\ allowing squared $C_{tG}$-contributions to 
$|{\cal M}_\text{dim-6}|^2$ in truncation option (b). However, no NLO QCD corrections to the 
squared $C_{tG}$-part are available.

In addition, there is an option for 4-top contributions to choose between the NDR scheme ({\tt GAMMA5BMHV=0})
and the BMHV scheme ({\tt GAMMA5BMHV=1}) with the definition of chiral vertices according to Eq.~\eqref{eq:chiral_vert}.
As described at the end of Section \ref{sec:ME4t}, this will only affect the dependence on
{\tt CQt} and {\tt CQt8}.

\section{Results}
\label{sec:results}

The  results presented in the following were obtained for a centre-of-mass energy of 
$\sqrt{s}=13.6$\,TeV 
using the PDF4LHC15{\tt\_}nlo{\tt\_}30{\tt\_}pdfas~\cite{Butterworth:2015oua}
parton distribution functions, interfaced to our code via
LHAPDF~\cite{Buckley:2014ana}, along with the corresponding value for
$\alpha_s$.  We used $m_h=125$\,GeV for the mass of the Higgs boson; the top quark mass has been fixed to  $m_t=173$\,GeV to be coherent with the virtual two-loop amplitude calculated numerically,  and the top quark and Higgs widths
have been set to zero.
Jets are clustered with the anti-$k_T$ algorithm~\cite{Cacciari:2008gp} as
implemented in the FastJet package~\cite{Cacciari:2005hq,
Cacciari:2011ma}, with jet radius $R=0.4$ and a minimum transverse momentum
$p_{T,\mathrm{min}}^{\rm{jet}}=20$\,GeV. We set the central renormalisation and factorisation
scales to $\mu_R=\mu_F=m_{hh}/2$. We use 3-point scale variations unless specified otherwise.

\subsection{Total cross sections and heat maps}
\label{sec:totalXSresults}
In this subsection we investigate the dependence of the total cross section on the contribution of 
subleading operators. 
Following the decomposition of the cross section in Eqs.~\eqref{eq:XS_expansion}, \eqref{eq:XSBorn_expansion} and \eqref{eq:XSNLO_expansion},
these contributions only enter linearly in interference terms; we postpone the discussion of quadratic contributions from $C_{tG}$ to Section~\ref{sec:lvsl+q_chromo}.
The first part demonstrates the effect of variations of pairs of Wilson coefficients 
with respect to the SM configuration, where all contributions are included at LO QCD. In the second 
part, we present values for the total cross section of the SM and benchmark point 6 of 
Refs.~\cite{Heinrich:2022idm,Alasfar:2023xpc}  at NLO QCD and their dependence on variations of a single
subleading Wilson coefficient. The definition of benchmark point 6 in terms of SMEFT Wilson coefficients is given in Table\ \ref{tab:benchmarks}.
\begin{table}[htb]
  \begin{center}
   \renewcommand*{\arraystretch}{1.3}
    \begin{footnotesize}
\begin{tabular}{ |c|c|c|c|c|c||c|c|c|c| }
\hline
\begin{tabular}{c}
benchmark \\
\end{tabular}  & $C_{H,\text{kin}}$ & $C_{H}$ & $C_{tH}$ & $C_{HG}$\\
\hline
SM & $0$ & $0$ & $0$ & $0$\\
\hline
$6$ &  $0.561$ & $3.80$ & $2.20$ & $0.0387$\\
\hline
\end{tabular}
\end{footnotesize}
\end{center}
  \caption{\label{tab:benchmarks}Definition of  benchmark scenarios considered here in terms of SMEFT Wilson coefficients. 
  Benchmark point 6 refers to the set in Refs.~\cite{Heinrich:2022idm,Alasfar:2023xpc}, which is an updated version of Ref.~\cite{Capozi:2019xsi}. 
  The benchmarks were originally derived in a non-linear theory (HEFT), where benchmark point 6 corresponds to 
  $c_{hhh}=-0.684$,  $c_{tth}=0.9$,  $c_{tthh}= -\frac{1}{6}$, $c_{ggh}=0.5$,  $c_{gghh}=0.25$.
  A value of $\Lambda=1$\,TeV is assumed for the translation between HEFT and SMEFT coefficients and $C_{HG}$ is determined using $\alpha_s(m_Z)=0.118$.}
\end{table}
The ranges for the variation of
$C_H$ are oriented at a translation of the limits on $\kappa_{\lambda}$ from Ref.~\cite{ATLAS:2022jtk}, 
the ranges for the other Wilson coefficients are 
taken from Ref.~\cite{Ethier:2021bye} based on ${\cal O}(\Lambda^{-2})$ individual bounds or ${\cal O}(\Lambda^{-2})$ 
marginalised fits over the other Wilson coefficients.
Meanwhile, constraints on 4-fermion operators in the 3rd generation
also have been derived from the measurement of 4-top-quark
production~\cite{ATLAS:2023ajo,CMS:2023ftu}, based on fits varying each
Wilson coefficient individually, however we use the more conservative ranges here.
Note that, besides a flavour assumption, no a priori assumptions on the Wilson coefficients were made for the derivation of those limits,
such that their ranges include values where the truncation at ${\cal O}(\Lambda^{-2})$ and/or our power 
counting may not be valid, i.e. the value of $C_{tG}$ is not suppressed by a factor of $(16\pi^2)^{-1}$ and
the ranges for the 4-top Wilson coefficients, with values ${\cal O}(100)$, may be too large.\footnote{
  Interestingly, the conservative limits from the marginalised fits
  have values below one for $C_{tG}$ and 
 values of $\mathcal{O}(100)$ for $C_{Qt}^{(1)}$, such that the contribution of the scheme translation in Eq.~\eqref{eq:scheme_translation} 
  can be by accident of the same order or even larger than the original coefficient, inserting the numbers naively.
} 
The presented results using these ranges from marginalised fits 
should not be understood as predictions motivated by realistic UV effects, 
but rather investigate the potential for improvement in global fits, as the process $gg\to hh$ (and also $gg\to h$) probe 
directions that are complementary to the data points included so far.

\begin{figure}[htb]
  \begin{center}
  \includegraphics[width=.47\textwidth,page=1]{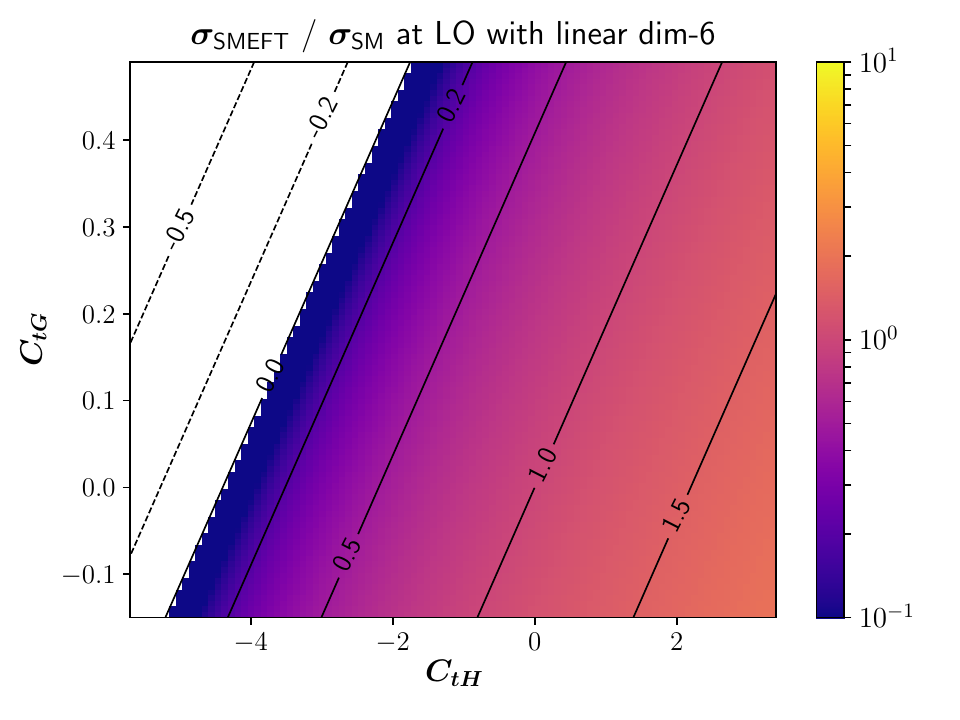}%
  \includegraphics[width=.47\textwidth,page=1]{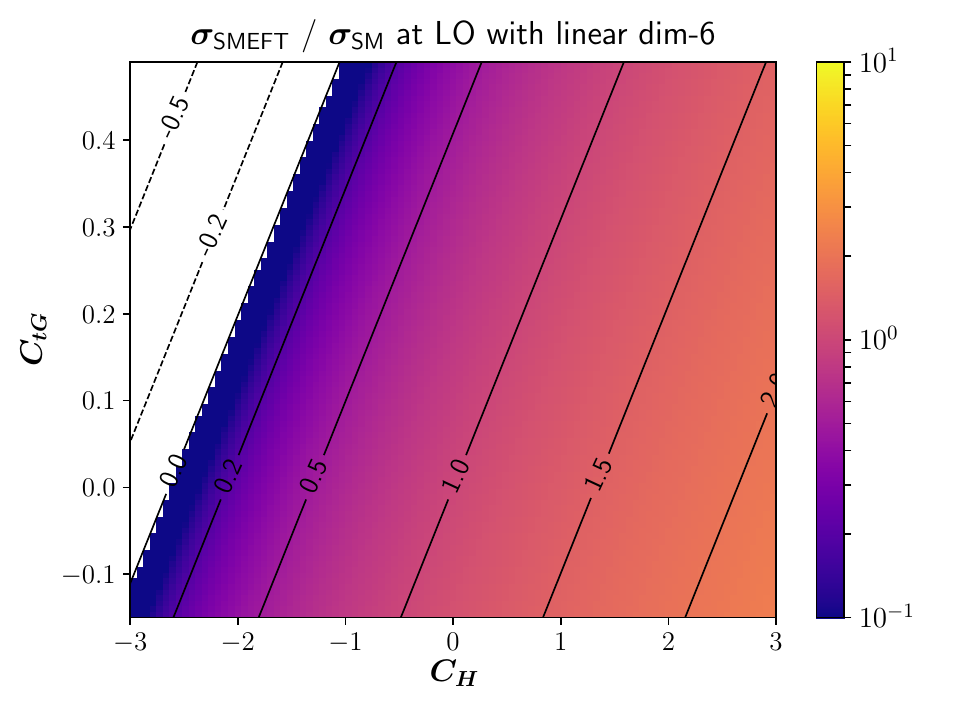}%
      \caption{\label{heatmaps_CtHCH_CtG}Heat maps showing
      the dependence of the LO cross section on the pair of Wilson coefficients $C_{tG}$, $C_{tH}$ (left) and $C_{tG}$, $C_{H}$ (right), respectively,  
     with $\Lambda=1$\,TeV for the linear dimension-6 truncation. 
     The ranges for $C_H$ are oriented at a translation of the limits on $\kappa_{\lambda}$ from Ref.~\cite{ATLAS:2022jtk},
     the ranges for the other Wilson coefficients are obtained at ${\cal O}\left(\Lambda^{-2}\right)$ constraints from Ref.~\cite{Ethier:2021bye} 
     (marginalised over the other coefficients).
      The white areas denote regions in parameter space where the corresponding cross
      section would be negative.}
  \end{center}
\end{figure}

Nonetheless, for the ranges of the Wilson coefficients in the following heat maps we use the marginalised ${\cal O}(\Lambda^{-2})$ bounds 
of Ref.~\cite{Ethier:2021bye} in order to cover a conservative parameter range.
In Fig.~\ref{heatmaps_CtHCH_CtG} we show heat maps illustrating the dependence of the LO QCD
cross section on the variation of $C_{tG}$ at the level of linear dimension-6 truncation (option (a)), compared 
to the leading couplings $C_{tH}$ and $C_{H}$, which corresponds to a comparison on equal footing.
The allowed ranges of Wilson coefficients are still quite large, such that a sizeable fraction
of the 2-dimensional parameter space leads to unphysical negative cross section values. As to be expected, the effect
of a variation of $C_{tG}$ within the given range is less
pronounced than the one from variations of the leading couplings $C_{tH}$ and $C_{H}$ within their range. 
From a power counting point of view, the allowed range for $C_{tG}$ should be 
much smaller, such that the difference of the impact on the cross section would be even more obvious.
Nevertheless, it is reasonable to derive bounds while being agnostic about the size of Wilson coefficients
as well as considering power counting arguments on the expected impact. The latter is the approach
we follow.

\begin{figure}[htb]
\begin{center}
\includegraphics[width=.47\textwidth,page=1]{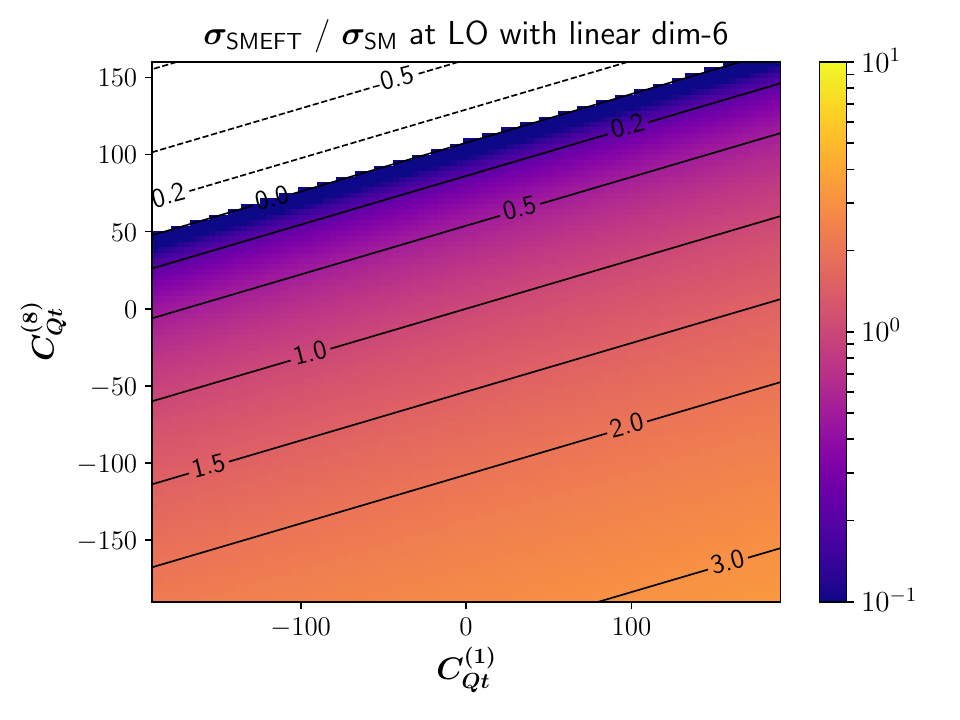}%
\includegraphics[width=.47\textwidth,page=1]{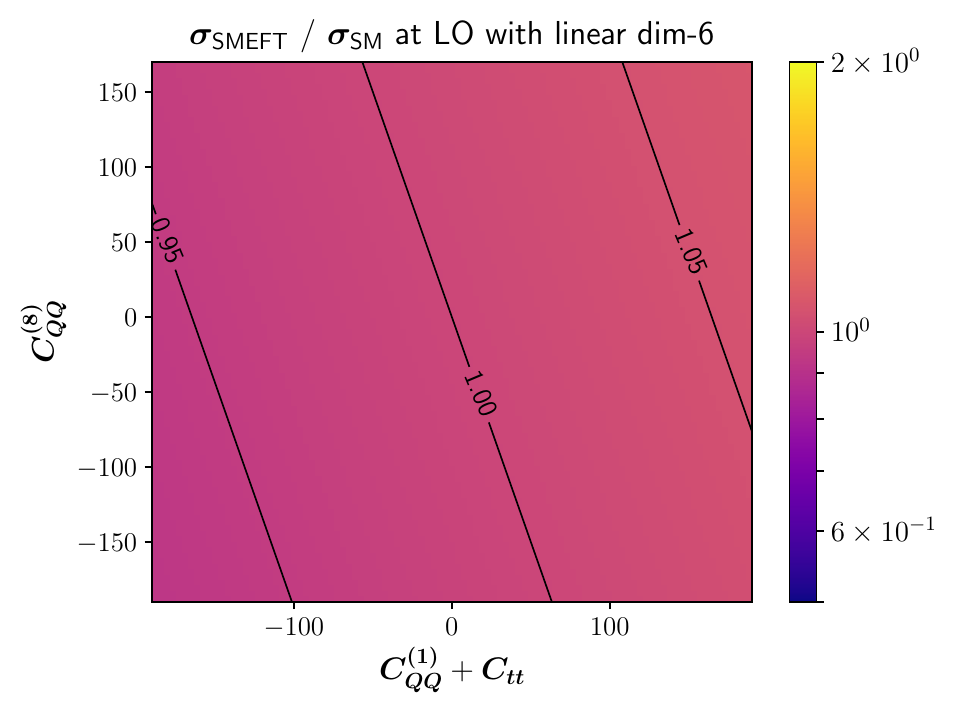}%
    \caption{\label{heatmaps_CLRCLR8_CLLRRCLLRR8}Heat maps showing
    the dependence of the cross section on the couplings $C_{Qt}^{(1)}$ and
    $C_{Qt}^{(8)}$ (left) and $C_{QQ}^{(1)}+C_{tt}$ and $C_{QQ}^{(8)}$ (right) with
    $\Lambda=1$\,TeV. 
    The ranges are taken from Ref.~\cite{Ethier:2021bye} based on an ${\cal O}(\Lambda^{-2})$ fit marginalised over the other Wilson coefficients.}
\end{center}
\end{figure}

In Fig.~\ref{heatmaps_CLRCLR8_CLLRRCLLRR8}, heat maps for the dependence of the cross section  on a variation
of (independent) 4-top operator pairs $C_{Qt}^{(1)}$, $C_{Qt}^{(8)}$ and $C_{QQ}^{(1)}+C_{tt}$, $C_{QQ}^{(8)}$ are shown.
Looking at the right plot it is apparent that the $(LL)(LL)$ and $(RR)(RR)$ operators of Ref.~\cite{Grzadkowski:2010es} 
with coefficients $C_{QQ}^{(1)}$, $C_{tt}$ and $C_{QQ}^{(8)}$
hardly affect the cross section. This can be understood by the very limited contribution to the amplitude, 
given only by the residual structure ${\cal M}_{\Delta QQ,tt,(8)}^\text{4-top}$ in Eq.~\eqref{eq:pair-ghgh}. 
On the other hand, the $(LL)(RR)$ operators, with coefficients $C_{Qt}^{(1)}$ and $C_{Qt}^{(8)}$,
(left plot of Fig.~\ref{heatmaps_CLRCLR8_CLLRRCLLRR8}) have a large impact on the cross section in the considered range of values,
leading to modifications of more than $100\%$ of the LO cross section.
The effect on the total cross section of $C_{Qt}^{(8)}$ is stronger than the effect of $C_{Qt}^{(1)}$ (in NDR), 
which is due to a large impact following from a sign change of the interference with the SM,  visible in 
the upper left diagram of Fig.~\ref{nlo_CLR_scheme}.

\begin{figure}[htb]
  \begin{center}
    \includegraphics[width=.47\textwidth,page=1]{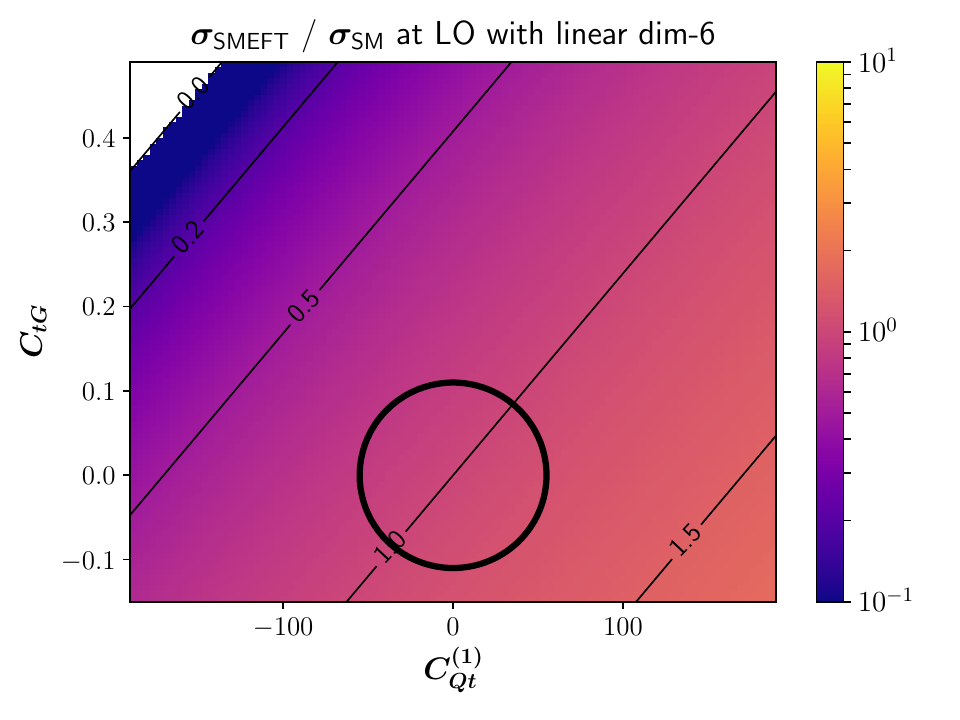}%
    \includegraphics[width=.47\textwidth,page=1]{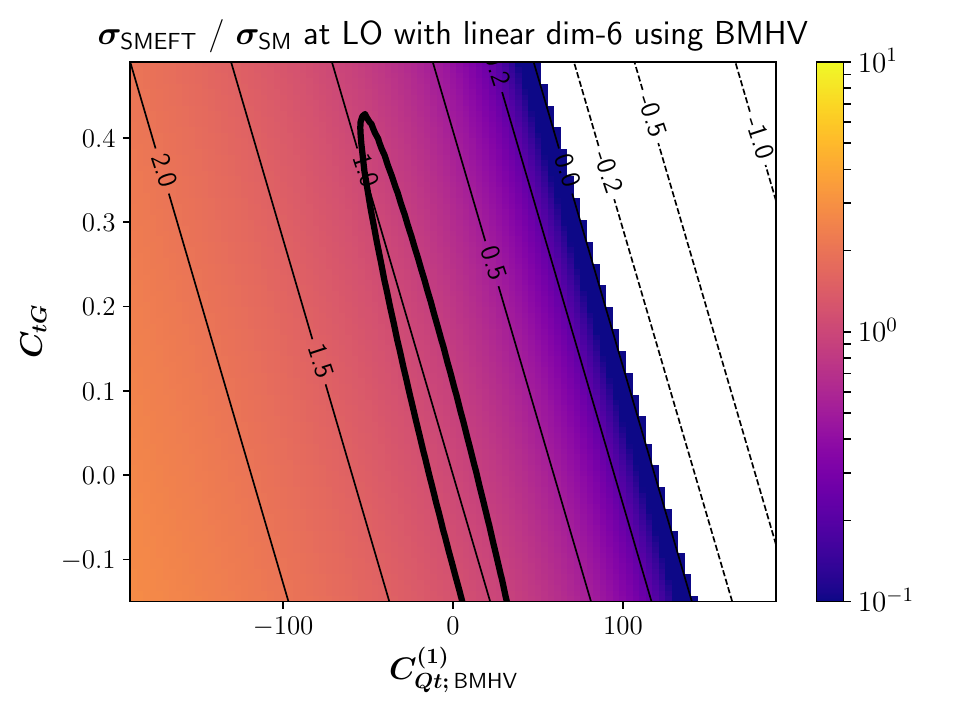}%
      \caption{\label{heatmaps_CLR_CtG_scheme}Heat maps demonstrating the effect of the $\gamma_5$-scheme choice
      on the dependence of the cross section on the couplings $C_{tG}$ and $C_{Qt}^{(1)}$ with $\Lambda=1$\,TeV. Left plot NDR, right plot BMHV.
      The ranges are taken from Ref.~\cite{Ethier:2021bye}, based on an ${\cal O}(\Lambda^{-2})$ fit marginalised over the other Wilson coefficients. 
     The areas within the black circle (left) and  within the ellipsis (right) demonstrate value pairs of Wilson coefficients
      that would be mapped into each other by using the relation for $C_{tG}$ in Eq.~\eqref{eq:scheme_translation}.}
  \end{center}
\end{figure}

Fig.~\ref{heatmaps_CLR_CtG_scheme} shows the dependence of the LO cross section on the variation of
$C_{tG}$ and $C_{Qt}^{(1)}$, comparing the NDR  and BMHV scheme choices for the chiral structure of the 
4-top operator.
We introduce $C_{Qt;\,\text{BMHV}}^{(1/8)}$ as a short-hand notation to specify that the corresponding {\em amplitude} is calculated in the BMHV scheme.
Hence, this does not mean that the value of $C_{Qt}^{(1/8)}$ itself is changed by the scheme choice.
This selection is an interesting showcase, since in Ref.~\cite{DiNoi:2023ygk} it has been demonstrated that the two Wilson coefficients are closely related, 
because part of the translation between the schemes is achieved by shifting $C_{tG}$, see Eq.~\eqref{eq:scheme_translation}. 
Supplementing SMEFT with a tree-loop classification of Wilson coefficients, 
these shifts are of equal order in the power counting as the original value of $C_{tG}$. In Fig.~\ref{heatmaps_CLR_CtG_scheme},
the gradient of the cross section in NDR (left) points in a completely different direction than the one in BMHV (right) 
and also the magnitude of the gradient changes significantly. 
The effect of the translation of $C_{tG}$ in Eq.~\eqref{eq:scheme_translation} is visualised by the areas surrounded by the black circle (left) and black ellipsis (right), respectively:
The relation for the scheme translation would map coefficient value pairs $(C_{Qt}^{(1)},C_{tG})$ from within the circle onto value pairs $(C_{Qt}^{(1)},C_{tG}^\text{BMHV})$ 
within the ellipsis and vice versa. 
Note that this does not describe the full scheme translation, as the shift in $C_{tH}$ of Eq.~\eqref{eq:scheme_translation} is not considered, 
however it is not as relevant as the shift in $C_{tG}$, as will become
clear in the discussion of Fig.~\ref{fig:rationalterms_smshifts}.
In addition, the shift of $C_{tG}$ depends on a scale dependent coupling $g_s$ which was set to a constant, thus these areas should be only understood as a qualitative visualisation. 
This clearly highlights that predictions using just $C_{Qt}$ would suffer from significant ambiguities if they are not considered 
in combination with $C_{tG}$, since the scheme differences can only be resolved if shifts of the form in Eq.~\eqref{eq:scheme_translation} are considered. 
Moreover, when the tree-loop classification of Wilson coefficients is applied, $C_{tG}$ would similarly suffer from ambiguities if $C_{Qt}$ was neglected.
In principle, this also holds for other processes where operators that are connected by similar relations enter at the same order.
This demonstrates that bounds set on these operators individually, without considering cancellations of the scheme dependence between different operator contributions, may not be very meaningful.

In Table~\ref{tab:sigmatot} we present values for the total cross section for the SM and benchmark point 6, 
using truncation options (a) and (b) at NLO QCD. 
We also demonstrate their dependence on the variation of a single subleading Wilson coefficient.
\begin{table}[htb]
    \begin{center}
    \begin{tabular}{| c || c | c | c |}
    \hline
    &&&\\[-10pt]
      BM & SM & 6 (a) & 6 (b)  \\[-10pt]
      &&&\\
        \hline
        \hline
        &&&\\[-10pt]
      $\sigma_{\rm{NLO}}$[fb] & 30.9$^{+14\%}_{-13\%}$ & 56.5$^{+22\%}_{-19\%}$ & 78.7$^{+18\%}_{-15\%}$
      \\[-10pt]
      &&&\\
      \hline
      \hline
      &&&\\[-10pt]
      $C_{tG}$ &&&\\
      $\scriptsize\begin{aligned}&\left[{0.0085}\right.,&\left.{0.14}\right]\\&\left[{-0.15}\right.,&\left.{0.49}\right]\end{aligned}$ 
    & 
    $\scriptsize\begin{aligned}&\left[{-0.63\%}\right.\hspace{-.25cm}&\left.{-10\%}\right]\\&\left[{+11\%}\right.\hspace{-.25cm}&\left.{-36\%}\right]\end{aligned}$ & 
    $\scriptsize\begin{aligned}&\left[{-0.34\%}\right.\hspace{-.25cm}&\left.{-5.6\%}\right]\\&\left[{+6.0\%}\right.\hspace{-.25cm}&\left.{-20\%}\right]\end{aligned}$ & 
    $\scriptsize\begin{aligned}&\left[{-0.26\%}\right.\hspace{-.25cm}&\left.{-4.3\%}\right]\\&\left[{+4.6\%}\right.\hspace{-.25cm}&\left.{-15\%}\right]\end{aligned}$ \\[-10pt] 
    &&&\\
      \hline
      \hline
      &&&\\[-10pt]
      $C_{Qt}^{(1)}$ &&&\\
      $\scriptsize\begin{aligned}&\left[{-200}\right.,&\left.{160}\right]\\&\left[{-190}\right.,&\left.{190}\right]\end{aligned}$ 
    & 
    $\scriptsize\begin{aligned}&\left[{-35\%}\right.\hspace{-.25cm}&\left.{+28\%}\right]\\&\left[{-34\%}\right.\hspace{-.25cm}&\left.{+34\%}\right]\end{aligned}$ & 
    $\scriptsize\begin{aligned}&\left[{-19\%}\right.\hspace{-.25cm}&\left.{+15\%}\right]\\&\left[{-18\%}\right.\hspace{-.25cm}&\left.{+18\%}\right]\end{aligned}$ & 
    $\scriptsize\begin{aligned}&\left[{+31\%}\right.\hspace{-.25cm}&\left.{-25\%}\right]\\&\left[{+30\%}\right.\hspace{-.25cm}&\left.{-30\%}\right]\end{aligned}$ \\[-10pt]
    &&&\\
      \hline
      \hline
      &&&\\[-10pt]
      $C_{Qt;\,\text{BMHV}}^{(1)}$ &&&\\
      $\scriptsize\begin{aligned}&\left[{-200}\right.,&\left.{160}\right]\\&\left[{-190}\right.,&\left.{190}\right]\end{aligned}$ 
    & 
    $\scriptsize\begin{aligned}&\left[{+101\%}\right.\hspace{-.25cm}&\left.{-81\%}\right]\\&\left[{+96\%}\right.\hspace{-.25cm}&\left.{-96\%}\right]\end{aligned}$ & 
    $\scriptsize\begin{aligned}&\left[{+55\%}\right.\hspace{-.25cm}&\left.{-44\%}\right]\\&\left[{+53\%}\right.\hspace{-.25cm}&\left.{-53\%}\right]\end{aligned}$ & 
    $\scriptsize\begin{aligned}&\left[{+88\%}\right.\hspace{-.25cm}&\left.{-71\%}\right]\\&\left[{+84\%}\right.\hspace{-.25cm}&\left.{-84\%}\right]\end{aligned}$ \\[-10pt]
    &&&\\
    \hline
    \hline
    &&&\\[-10pt]
      $C_{Qt}^{(8)}$ &&&\\
      $\scriptsize\begin{aligned}&\left[{-5.6}\right.,&\left.{20}\right]\\&\left[{-190}\right.,&\left.{160}\right]\end{aligned}$ 
    & 
    $\scriptsize\begin{aligned}&\left[{+3.2\%}\right.\hspace{-.25cm}&\left.{-11\%}\right]\\&\left[{+106\%}\right.\hspace{-.25cm}&\left.{-89\%}\right]\end{aligned}$ & 
    $\scriptsize\begin{aligned}&\left[{+1.7\%}\right.\hspace{-.25cm}&\left.{-6.1\%}\right]\\&\left[{+58\%}\right.\hspace{-.25cm}&\left.{-49\%}\right]\end{aligned}$ & 
    $\scriptsize\begin{aligned}&\left[{+3.1\%}\right.\hspace{-.25cm}&\left.{-11\%}\right]\\&\left[{+105\%}\right.\hspace{-.25cm}&\left.{-88\%}\right]\end{aligned}$ \\[-10pt]
    &&&\\
    \hline
    \hline
    &&&\\[-10pt]
      $C_{Qt;\,\text{BMHV}}^{(8)}$ &&&\\
      $\scriptsize\begin{aligned}&\left[{-5.6}\right.,&\left.{20}\right]\\&\left[{-190}\right.,&\left.{160}\right]\end{aligned}$ 
    & 
    $\scriptsize\begin{aligned}&\left[{+3.8\%}\right.\hspace{-.25cm}&\left.{-13\%}\right]\\&\left[{+127\%}\right.\hspace{-.25cm}&\left.{-107\%}\right]\end{aligned}$ & 
    $\scriptsize\begin{aligned}&\left[{+2.1\%}\right.\hspace{-.25cm}&\left.{-7.3\%}\right]\\&\left[{+69\%}\right.\hspace{-.25cm}&\left.{-58\%}\right]\end{aligned}$ & 
    $\scriptsize\begin{aligned}&\left[{+3.4\%}\right.\hspace{-.25cm}&\left.{-12\%}\right]\\&\left[{+114\%}\right.\hspace{-.25cm}&\left.{-96\%}\right]\end{aligned}$ \\[-10pt]
    &&&\\
    \hline
    \hline
    &&&\\[-10pt]
      $C_{QQ}^{(1)}+C_{tt}$ &&&\\
      $\scriptsize\begin{aligned}&\left[{-6.1}\right.,&\left.{23}\right]\\&\left[{-190}\right.,&\left.{190}\right]\end{aligned}$ 
    & 
    $\scriptsize\begin{aligned}&\left[{-0.11\%}\right.\hspace{-.25cm}&\left.{+0.42\%}\right]\\&\left[{-3.5\%}\right.\hspace{-.25cm}&\left.{+3.5\%}\right]\end{aligned}$ & 
    $\scriptsize\begin{aligned}&\left[{-0.061\%}\right.\hspace{-.25cm}&\left.{+0.23\%}\right]\\&\left[{-1.9\%}\right.\hspace{-.25cm}&\left.{+1.9\%}\right]\end{aligned}$ & 
    $\scriptsize\begin{aligned}&\left[{+0.094\%}\right.\hspace{-.25cm}&\left.{-0.36\%}\right]\\&\left[{+2.9\%}\right.\hspace{-.25cm}&\left.{-2.9\%}\right]\end{aligned}$ \\[-10pt]
    &&&\\
    \hline
    \hline
    &&&\\[-10pt]
      $C_{QQ}^{(8)}$ &&&\\
      $\scriptsize\begin{aligned}&\left[{-26}\right.,&\left.{58}\right]\\&\left[{-190}\right.,&\left.{170}\right]\end{aligned}$ 
    & 
    $\scriptsize\begin{aligned}&\left[{-0.16\%}\right.\hspace{-.25cm}&\left.{+0.35\%}\right]\\&\left[{-1.2\%}\right.\hspace{-.25cm}&\left.{+1.0\%}\right]\end{aligned}$ & 
    $\scriptsize\begin{aligned}&\left[{-0.087\%}\right.\hspace{-.25cm}&\left.{+0.19\%}\right]\\&\left[{-0.63\%}\right.\hspace{-.25cm}&\left.{+0.57\%}\right]\end{aligned}$ & 
    $\scriptsize\begin{aligned}&\left[{+0.13\%}\right.\hspace{-.25cm}&\left.{-0.30\%}\right]\\&\left[{+0.98\%}\right.\hspace{-.25cm}&\left.{-0.87\%}\right]\end{aligned}$ \\[-10pt]
    &&&\\
    \hline
    \end{tabular}
    \end{center}
    \caption{Total cross sections for Higgs-boson pair production at NLO QCD for the SM and benchmark point 6 
    using truncation option (a) or (b) at $13.6$~TeV.
      The modification of the cross section due to a variation of the subleading Wilson coefficients is given as relative change to the base value in the second row.
      The uncertainties in the second row are scale uncertainties based on 3-point scale variations.
      The ranges of the subleading Wilson coefficients are oriented at ${\cal O}\left(\Lambda^{-2}\right)$ constraints from Ref.~\cite{Ethier:2021bye} (Upper values: individual bounds, lower values: marginalised over the other coefficients).
      The effect of the Wilson coefficients $C_{Qt}^{(1)}$ and  $C_{Qt}^{(8)}$ is also shown for the BMHV scheme, 
      which is denoted by $C_{Qt;\,\text{BMHV}}^{(1)}$ and $C_{Qt;\,\text{BMHV}}^{(8)}$.
      \label{tab:sigmatot}}
\end{table}
In general, the relative difference due to the variation of these
Wilson coefficients  is more pronounced for the SM cross section than for benchmark point 6.

Due to the asymmetric range of $C_{tG}$, its variation tends to a damping of the cross section, with
up to $-36\%$ relative to the SM. 
For benchmark point 6, truncation (a) leads to a larger relative effect of $C_{tG}$ on the cross section than truncation (b).

The variation of single 4-top Wilson coefficients, on the other hand, is fairly symmetric for the marginalised limits 
and has larger relative impact for truncation option (b) than for truncation option (a). 
The cross section difference for a variation of $C_{Qt}^{(1)}$ or $C_{Qt}^{(8)}$ is larger when working in the BMHV scheme than in NDR, 
and the scheme difference is much more visible for $C_{Qt}^{(1)}$.
The $C_{Qt}^{(1)}$ variation leads to up to $\sim 35\%$ effects on the cross section in the NDR scheme 
and up to $\sim 100\%$ in BMHV, whereas for $C_{Qt}^{(8)}$ the maximum difference is in both schemes $\gtrsim 100\%$.
As already indicated by the heat map on the right of Fig.~\ref{heatmaps_CLRCLR8_CLLRRCLLRR8},
the effect of $C_{QQ}^{(1)}$, $C_{tt}$ and $C_{QQ}^{(8)}$ variation is very small, with a relative difference of less than $4\%$ 
and being only a fraction of the uncertainty due  to 3-point scale variations.
The effects of $C_{Qt}^{(1)}$ or $C_{Qt}^{(8)}$ on the difference $\Delta C_{tG}:=C_{tG}^\text{BMHV}-C_{tG}$ and $\Delta C_{tH}:=C_{tH}^\text{BMHV}-C_{tH}$ are illustrated later at distribution level in Fig.~\ref{fig:rationalterms_smshifts}.

\FloatBarrier
\subsection{Higgs boson pair invariant mass distributions}
In this section we present differential distributions depending on the invariant mass of 
the Higgs boson pair, $m_{hh}$, combining
NLO QCD results and subleading operator contributions at LO QCD. Each plot 
demonstrates the variation of a single subleading Wilson coefficient
w.r.t.\ either the SM or benchmark point 6 
for truncations (a) (linear dimension-6 only) and (b) (linear+quadratic dimension-6). 
The ranges we used are oriented at the ${\cal O}(\Lambda^{-2})$ marginalised fits of Ref.~\cite{Ethier:2021bye}.

\begin{figure}[htb]
\begin{center}
\includegraphics[width=.47\textwidth,page=1]{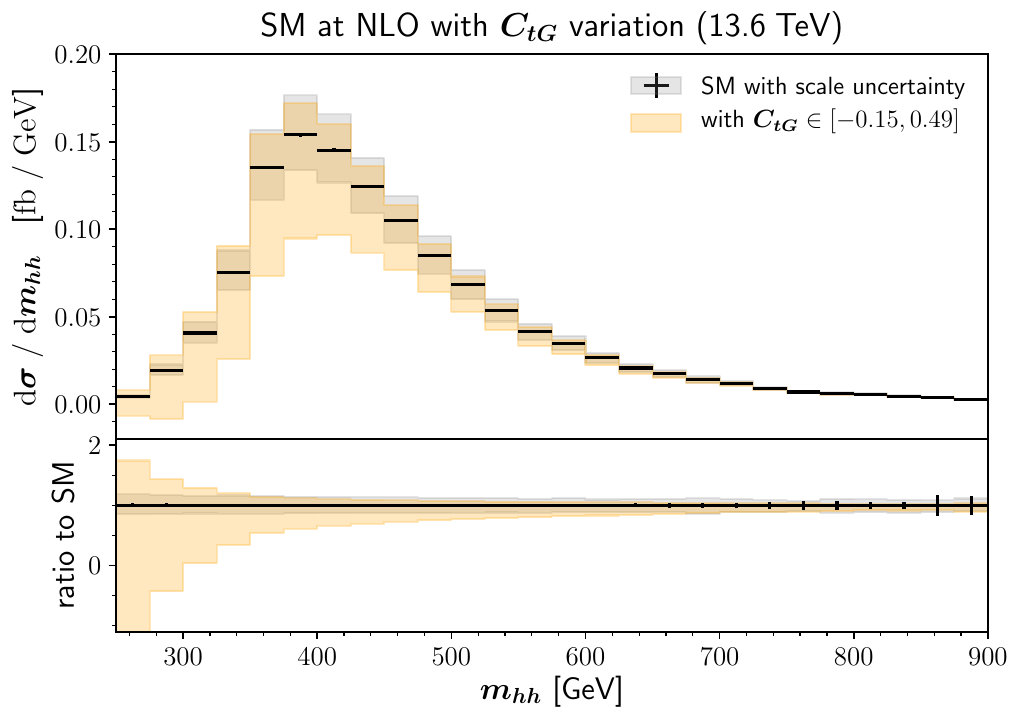}%
\includegraphics[width=.47\textwidth,page=1]{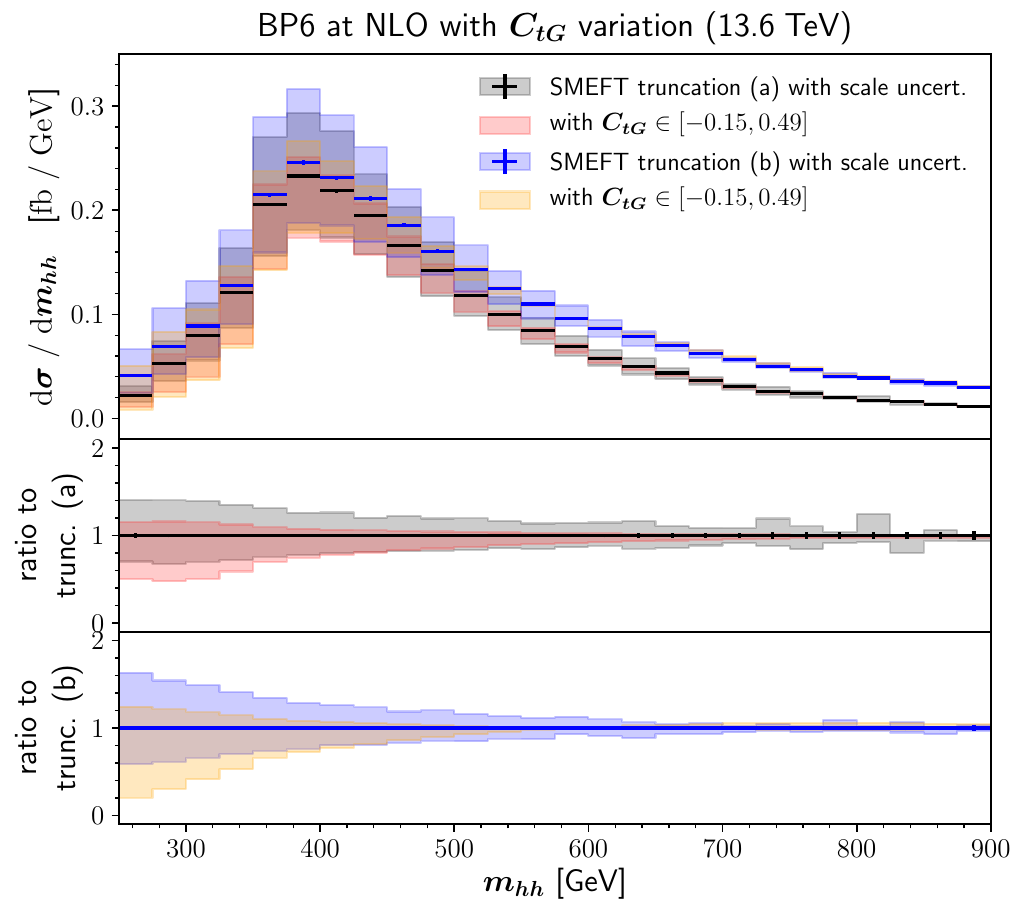}%
    \caption{\label{nlo_CtGvar}Effects of $C_{tG}$-variations on $\mhh$-distributions. 
    Left: variation w.r.t. the SM, right: variation w.r.t. benchmark point 6 (BP6) for truncation options (a) and (b).}
\end{center}
\end{figure}

In Fig.~\ref{nlo_CtGvar} the variation of the chromomagnetic operator coefficient $C_{tG}$ in the ranges specified in Table~\ref{tab:sigmatot} is shown.
In the  low $m_{hh}$-region, the effects can noticeably exceed the scale uncertainty band. 
Note that the  $C_{tG}$-variation range is asymmetric around zero and that the interference of the $C_{tG}$-term with the SM contribution tends to decrease the cross section.

\begin{figure}[htb]
\begin{center}
\includegraphics[width=.47\textwidth,page=1]{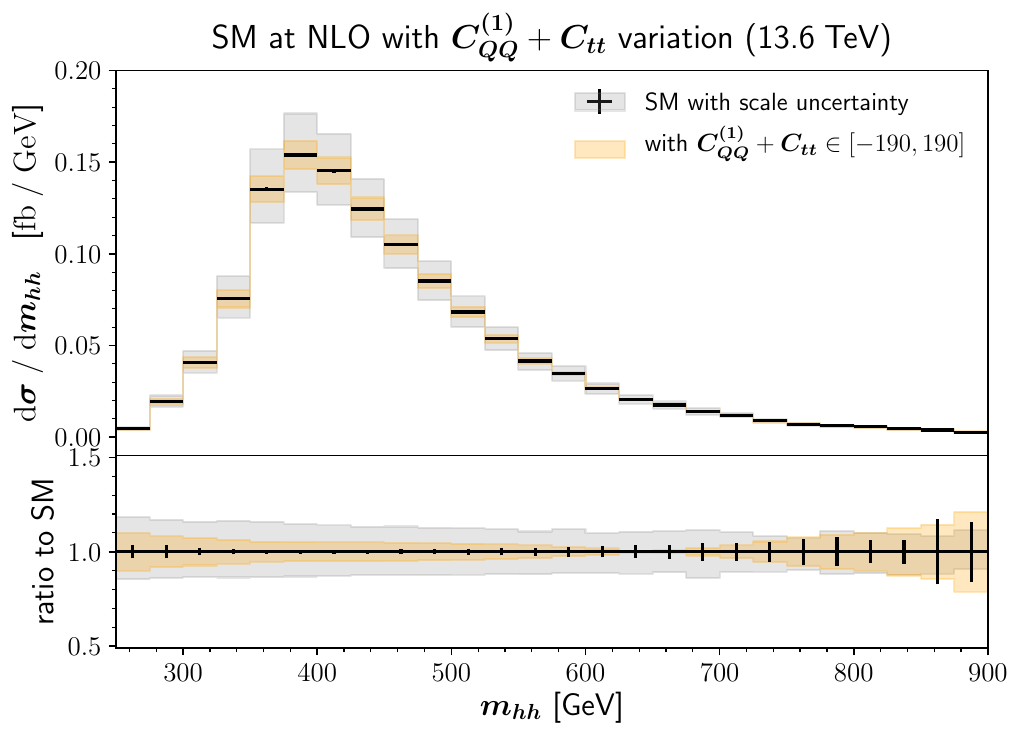}%
\includegraphics[width=.47\textwidth,page=1]{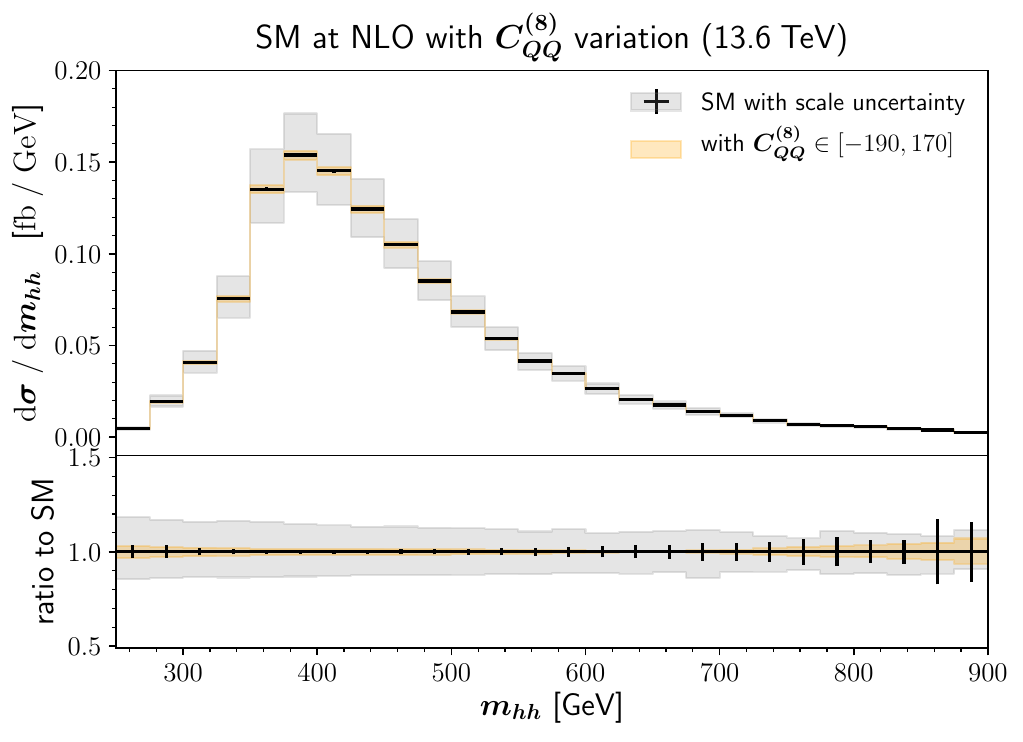}%
\\
\includegraphics[width=.47\textwidth,page=1]{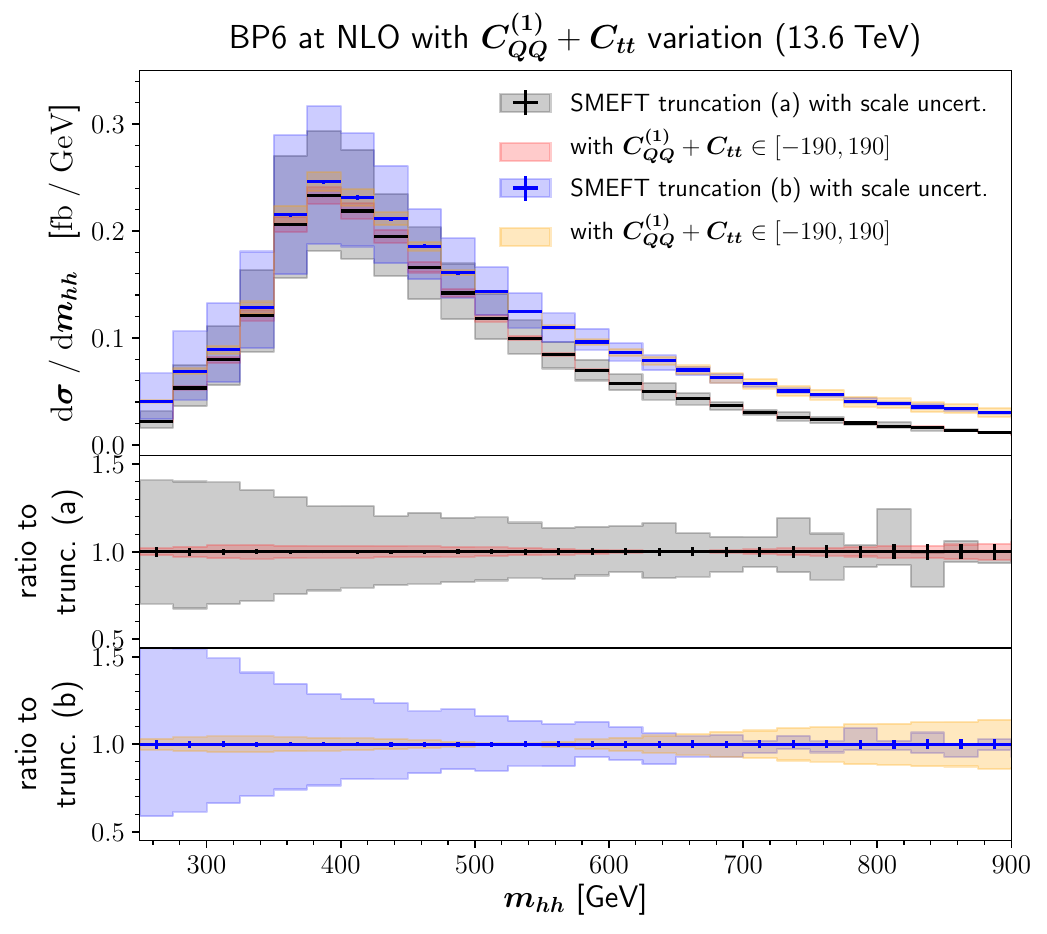}%
\includegraphics[width=.47\textwidth,page=1]{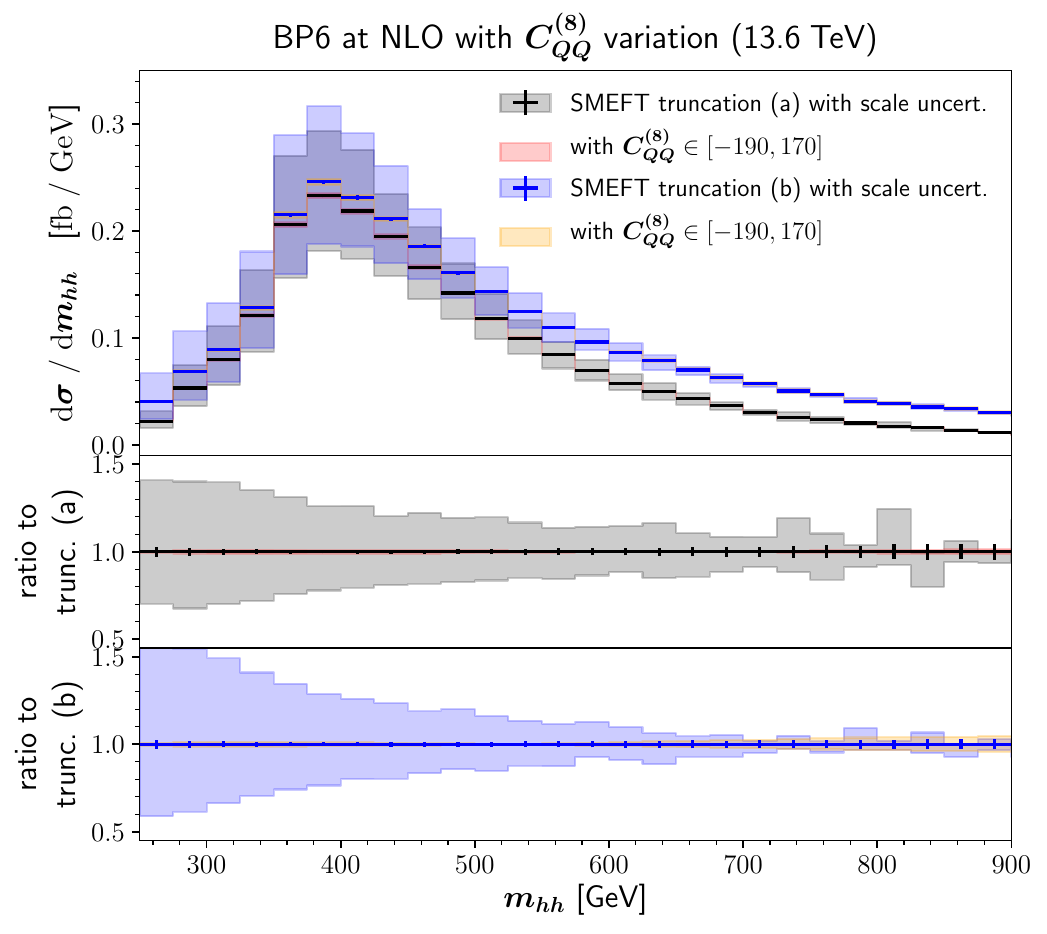}%
    \caption{\label{nlo_CLLRR_CLLRR8}Effects of $C_{QQ}^{(1)}+C_{tt}$ or $C_{QQ}^{(8)}$-variations on $\mhh$-distributions. 
    Left: variation of $C_{QQ}^{(1)}+C_{tt}$, right: variation of $C_{QQ}^{(8)}$; upper: SM baseline scenario, lower: benchmark point 6 for truncation options (a) and (b).}
\end{center}
\end{figure}

In Fig.~\ref{nlo_CLLRR_CLLRR8} we present the variation of the 4-top operator coefficient $C_{QQ}^{(8)}$ and the combination 
$C_{QQ}^{(1)}+C_{tt}$.
%
As observed at the level of total cross sections in Section \ref{sec:totalXSresults}, the contribution of these operators
remains within the scale uncertainties, except for small deviations in the tails 
for the case of $C_{QQ}^{(1)}+C_{tt}$.
Thus the process $gg\to hh$ is not sensitive to those operators even if the coefficients are varied in ranges as large as $ [-190,190]$.
The situation is different for the operators $C_{Qt}^{(1)}$ and $C_{Qt}^{(8)}$, as we will show below.
However, 
the contribution of these Wilson coefficients
depends on the chosen $\gamma_5$-scheme in dimensional regularisation, as explained in Section \ref{sec:ME4t}.

\begin{figure}[htb]
\begin{center}
\includegraphics[width=.47\textwidth,page=1]{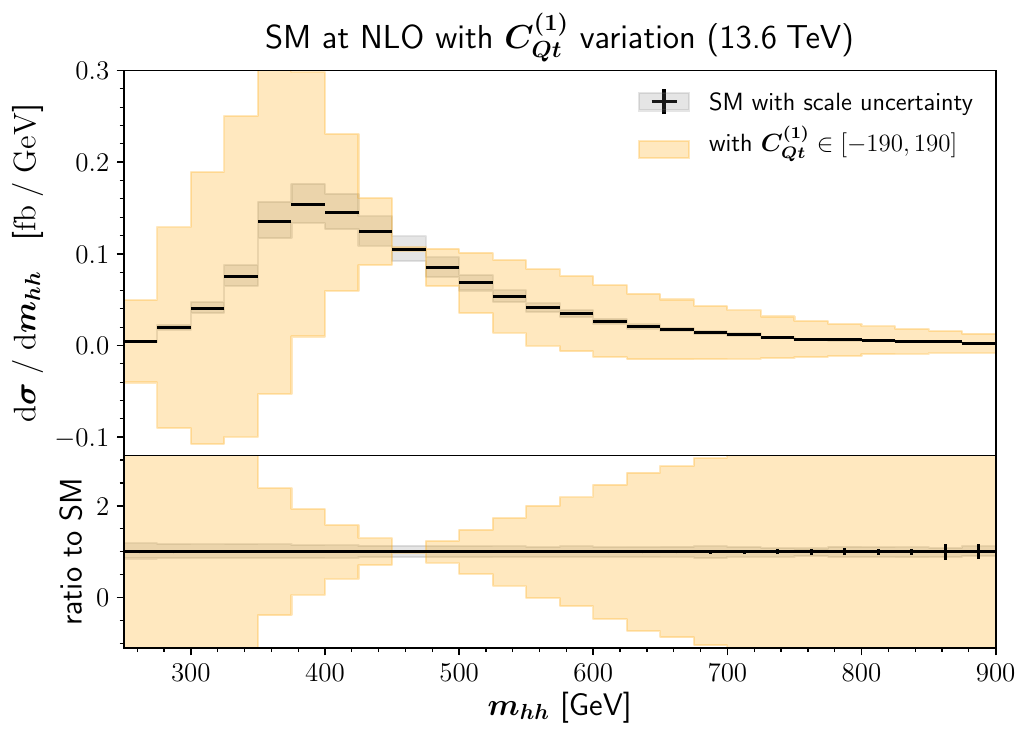}%
\includegraphics[width=.47\textwidth,page=1]{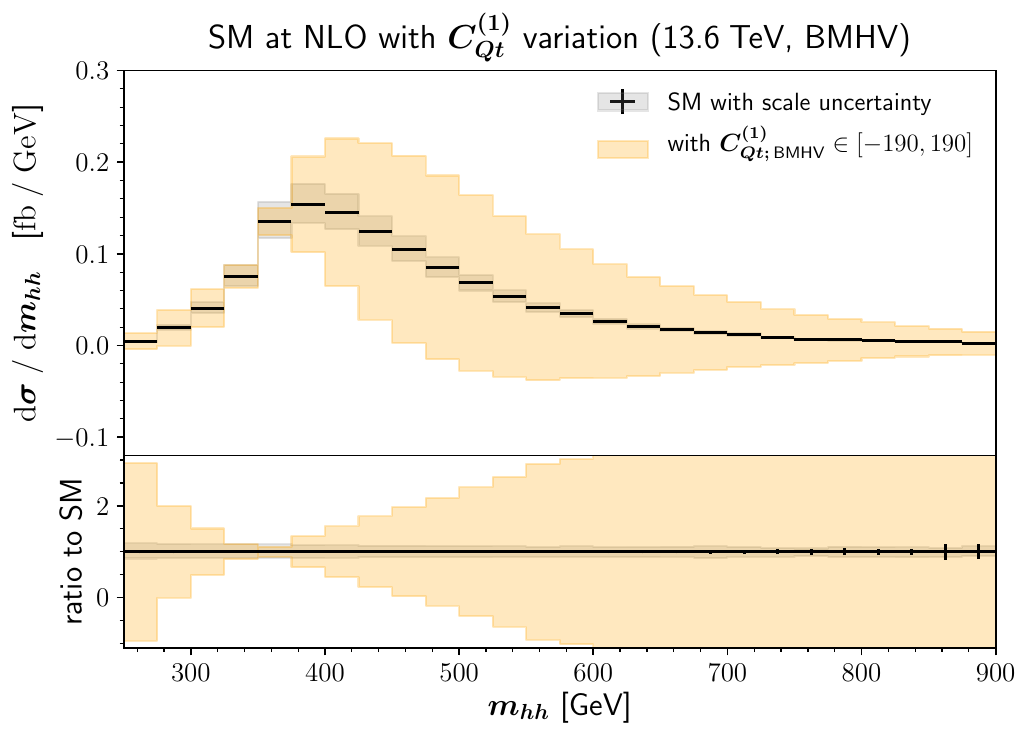}%
\\
\includegraphics[width=.47\textwidth,page=1]{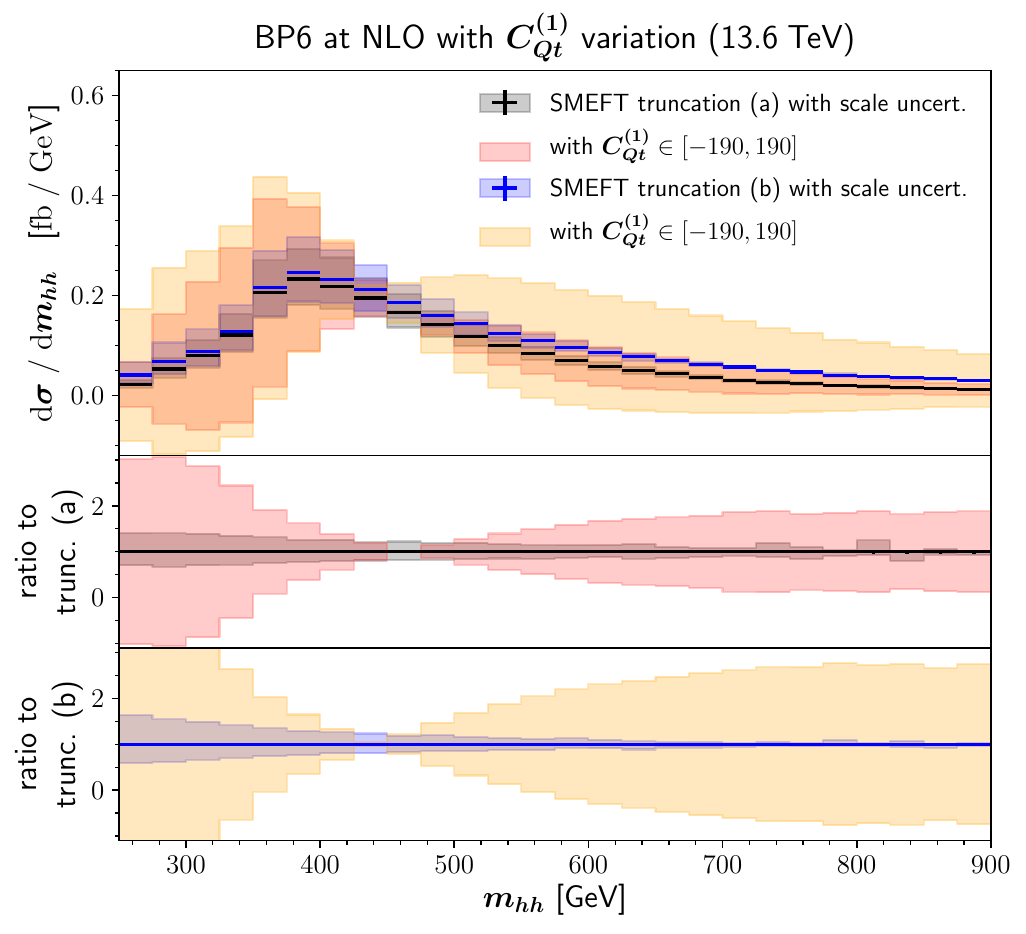}%
\includegraphics[width=.47\textwidth,page=1]{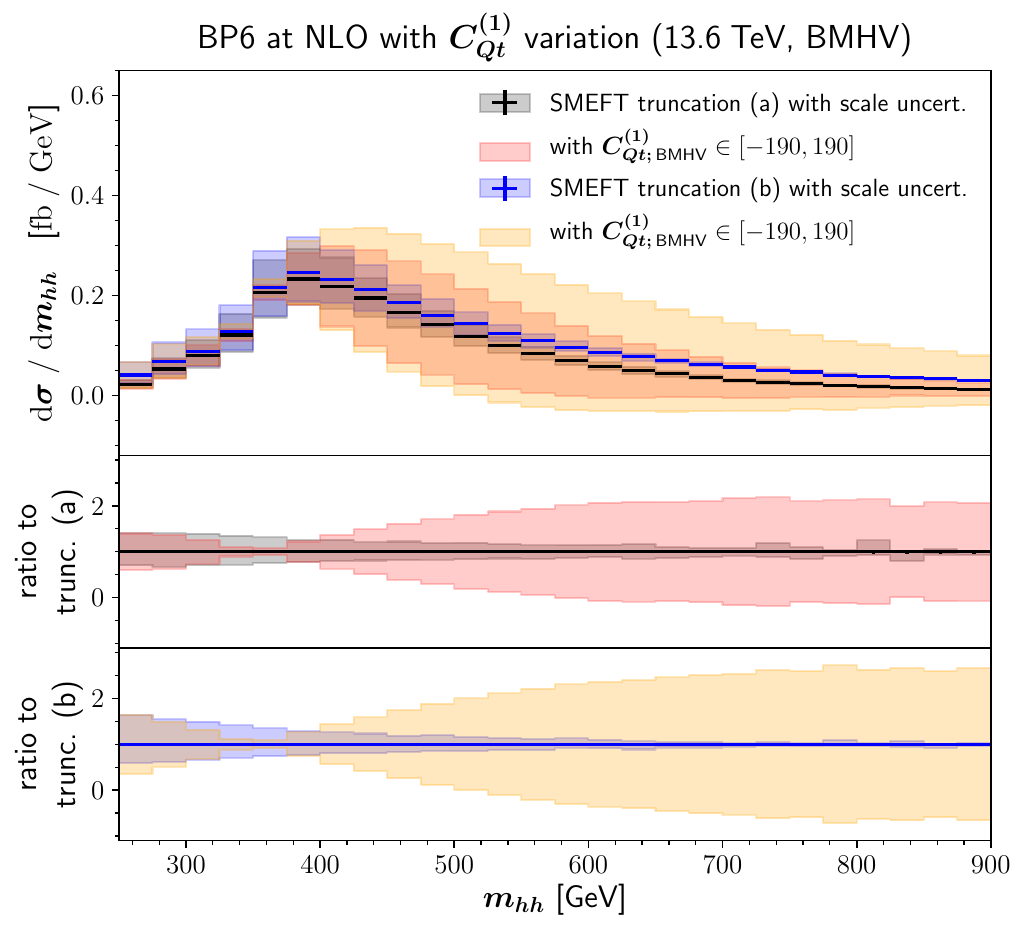}%
    \caption{\label{nlo_CLR_scheme}Effects of $C_{Qt}^{(1)}$-variations on $\mhh$-distributions comparing $\gamma_5$-schemes. 
    Left: NDR scheme, right: BMHV scheme; upper: SM baseline scenario, lower:  benchmark point 6 for truncation options (a) and (b).}
\end{center}
\end{figure}

We begin with Fig.~\ref{nlo_CLR_scheme} which demonstrates the effect of varying $C_{Qt}^{(1)}$.
%
We observe sizeable effects, differing from the baseline prediction (SM or benchmark 6) by more than $100\%$ for some regions,
which also leads to negative cross section values. 
In NDR, the low- and high $\mhh$-regions exhibit large differences beyond the scale uncertainty, 
with unphysical cross sections at low $\mhh$ values and a sign change around $m_{hh}\sim 460$\,TeV.
This behaviour changes significantly in BMHV: there are visible, but weaker effects in the low $m_{hh}$-region, 
the sign change occurs around $m_{hh}\sim 360$\,TeV and the deviation in the high $m_{hh}$-region begins for lower invariant masses and is also more pronounced. 

\begin{figure}[htb]
\begin{center}
\includegraphics[width=.47\textwidth,page=1]{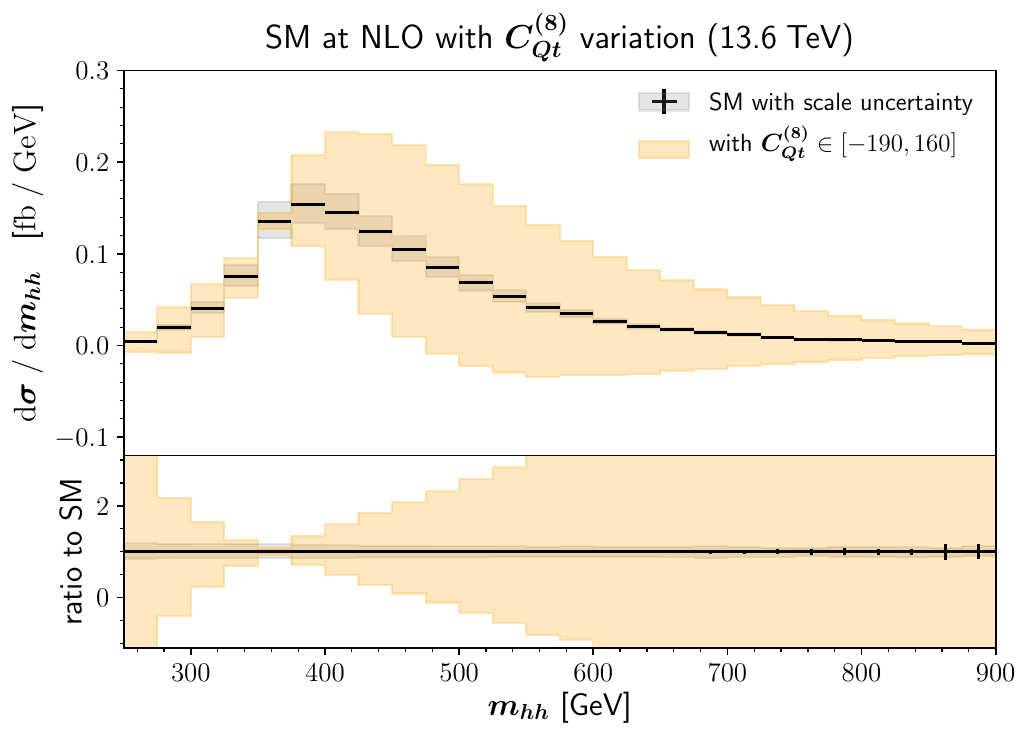}%
\includegraphics[width=.47\textwidth,page=1]{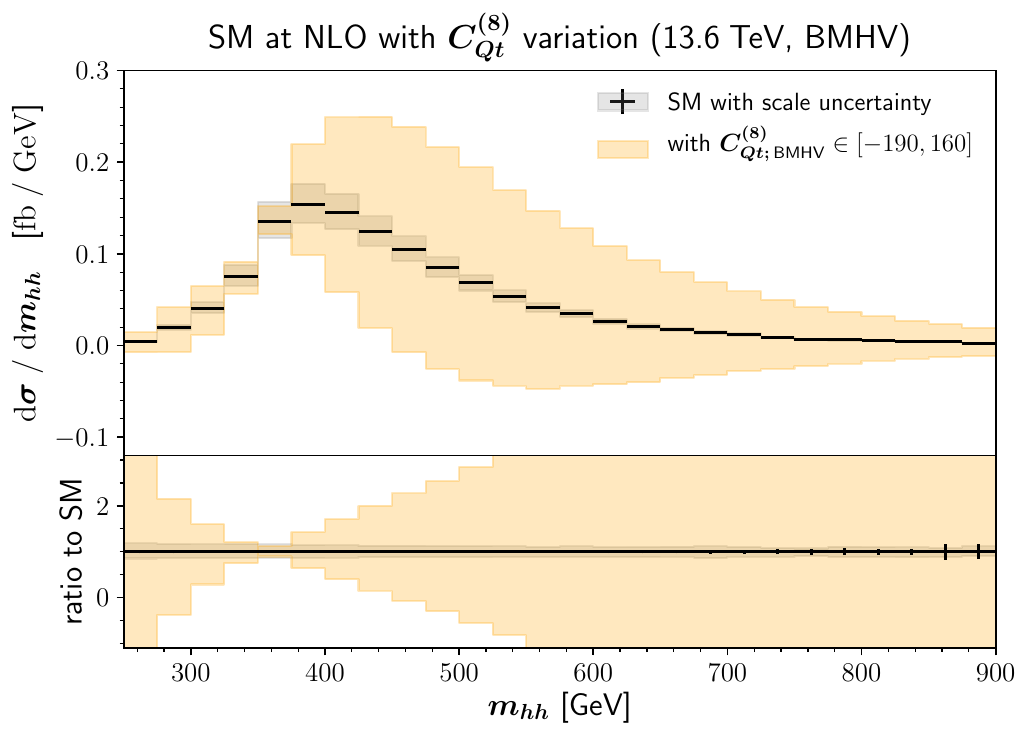}%
\\
\includegraphics[width=.47\textwidth,page=1]{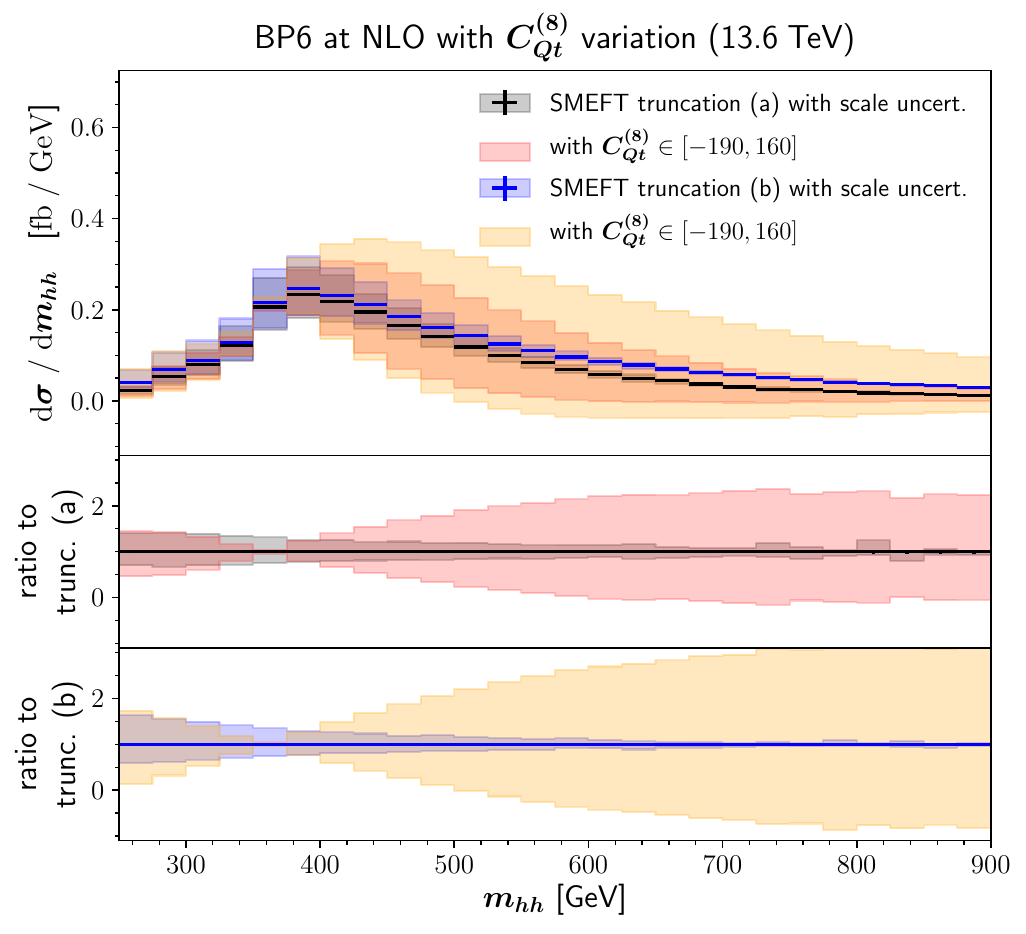}%
\includegraphics[width=.47\textwidth,page=1]{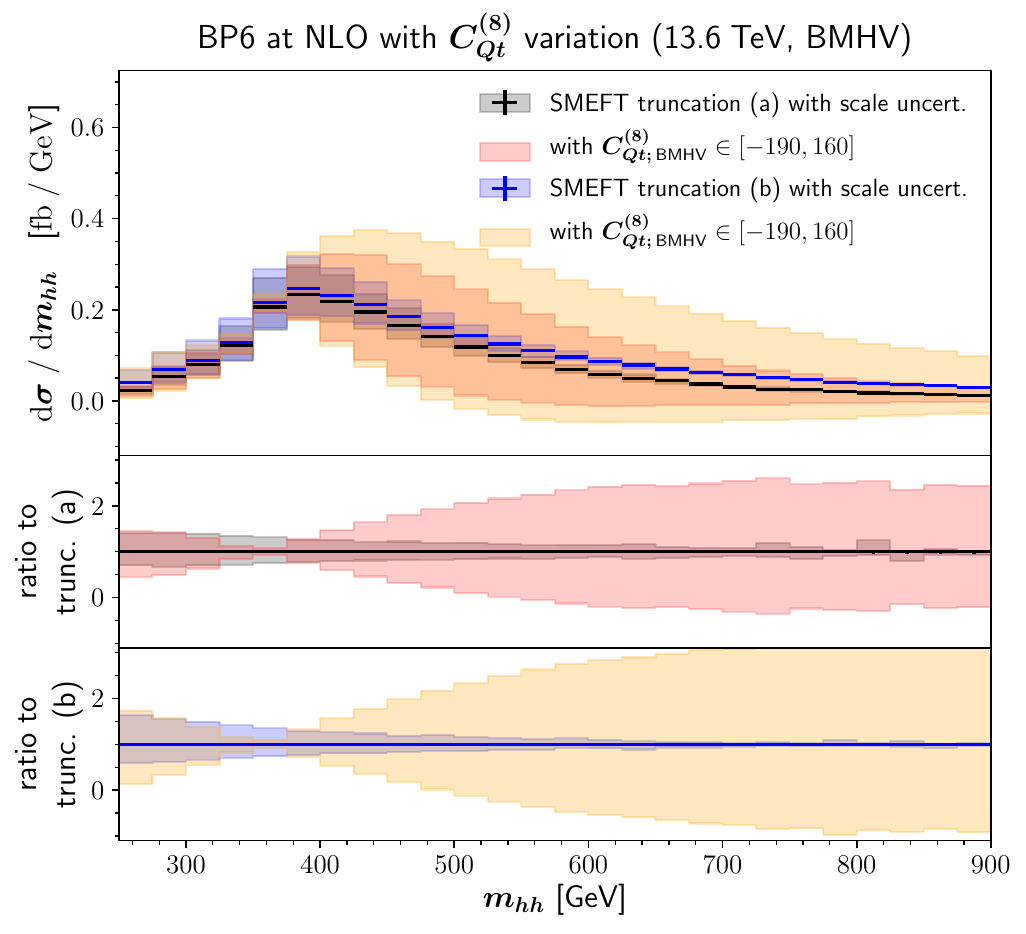}%
\caption{\label{nlo_CLR8_scheme}Effects of $C_{Qt}^{(8)}$-variations on $\mhh$-distributions comparing $\gamma_5$-schemes. 
Left: NDR scheme, right: BMHV scheme; upper: SM baseline scenario, lower:  benchmark point 6 for truncation options (a) and (b).}
\end{center}
\end{figure}

The scheme dependent behaviour  of $C_{Qt}^{(8)}$ is shown in Fig.~\ref{nlo_CLR8_scheme}.
%
For both schemes we observe small effects in the low $m_{hh}$-region, a sign change of the contribution around 
$m_{hh}\sim360$\,TeV and a pronounced effect in the high $m_{hh}$-region.
Overall, the difference between the schemes is not as significant as in the case of $C_{Qt}^{(1)}$.
The contribution to the $\mhh$-distribution in the BMHV scheme (right column of Fig.~\ref{nlo_CLR8_scheme}) is qualitatively very similar to the case of $C_{Qt}^{(1)}$ shown in Fig.~\ref{nlo_CLR_scheme}.

\begin{figure}[htb!]
  \begin{center}
  \includegraphics[width=.47\textwidth,page=1]{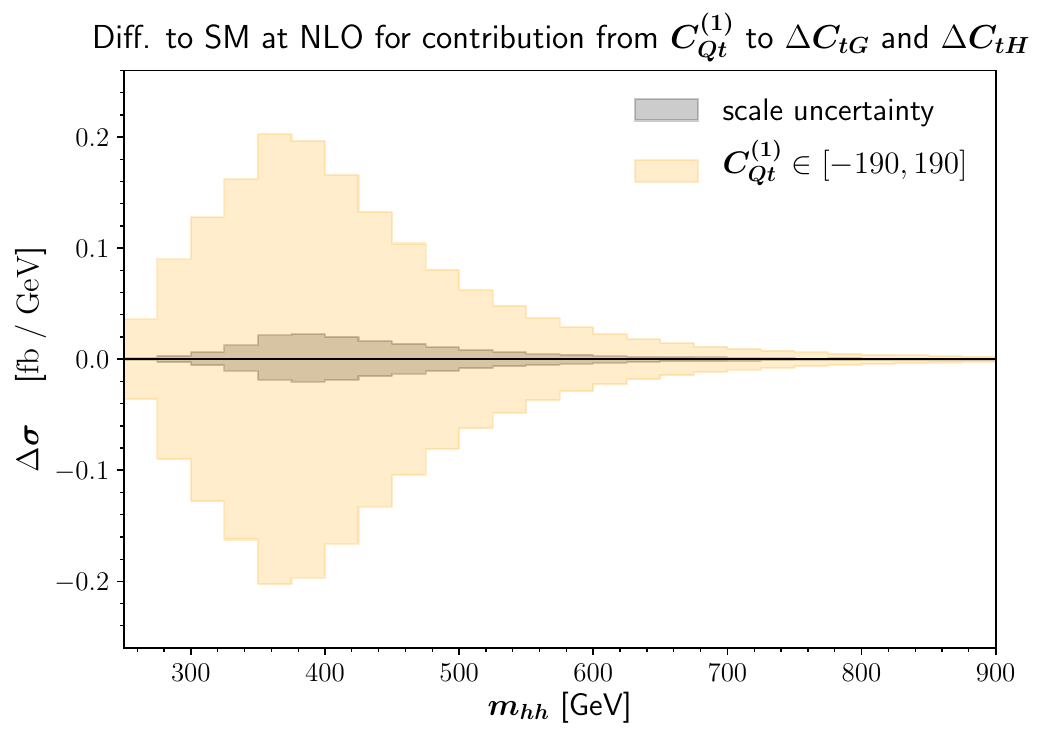}%
  \includegraphics[width=.47\textwidth,page=1]{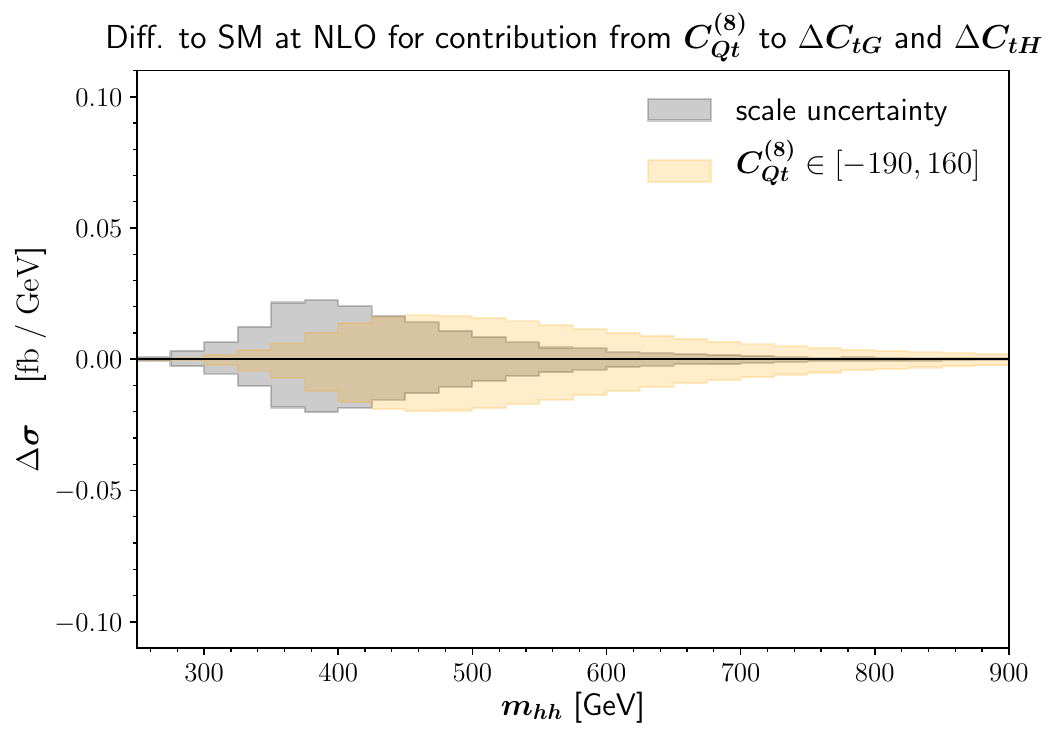}%
  \\
  \includegraphics[width=.47\textwidth,page=1]{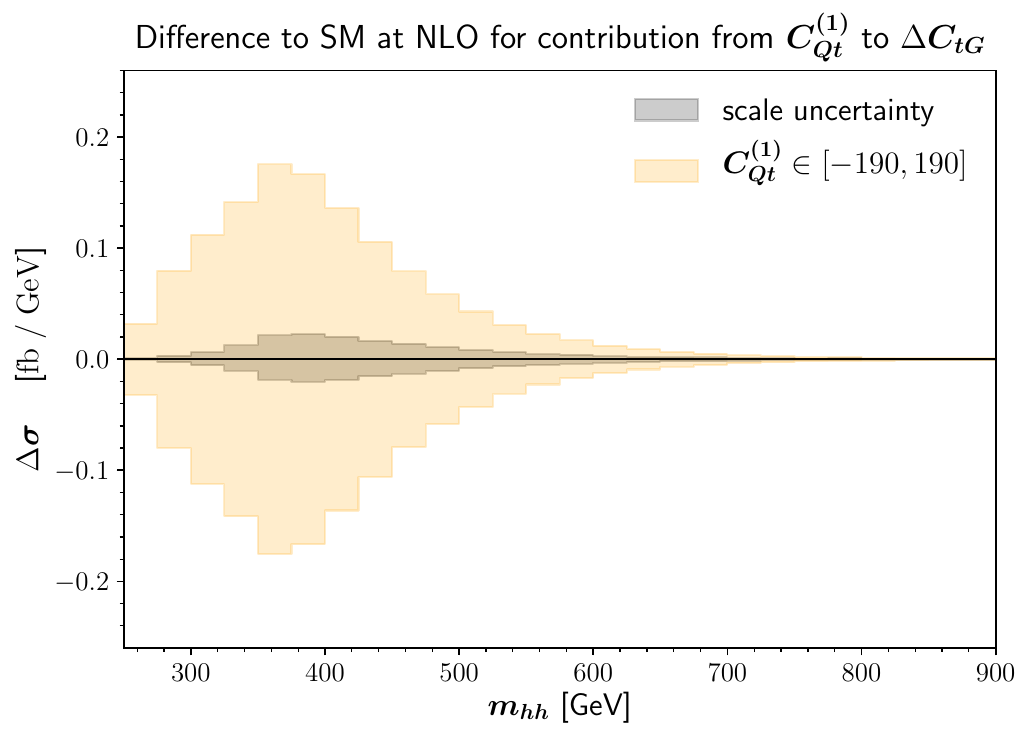}%
  \includegraphics[width=.47\textwidth,page=1]{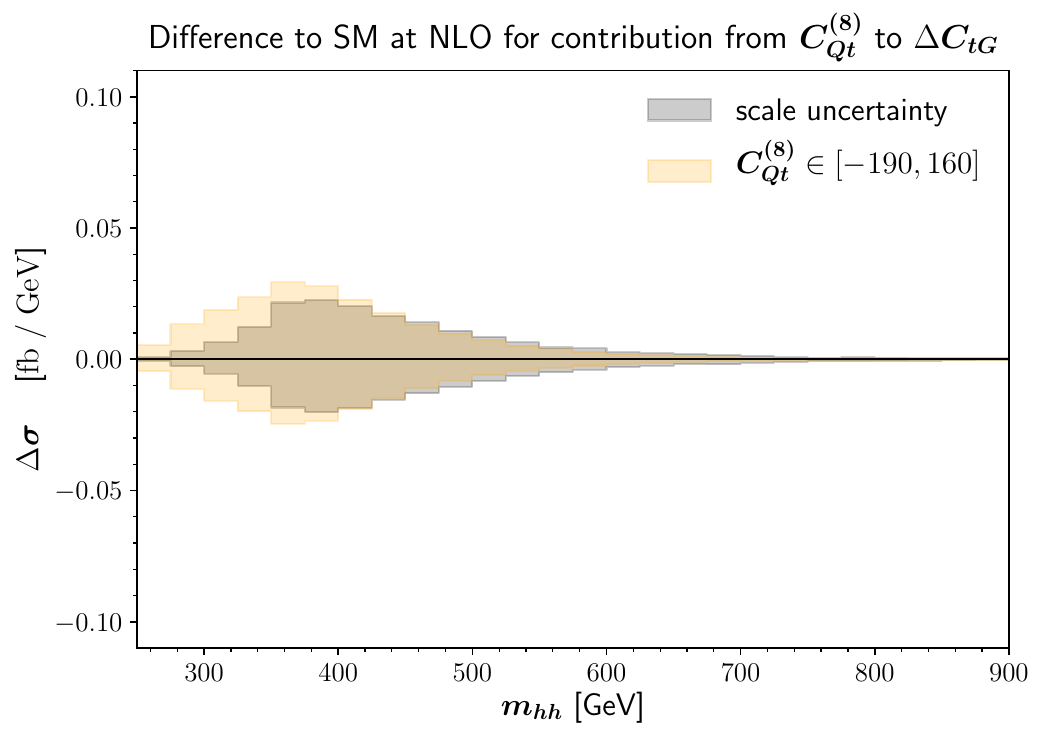}%
  \\
  \includegraphics[width=.47\textwidth,page=1]{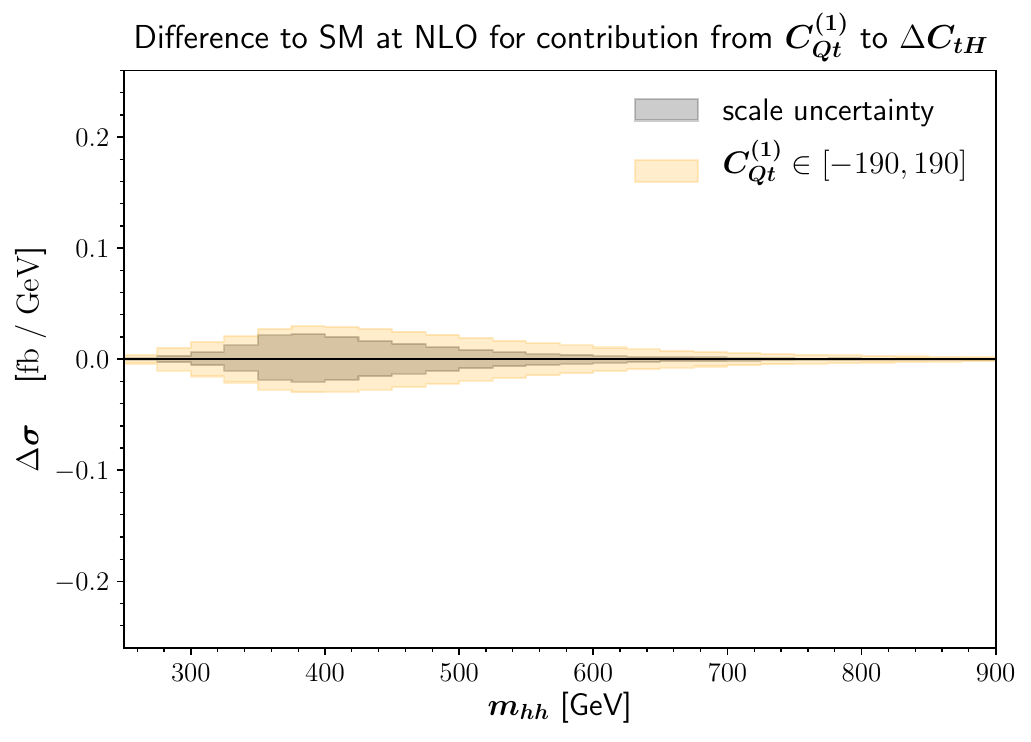}%
  \includegraphics[width=.47\textwidth,page=1]{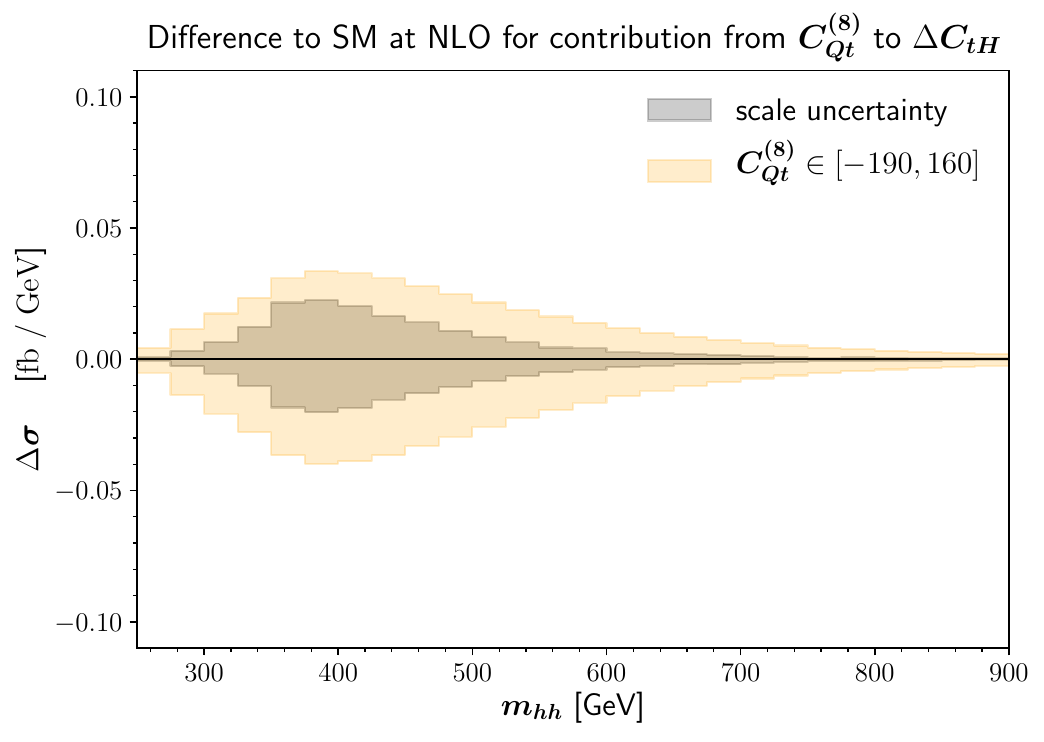}%
  \caption{\label{fig:rationalterms_smshifts}Demonstration of the difference $\Delta\sigma=\frac{d\sigma}{d\mhh}-\frac{d\sigma_\text{SM}}{d\mhh}$ 
      to the SM invariant mass distribution only including
      contributions of the scheme dependent terms, $\Delta C_{tG}:=C_{tG}^\text{BMHV}-C_{tG}$ 
      and $\Delta C_{tH}:=C_{tH}^\text{BMHV}-C_{tH}$, for individual variations of $C_{Qt}^{(1)}$ and $C_{Qt}^{(8)}$, respectively. 
      Left: contribution from a $C_{Qt}^{(1)}$ variation, right:
      contribution from a $C_{Qt}^{(8)}$ variation.
      Upper: sum of scheme dependent terms ($\Delta C_{tG}$ and $\Delta C_{tH}$), middle: only $\Delta C_{tG}$, 
      lower: only $\Delta C_{tH}$. 
      The gray bands denote the SM 3-point scale uncertainty for reference.
     }
  \end{center}
\end{figure}

In order to better understand the qualitative difference between the $C_{Qt}^{(1)}$ and $C_{Qt}^{(8)}$
variations in NDR, we investigate the effect of those rational terms contributing in NDR which
are responsible for the scheme difference and eventually the translation relation Eq.~\eqref{eq:scheme_translation}.
We distinguish in the following between the scheme dependent parts $\Delta C_{tG}:=C_{tG}^\text{BMHV}-C_{tG}$, 
leading to the shift of $C_{tG}$, and $\Delta C_{tH}:=C_{tH}^\text{BMHV}-C_{tH}$, leading to the shift of $C_{tH}$.
In Fig.~\ref{fig:rationalterms_smshifts} we present the difference to
the SM $\mhh$-distribution originating from those scheme dependent terms, where we
individually vary $C_{Qt}^{(1)}$ or $C_{Qt}^{(8)}$, respectively.
Considering all scheme dependent terms, there is a prominent contribution from $C_{Qt}^{(1)}$,
which is much larger than the scale uncertainty of the SM result for the whole $\mhh$-range, especially apparent 
in the low to intermediate $\mhh$-regime.
Investigating the constituents, we notice that  
$\Delta C_{tG}$ is much more relevant than $\Delta C_{tH}$ when considering the contributions from $C_{Qt}^{(1)}$ to the shift. 
Comparing the change on the distribution related to $\Delta C_{tG}$
and $\Delta C_{tH}$ separately (middle and bottom left panels in Fig.~\ref{fig:rationalterms_smshifts}) to the effect of the sum of both contributions (top left panel in Fig.~\ref{fig:rationalterms_smshifts}), 
we observe that the range of the band in the top left panel is given by the
sum of the ranges observed for $\Delta C_{tG}$ and $\Delta C_{tH}$ individually.
For $C_{Qt}^{(8)}$, the structure of the contributions from the scheme
dependent terms is different. Here the effect is larger for the case of $\Delta C_{tH}$  than for $\Delta C_{tG}$, see middle and bottom right panels of Fig.~\ref{fig:rationalterms_smshifts}.
In addition, there is a clear cancellation between  individual
contributions from $\Delta C_{tG}$ and $\Delta C_{tH}$, as can be seen
from the effect on the sum of all rational terms (top right panel of Fig.~\ref{fig:rationalterms_smshifts}),
thus leading to an almost vanishing contribution in the low $\mhh$-region.
Comparing the left and right columns of Fig.~\ref{fig:rationalterms_smshifts}, we observe that the individual shifts due to  $C_{Qt}^{(8)}$ versus $C_{Qt}^{(1)}$ behave quite differently. This difference is related to the different colour structures of the relevant scheme dependent terms. 
On the one hand, the terms contributing to the shift $\Delta C_{tH}$  include a factor of 
$\Delta C_{tH}\sim \left(C_{Qt}^{(1)}+\frac{4}{3}C_{Qt}^{(8)}\right)$ (inserting explicit  $SU(3)_{QCD}$ colour factors),
such that the contribution from $C_{Qt}^{(8)}$ is slightly enhanced.
The terms contributing to the shift $\Delta C_{tG}$, on the other hand, include a factor of 
$\Delta C_{tG}\sim \left(C_{Qt}^{(1)}-\frac{1}{6}C_{Qt}^{(8)}\right)$, thus this effect is larger 
for $C_{Qt}^{(1)}$ and the sign of the contribution from $C_{Qt}^{(8)}$ is opposite to the one from $C_{Qt}^{(1)}$.


We should emphasise again that the observed $\gamma_5$-scheme dependence of individual Wilson coefficients does not lead to a scheme dependence of the full amplitude.  
Both schemes represent equivalent parametrisations of the amplitude and of the renormalisation group flow, the translation has been worked out in Ref.~\cite{DiNoi:2023ygk}.
However, fits to constrain  these Wilson coefficients should take into account that they are not individually scheme-independent. 
For example, constraints on  $C_{tG}$ either come with a scheme uncertainty or should be derived in combination with  $C_{Qt}^{(1)}$ and $C_{Qt}^{(8)}$, calculated in the same scheme.

\FloatBarrier
\subsection{Linear versus linear+quadratic contributions from the chromomagnetic operator}
\label{sec:lvsl+q_chromo}
So far, the results involving the chromomagnetic operator only include its linear contribution, 
as it is classified subleading in the scenario of weakly coupling and renormalisable new physics such that its square would be beyond the order we consider. 
In this subsection, however, we step back from this assumption and assess the effect of the quadratic chromomagnetic contribution at LO QCD.
As in the previous subsections we vary $C_{tG}$ within the ranges from ${\cal O}(\Lambda^{-2})$ marginalised fits of Ref.~\cite{Ethier:2021bye}.\footnote{
  When the linear and quadratic contributions are included, we should in principle consider ${\cal O}\left(\Lambda^{-4}\right)$ constraints from Ref.~\cite{Ethier:2021bye} 
  for the limits of the variation. Nevertheless we use the same bound in both cases for comparability.
}

\begin{figure}[htb]
  \begin{center}
  \includegraphics[width=.47\textwidth,page=1]{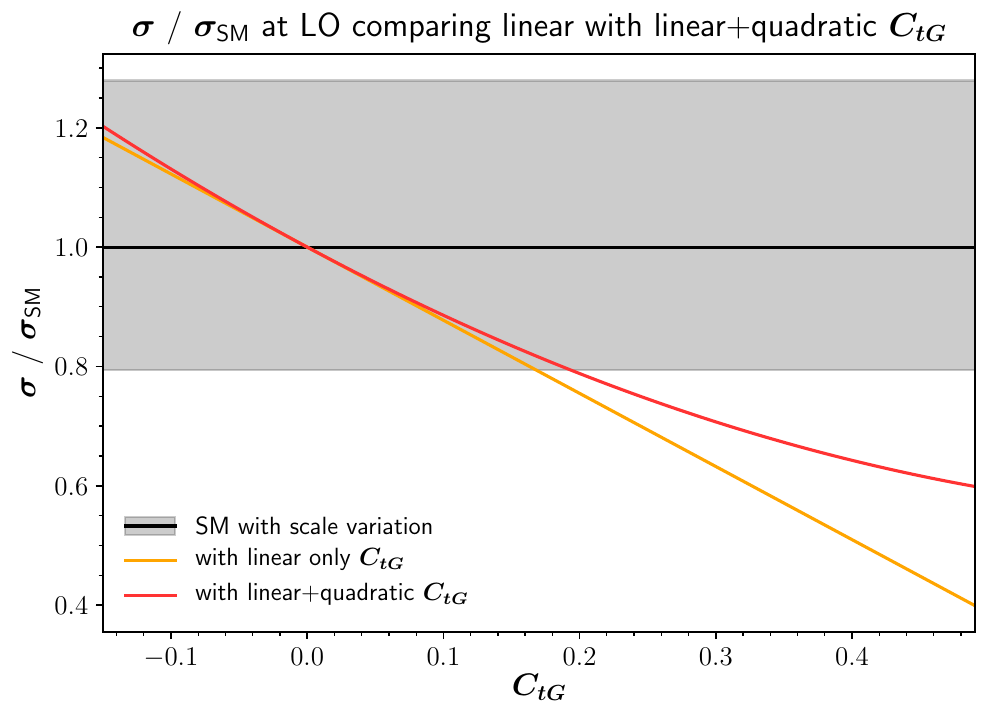}%
  \includegraphics[width=.47\textwidth,page=1]{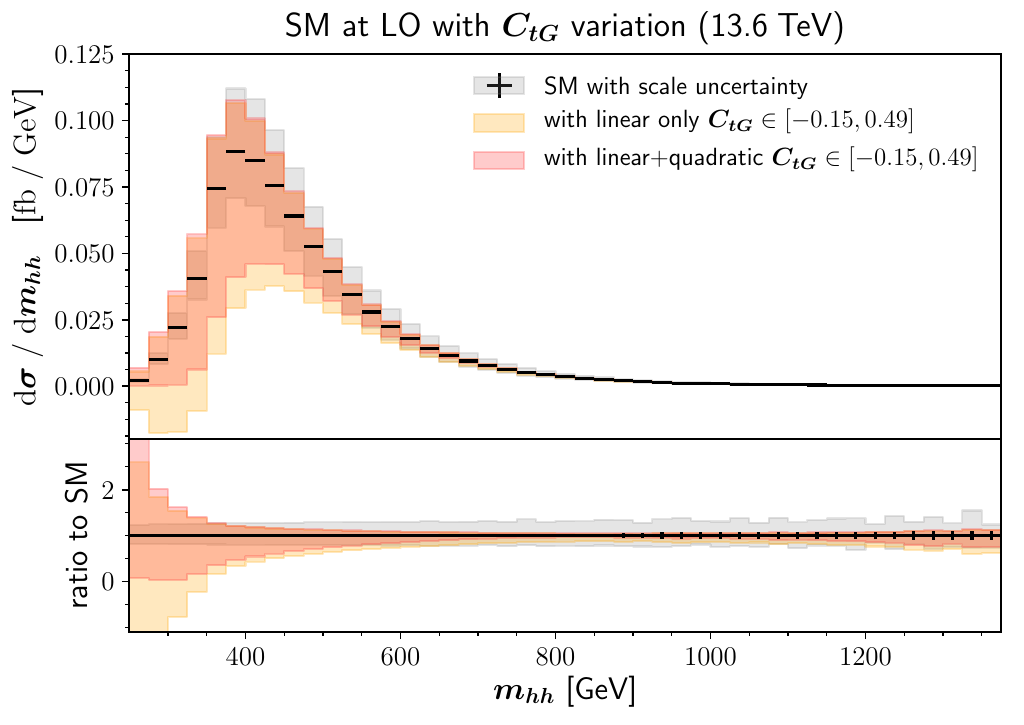}%
      \caption{\label{diagrams_LO_CtG}
      Diagrams comparing linear only with linear+quadratic contribution of $C_{tG}$ (with $\Lambda=1$\,TeV) using variations within the 
      marginalised ${\cal O}\left(\Lambda^{-2}\right)$ constraints from Ref.~\cite{Ethier:2021bye}.  
      (Left) Total cross section normalised to the SM at LO QCD as a function of $C_{tG}$,  
      (right) envelope of $C_{tG}$-variations on the SM $\mhh$-distributions at LO QCD. 
     }
  \end{center}
\end{figure}

In Fig.~\ref{diagrams_LO_CtG} (left) we present the total cross section normalised to the SM value as a function of $C_{tG}$. 
For moderate values of $C_{tG}$ the quadratic contribution is less dominant than the linear when comparing the distance between the two lines. 
Beyond $C_{tG}\sim0.2$ the quadratic piece gets relevant and has an effect of $5\%$ of the SM cross section. 
However, this only reduces the destructive interference of the linear contribution due to the asymmetric range, making the overall difference to the SM smaller. 

Fig.~\ref{diagrams_LO_CtG} (right) demonstrates the SM distribution at LO QCD together with a variation of $C_{tG}$ 
comparing the linear and linear+quadratic insertions. 
Similar to the observations in Fig.~\ref{diagrams_LO_CtG} (left) on the level of the total cross section, 
the quadratic terms are most relevant for the largest values of $C_{tG}$, which however leads to a reduction of the destructive interference with the SM, 
thus reducing the overall effect on the distribution. 
Note that the tails do not yet reach the energy range where,
as predicted by the high energy expansion of the helicity amplitudes in
Ref.~\cite{Rossia:2023hen}, the quadratic contribution takes over.

\section{Conclusions}
\label{sec:conclusions}

We have calculated the matrix elements including the chromomagnetic operator and 4-top operators contributing 
to Higgs boson pair production in gluon fusion and demonstrated that
these operators both appear at the same subleading order in a power counting scheme that takes into account a tree-loop classification of dimension-6 SMEFT operators.
We emphasize again that this classification is based on the generic assumption of a renormalisable and weakly coupling new physics sector, so does not represent all potential UV effects in full generality.
These subleading contributions, entering the cross section at LO QCD, 
have been combined with the leading SMEFT operators including NLO QCD corrections as described in Ref.~\cite{Heinrich:2022idm}, in the form of Eqs.~\eqref{eq:XS_expansion}-\eqref{eq:XSNLO_expansion}.
This combination is provided as an extension to the public {\tt ggHH\_SMEFT} code as
part of the {\tt POWHEG-Box-V2}. We have also described the usage of the new features.

The matrix elements of the 4-top contributions have been decomposed analogous to the case of $gg\to h$
described in Refs.~\cite{Alasfar:2022zyr,DiNoi:2023ygk}. In particular, the parts depending on the $\gamma_5$-scheme 
in dimensional regularisation have been identified, such that we
found a similar scheme dependence as in the $gg\to h$ case, which can be understood as a finite shift of Wilson coefficients, 
see Eq.~\eqref{eq:scheme_translation} and Ref.~\cite{DiNoi:2023ygk}.  

The effect of the subleading operators on the total cross section and on the Higgs boson pair invariant mass distribution has been studied in detail, 
both with respect to the SM and for benchmark point 6.
We observed that the operators $\mathcal{O}_{QQ}^{(1)}$, $\mathcal{O}_{tt}$ and $\mathcal{O}_{QQ}^{(8)}$ only marginally
contribute, therefore $gg\to hh$ is not an adequate process to probe those coefficients.
The cross section is noticeably affected by a variation of the Wilson coefficient $C_{tG}$ within current conservative bounds, 
which can lead to a damping of the invariant mass distribution in the low to intermediate $\mhh$-region.
However, the highest sensitivity is observed by a variation of $C_{Qt}^{(1)}$ and $C_{Qt}^{(8)}$ within current bounds.
Since the  limits on the 4-top Wilson coefficients from marginalised fits are very loosely constrained so far, 
the inclusion of processes like $gg\to h$ and $gg\to hh$, where the operators enter at higher orders,
could potentially improve the global determination of bounds on $C_{Qt}^{(1)}$ and $C_{Qt}^{(8)}$.

As has been investigated for single Higgs production in Ref.~\cite{DiNoi:2023ygk} and confirmed in this work, contributions of those Wilson coefficients
are precisely the ones which, when considered individually, depend on the chosen $\gamma_5$-scheme. 
Therefore, bounds for individual coefficients  can turn out to be significantly different due to a (more or less arbitrary) calculational scheme choice, 
which makes their interpretation difficult. 
In general, this scheme dependence enters as soon as the calculation is performed at an order 
at which loop contributions of such chiral current-current operators are to be considered. 
This statement is of particular relevance when the effects of these chiral current-current operators are investigated in processes where they only enter at loop level, 
as in this case contributions of Wilson coefficients entering at lower loop order are necessary to resolve $\gamma_5$-scheme ambiguities.   
Considering a tree-loop classification of Wilson coefficients, the requirement for $\gamma_5$-scheme independence has even stronger impications: 
In case of a clear hierarchy, e.g. by the loop suppression of the shift translating $C_{tH}$ in Eq.\eqref{eq:scheme_translation}, the shift would only be a higher order effect. 
For loop-induced Wilson coefficients, this would however not be the case, as the shifts can be of the same order in the power counting. 
This holds, for example,
for the Wilson coefficient $C_{tG}$, which, at the same order in the power counting, can contain
a contribution from $C_{Qt}^{(1)}$ and $C_{Qt}^{(8)}$, depending on the scheme choice.
Inserting numerical values for current  bounds on these Wilson coefficients~\cite{Ethier:2021bye} into Eq.~\eqref{eq:scheme_translation} illustrates 
that the shift induced by a scheme change can even be larger than the interval given by the original bounds.
To obtain more meaningful results, it is therefore recommended  to study  those
Wilson coefficients which are connected through the scheme translation relations together,
such that their combination is a scheme independent parametrisation of BSM physics at the studied order in the power counting.


In the future it would be desirable to have QCD corrections 
to those subleading operators as well, in order to compare on equal footing with the leading operators, at NLO QCD.
However, including NLO corrections to the 4-top operators would require a 3-loop calculation involving Higgs and top-quark masses. 
The two-loop contributions to the chromomagnetic operator would be more feasible, but also challenging due to the high tensor rank induced by this operator. 
Therefore these calculations would be clearly beyond the scope of this paper.
Furthermore, operators of the class $\psi^2\phi^2D$ have not been considered in this work, even though they would enter at the same power counting order, because they are considered as electroweak-type.
However, this indicates that the strict separation between QCD and electroweak contributions becomes ambiguous once SMEFT operators  beyond the leading contributions are included and combined with higher order corrections.

Finally, we note that renormalisation group running effects have not been included in the present study, even though they may lead to sizeable effects.
This is left to upcoming work.

\section*{Acknowledgements}

We would like to thank Stephen Jones, Matthias Kerner and Ludovic Scyboz for
collaboration related to the $ggHH$@NLO project and Gerhard Buchalla,
Stefano Di Noi, Ramona Gr\"ober, Christoph M\"uller-Salditt, Michael Trott and Marco Vitti for useful discussions. 
This research was supported by the Deutsche Forschungsgemeinschaft (DFG, German Research Foundation) under grant 396021762 - TRR 257.


\bibliographystyle{JHEP}
\bibliography{main_4top}

\end{document}